\documentclass[preprint,12pt,3p]{elsarticle}

\pdfoutput=1
\usepackage{xspace}
\usepackage{color}
\usepackage{epstopdf}
\usepackage{hyperref}
\usepackage{colortbl}
\usepackage{arydshln}
\usepackage{braket}
\usepackage{url}     
\usepackage{color}\definecolor{darkblue}{rgb}{0,0,1}
\hypersetup{
    unicode=false,          
    pdftoolbar=true,        
    pdfmenubar=true,        
    pdffitwindow=true,      
    pdftitle={My title},    
    pdfauthor={Author},     
    pdfsubject={Subject},   
    pdfnewwindow=true,      
    pdfkeywords={keywords}, 
    colorlinks=true,       
    linkcolor=darkblue,          
    citecolor=darkblue,        
    filecolor=magenta,      
    urlcolor=cyan           
}

\usepackage{graphicx}
\graphicspath{ {Figures/} }
\usepackage{tablefootnote}

\usepackage{amssymb}

\usepackage{tikz}
\usepackage{lineno}
\usepackage{comment}
\usepackage{booktabs}
\usepackage{mathrsfs}
\usepackage{amsmath}
\usepackage{verbatim}
\usepackage{siunitx}
\journal{JINST}

\newcommand{\three}{\SI[product-units=power]{3x1x1}{\meter}\xspace}
\newcommand{\units}[2]{\ensuremath{\mbox{#1}^{#2}}}
\newcommand{\fifty}{\SI[product-units=power]{50x50}{\cm}\xspace}

\newcommand{\Geff}{\ensuremath{G_{eff}}\xspace}

\newcommand{\Geffz}{\ensuremath{G_{eff}^0}\xspace}
\newcommand{\Geffinfty}{\ensuremath{G_{eff}^\infty}\xspace}

\newcommand{\figref}[1]{\figurename~\ref{#1}}
\newcommand{\secref}[1]{Section~\ref{#1}}

\newcommand{\tabref}[1]{Table~\ref{#1}}

\begin{document}
\begin{frontmatter}

\title{A 4 tonne demonstrator for large-scale dual-phase liquid argon time projection chambers}

\author[i]{B.\,Aimard}
\author[a]{Ch.\,Alt}
\author[u]{J.\,Asaadi}
\author[b]{M.\,Auger}
\author[w]{V.\,Aushev}
\author[c]{D.\,Autiero}
\author[r]{M.M.\,Badoi}
\author[r]{A.\,Balaceanu}
\author[i]{G.\,Balik}
\author[c]{L.\,Balleyguier}
% ...
\address[b]{University of Bern, Albert Einstein Center for Fundamental Physics, Laboratory for High Energy Physics (LHEP), Bern, Switzerland}
% ...
\address[a]{ETH Zurich, Institute for Particle Physics, Zurich, Switzerland}
% ...
\address[r]{Horia Hulubei National Institute of R\&D for Physics and Nuclear Engineering - IFIN-HH, Magurele, Romania}
% ...
\address[w]{Kyiv National University, Kyiv, Ukraine}
\address[i]{LAPP, Universit\'e de Savoie, CNRS/IN2P3, Annecy-le-Vieux, France}
% ...
\address[c]{Universit\'e de Lyon, Universit\'e Claude Bernard Lyon 1, IPN Lyon (IN2P3), Villeurbanne, France}
% ...
\address[u]{University of Texas Arlington, Arlington, USA}

\fntext[nowcern]{Now at CERN}





\begin{abstract}

A 10 kilo-tonne dual-phase liquid argon TPC is one of the detector options considered for the Deep Underground Neutrino Experiment (DUNE). The detector technology relies on amplification of the ionisation charge in ultra-pure argon vapour and offers several advantages compared to the traditional single-phase liquid argon TPCs. A 4.2 tonne dual-phase liquid argon TPC prototype, the largest of its kind, with an active volume of \three has been constructed and operated at CERN. In this paper we describe in detail the experimental setup and detector components as well as report on the operation experience. We also present the first results on the achieved charge amplification, prompt scintillation and electroluminescence detection, and purity of the liquid argon from analyses of a collected sample of cosmic ray muons. 


\end{abstract}

\begin{keyword}
Neutrinos, liquid argon TPC, tracking, dual-phase, DUNE
\end{keyword}

  \end{frontmatter}

\tableofcontents

\section{Introduction}
\label{sec_intro}
Liquid Argon Time Projection Chambers (LAr TPCs) \cite{Rubbia:1977zz} have been under development for many decades. They provide a dense target medium and offer both unprecedented 3D imaging capabilities and the functionality of a homogeneous calorimeter. Thus giant LAr TPCs at the multi-ktonne scale represent unique apparatuses to detect and study neutrinos and other rare phenomena such as proton decay where high-resolution imaging is the key to efficient background rejection.
The Deep Underground Neutrino Experiment (DUNE) envisages to deploy four 10 ktonne LAr TPC modules in the Sanford Underground Research Facility in South Dakota, USA \cite{Acciarri:2016ooe}. Using a powerful neutrino beam originating at Fermilab, DUNE aims to search for the leptonic CP violation and determine the ordering of the neutrino masses. In addition, the experiment plans to extend the sensitivity to the proton lifetime in a variety of possible decay channels and collect high statistics neutrino samples from atmospheric and astronomical sources.

Numerous R\&D efforts throughout the world have been focused at studying the feasibility to construct and operate giant LAr TPCs underground. One proposed solution consists on a single-phase liquid argon detector readout by wires immersed in the liquid. This solution, similar to the one implemented in other previous experiments \cite{Amerio:2004ze,Anderson:2012mra,Fleming:2012gvl}, does not require the amplification of the ionization electron charge. The charge is localized using alternative induction and collection wire planes with different orientations. This technology approach is being tested at CERN by ProtoDune-SP\cite{Abi:2017aow}.

An alternative solution has been developed within the LAGUNA-LBNO European design study for a dual-phase LAr TPC with a mass between 20 and 50 ktonne \cite{1742-6596-171-1-012020,Stahl:2012exa} and modular detector readout stages to facilitate their deployment underground. The dual-phase LAr TPC concept relies on the amplification of ionisation charges in the ultra-pure cold argon vapour layer above the liquid, to obtain low-energy detection thresholds with high signal-to-noise ratio over drift distances exceeding 10 meters. Chambers with active volumes varying in scale from 3 litres to 250 litres have been constructed and successfully operated \cite{Badertscher:2008rf, Badertscher:2009av, Badertscher:2010zg,Badertscher:2013wm,devis-thesis,filippo-thesis}. More recent efforts have been targeting the optimisation of the charge readout, both in terms of performance and cost, to produce viable solutions for TPCs with $\mathcal{O}(\si{\meter\squared})$ active readout areas \cite{Cantini:2013yba,shuoxing-thesis}. The concepts developed within the LAGUNA-LBNO design study and the surrounding R\&D activities culminated in a proposal for the large-scale prototype of a dual-phase LAr TPC \cite{DeBonis:1692375}. This detector, presently considered as a prototype for a DUNE module (ProtoDUNE-DP), consisting of a \SI[product-units=power]{6x6x6}{\meter} (\SI{300}{\tonne}) active volume dual-phase LAr TPC, is currently under construction at CERN. To validate a new commercial technology for a non-evacuated industrial cryostat, test several key sub-systems for ProtoDUNE-DP, and demonstrate for the first time the capabilities of the dual-phase LAr TPC technology on a tonne scale, a prototype detector with an active volume of \three (\SI{4.2}{\tonne}) was built in 2016 and operated in 2017 at CERN. This publication covers the design of this detector as well as its construction, operation, and performance.

The manuscript is organised as follows. An overview of the experimental apparatus with emphasis on the dual-phase readout principle is given in \secref{sec_overview}. In \secref{sec_cryostat-cryo} the cryostat and cryogenic system are described. \secref{sec_tpc} presents the detailed description of the TPC elements and discusses some of the quality assurance steps during construction. \secref{sec_sc} describes the auxiliary instrumentation and the detector slow control system. The charge and light readout, analogue and digital electronics, data acquisition, as well as the online storage and processing are discussed in \secref{sec_elec}. Finally \secref{sec_operations} shows some of the initial results from the detector commissioning and operation. 


\section{Overview of the experimental apparatus}
\label{sec_overview}
\subsection{Experimental setup}

 \begin{figure}[ht!]
  \centering
  \includegraphics[width=.9\textwidth]{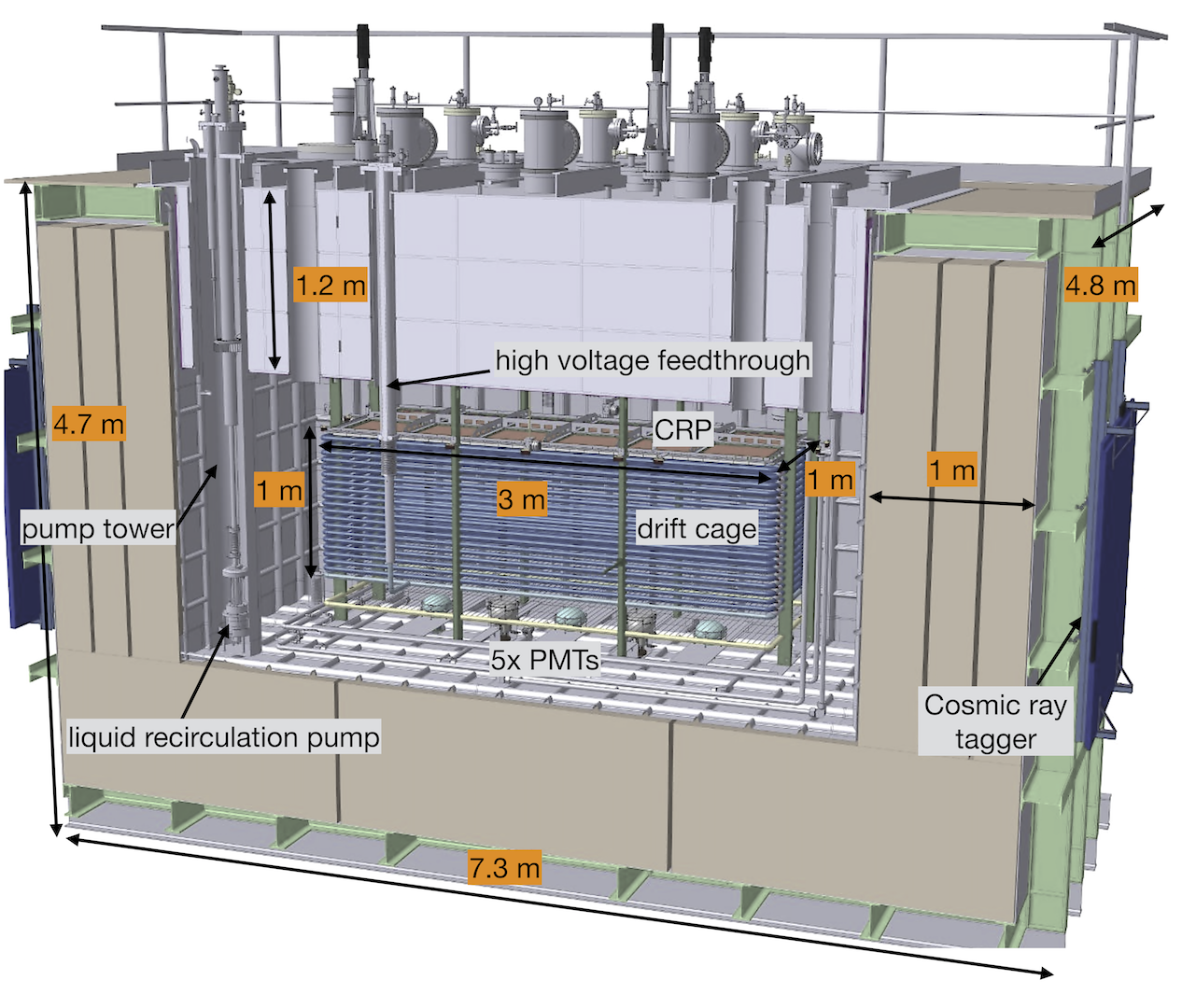}
  \caption{Drawing of the \three dual-phase LAr TPC in the cryostat.}
     \label{fig:311-full} 
 \end{figure}

\begin{table}[ht!]
\begin{center}
\begin{tabular}{ll}
\hline\hline
Parameter & Value \\
\hline
TPC fiducial mass & \SI{4.2}{\tonne} \\
TPC dimensions ($\mbox{l}\times\mbox{w}\times\mbox{h}$) & \SI[product-units=power]{3x1x1}{\meter} \\
Readout area coverage & \SI[product-units=power]{3x1}{\meter} \\
Area of basic readout unit & \SI[product-units=power]{0.5x0.5}{\meter} \\
Number of charge readout channels & 1280 \\
Number of light readout channels & 5 \\
\hline\hline
\end{tabular}
\caption{Principal parameters of the detector.}
\label{tab:detector-spec}
\end{center}
\end{table}

\begin{figure}[ht!]
  \centering
  \includegraphics[width=\textwidth]{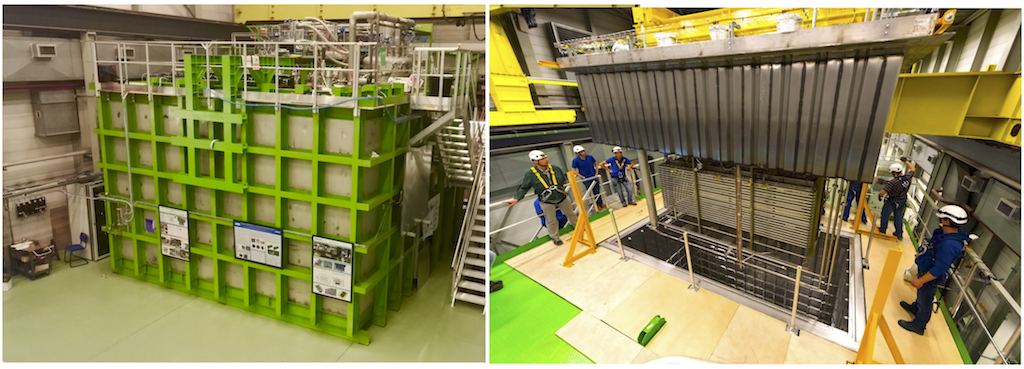}
  \caption{Pictures of the experimental setup. Left: overview of the cryostat. Right: the TPC suspended under the top-cap during insertion in the cryostat. }
     \label{fig:311-setup-pics} 
   \end{figure}

The experimental apparatus is illustrated in \figref{fig:311-full}; some pictures are also provided in \figref{fig:311-setup-pics}. It consists of a \three (\SI{4.2}{\tonne}) active volume dual-phase LAr TPC inside a  passively insulated cryostat with internal volume of \SI{\sim23}{\cubic\meter}. The principal parameters of the TPC are given in \tabref{tab:detector-spec}. The detector is suspended under a \SI{1.2}{m} thick insulating lid called \textit{top-cap}. The top-cap is part of the cryostat structure providing the functionality of reducing heat input and minimising the liquid and gas argon convection inside the tank. It hosts the necessary feedthroughs as well as the interfaces to the cryogenic system. The TPC was pre-assembled under the top-cap in a custom built clean room ``tent" and then inserted in the cryostat, see \figref{fig:311-setup-pics}. Subsequent access to the detector inside the cryostat is performed through the top-cap via a \SI{600}{\mm} diameter manhole. 
Given the size of the \SI[product-units=power]{3x1x1}{\meter} prototype, the top cap solution was adopted for its closure. For larger detectors such as ProtoDUNE-DP with lateral access to the inner volume, it is possible to build a cryostat as a single structure with uniform membrane coverage including the roof. 
%

Ionisation charges drift vertically towards the liquid-vapour boundary where they are extracted into the gas phase, amplified by Large Electron Multipliers (LEMs) \cite{Badertscher:2008rf}, and collected on finely segmented anodes. The electron extraction, amplification, and collection are performed inside a \SI[product-units=power]{3x1}{\meter} frame called charge readout plane (CRP). The CRP is electrically and mechanically independent from the drift cage and can be remotely adjusted to the liquid level. 

Five photo-multiplier tubes (PMTs) are mounted underneath the TPC drift cage. They detect the scintillation light from argon excimer states formed by charged particles crossing the liquid volume (primary scintillation, S1), as well as the secondary scintillation light (S2) from the electroluminescence of the electrons extracted in the argon vapour \cite{Monteiro:2008zz}. A wavelength shifter, 1,1,4,4-Tetraphenyl-1,3-butadiene (TPB) \cite{Burton:73}, is used to convert the deep ultra violet photons of the argon scintillation peaked at \SI{128}{\nano\meter} to the visible spectrum within the sensitivity of the PMT photo-cathode. The PMT signals give the absolute time on an event, $T_0$, and provide a trigger for the data acquisition system. The amount of detected light can also be useful for calorimetric measurements of the deposited energy. 
 
\subsection{Key concepts of dual-phase readout}
\label{ssec_overview_dpconcept}
\begin{figure}[ht!]
  \centering
  \includegraphics[width=\textwidth]{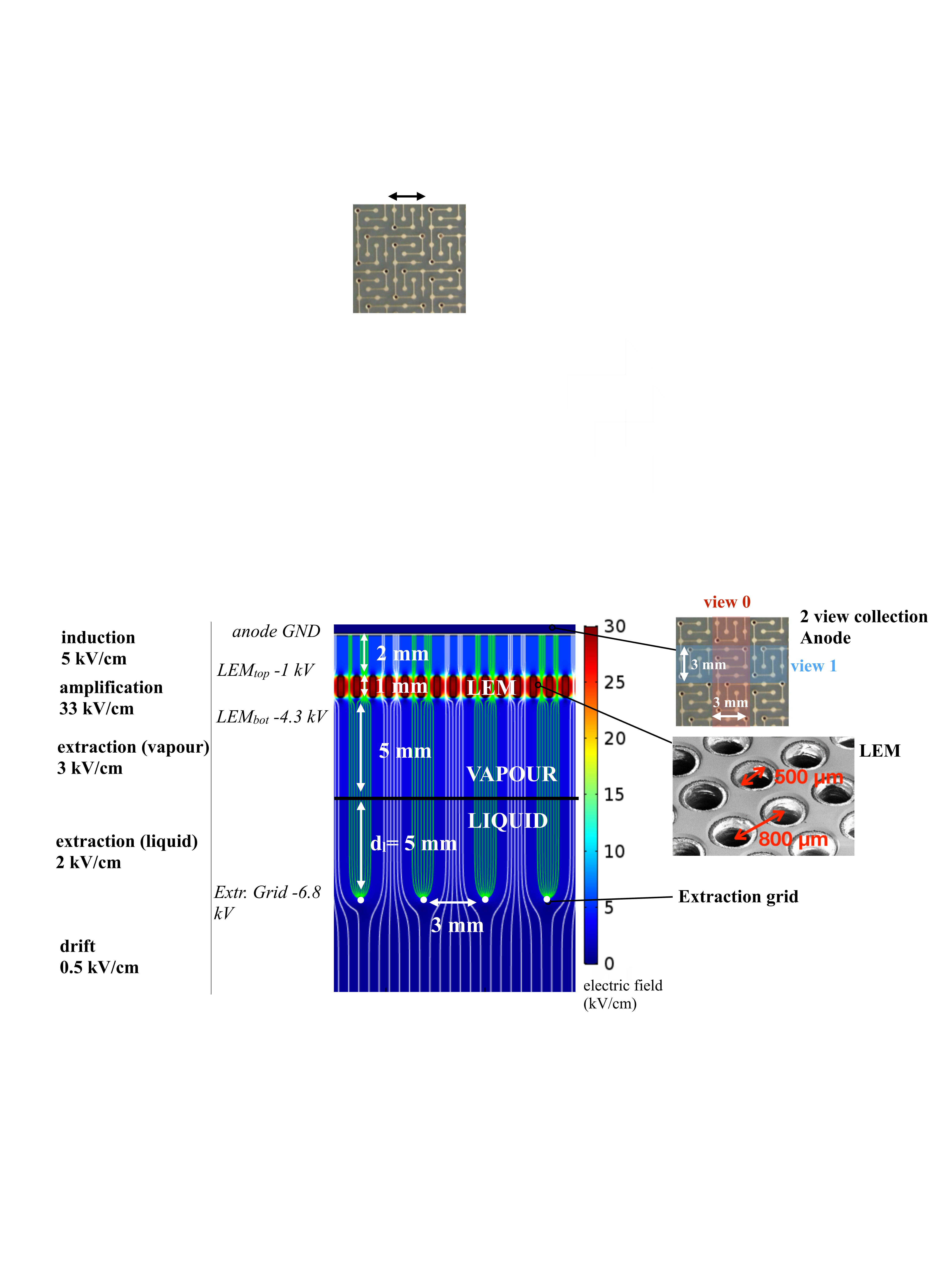}
  \caption{Left: illustration of the extraction, amplification and readout regions in a dual-phase LAr TPC. The simulated field lines in white are an indication of those followed by the drifting charges (without diffusion). The quoted electric fields (in bold) and electrode potentials (in italic) correspond to a \SI[product-units=power]{10x10}{\centi\meter} TPC operation at a stable effective gain of around 20 \cite{Cantini:2013yba}.}
     \label{fig:DP-principle} 
   \end{figure}
The principle of the extraction, amplification, and collection of the ionisation charge in a dual-phase TPC is schematically illustrated in \figref{fig:DP-principle}. The electrons are efficiently extracted from the liquid to the vapour by applying an electric field in the liquid above \SI{2}{\kilo\volt/\cm} \cite{Gushchin:1982}. This electric field is provided by a \SI{3}{\mm} pitch extraction grid positioned \SI{5}{mm} below the liquid argon surface. Once amplified inside the LEM holes, the charge is collected on a two-dimensional segmented anode which consists of a set of independent strips that provide the $x$ and $y$ coordinates of an event with a \SI{\sim3}{\mm} pitch. The LEMs and anodes have been developed towards unit sizes of \fifty. They are precisely assembled side by side to cover the \SI[product-units=power]{3x1}{\meter} area. The anodes are electrically bridged together so as to provide two orthogonal sets of three and one metre long readout strips called \textit{views}. 

The LEMs, also called Thick Gaseous Electron Multipliers (Th-GEMs), are one millimetre thick copper cladded epoxy plates with mechanically drilled holes of \SI{500}{\micro\meter} diameter spaced \SI{800}{\micro\meter} apart in a honeycomb pattern. They are robust detectors that have been demonstrated to work in cryogenic conditions (see for instance \cite{Badertscher:2008rf,Badertscher:2010fi}) and can be economically manufactured in the Printed Circuit Board (PCB) industry. Each \fifty LEM is independently biased with its own high voltage supply. An applied electric field of around \SI{30}{\kilo\volt/\cm} between the top and the bottom copper electrode leads to the multiplication of electrons inside the holes via Townsend avalanche. The UV photons, also produced in the avalanches, are largely confined inside the holes making LEMs suitable for the operation in pure argon vapour without the use of a quencher gas. 

 
 The performance of the extraction, amplification, and collection stage is characterised by a parameter called the effective gain, \Geff, which takes into account the multiplication of the electrons inside the LEM holes as well as the overall electron transparency of the extraction grid, LEM, and anode. From repeated operations of a prototype dual-phase TPC equipped with a \SI[product-units=power]{10x10}{\cm} readout and exposed to cosmic rays, it was observed that the effective gain of the chamber, $G_{eff}(t)$, relaxes from an initial value of \Geffz to a stable value of $\Geffinfty\approx\Geffz/3$, after a characteristic time of about 1.5 days for values of \Geffz near 100 \cite{Cantini:2014xza,shuoxing-thesis}. This relaxation is interpreted as arising from the charging up of the dielectric inside the LEM holes. Maximal effective gains of \Geffz, of around 150 were also achieved by polarising the LEM at \SI{35}{\kilo\volt/\cm}. The high voltage settings, referred to as \textit{nominal}, necessary to operate the \three TPC with CRP $\Geffinfty\approx20$ and the drift field strength of \SI{0.5}{\kilo\volt/\cm} are shown in \figref{fig:DP-principle}.
 
Another key feature of the dual-phase TPC is the ability to operate with two finely segmented \textit{collection views}. The amplification inside the LEMs provides a sufficient number of electrons on the readout for a splitting of the signal on both views. Furthermore it allows to construct readouts with strips of only \SI{\sim3}{\mm} pitch, thereby providing a remarkably high-resolution image of the event. The anodes, made from standard multilayer PCBs, are carefully designed to minimise the capacitance per unit length of one strip to ground (dC$_{det}$/d$l$) while ensuring that charge is evenly and symmetrically shared between the two views. More precisely, the dC$_{det}$/d$l$ quantity is measured for a given anode strip while the other strips are grounded. Tests have shown that it is predominantly determined by the capacitive couplings between strips. The influence of neighbouring electrodes, such as for instance the LEMs, is negligible. The anode PCB design, described in \cite{Cantini:2013yba}, features a dC$_{det}$/d$l$ within \SI{160}{\pico\farad/\m} and hence provides a total capacitance at the input of the preamplifiers of about \SI{480}{\pico\farad} for 3 meter long strips. In previous dual-phase TPCs \cite{Badertscher:2013wm,devis-thesis} such magnitude of the preamplifier input capacitance led to the measured Equivalent Noise Charge (ENC) of about \num{1400} electrons.

The operation with gain combined with the charge readout using two collection views are the key ingredients towards efficient hit finding and high performance imaging and constitutes a unique property of a dual-phase TPC.


The stages of the CRP (extraction grid, LEMs, anodes) are maintained at given electric potentials. These voltages, the electrode geometry, and the relative distance between pairs of electrodes (referred to as \textit{inter-stage distances}) define the three fields (extraction, amplification, and induction) necessary for effective charge extraction from the liquid, its amplification, and signal induction on the anode strips. In addition, the liquid argon level between the grid and the LEM has some influence on the strength of the extraction field. Below we outline some considerations concerning the impact each field has on the CRP effective gain in order to quantify the required precision for the inter-stage distances and the knowledge of the liquid argon level.

\begin{itemize}
\item \textit{Extraction of the charge from the liquid.} The value of the extraction electric field in the liquid, $E_{extr}$, is determined by the difference in potential applied between the grid and the bottom LEM electrodes ($\Delta V_{grid-LEM_{bot}}$) as well as the height of the liquid argon above the grid $d_l$:
\begin{equation}
E_{extr}=\frac{\Delta V_{grid-LEM_{bot}}}{d_l \left(1-\frac{\epsilon_l}{\epsilon_g}\right)+ D \frac{\epsilon_l}{\epsilon_g}}
\end{equation}
where $D$ is the distance between the bottom LEM electrodes and the grid (nominally \SI{10}{\mm}), and $\epsilon_l$ and $\epsilon_g$ are the dielectric constants of liquid and gas argon, respectively. In \figref{fig:extraction_vs_level}, $E_{extr}$ is plotted as a function of $d_l$ for various potential difference values applied between the grid and the bottom LEM electrodes. From past publications \cite{Gushchin:1982} and more recent experience \cite{filippo-thesis,shuoxing-thesis}, it is known that electrons are extracted from the liquid to the gas phase with an efficiency better than $\sim$90\%  for $E_{extr} \geq2$ kV/cm. From the figure we see that a $\Delta V_{grid-LEM_{bot}}\geq 2.5$ kV guarantees an electric field in the liquid of at least 2 kV/cm when the LAr level is 5 mm above the grid. If the liquid level decreases the extraction efficiency nevertheless remains above 80\%. Therefore applying a potential difference between the LEM bottom electrode  and the grid of 2.5 kV ensures a high extraction efficiency independently of the liquid level fluctuations. Our requirements on the liquid position are instead driven by the boundary conditions that on one hand, the LEMs should not be immersed and on the other hand, the extraction grid remains inside the liquid. Fluctuations of the liquid level (due for instance to temperature, pressure, cryostat rigidity, etc.)  should therefore be controlled within a few millimetres to satisfy the above requirement.

 \begin{figure}[ht!]
 \centering
 \includegraphics[width=.68\textwidth]{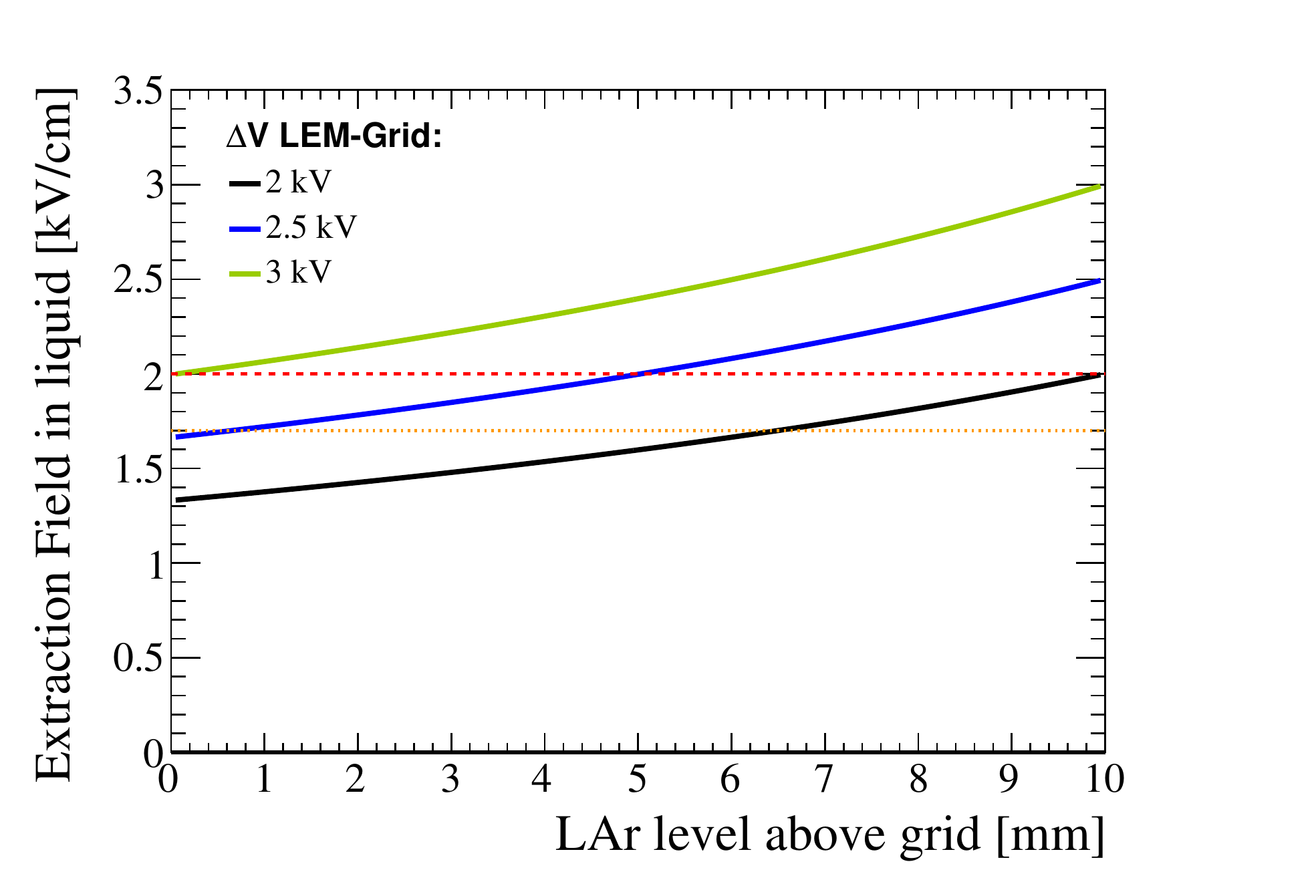}
  \caption{Extraction field in the liquid as a function of the liquid argon level above the extraction grid for different voltages applied across the grid-LEM planes. The orange and red dotted lines indicate, respectively, 80\% and 90\% electron extraction efficiency according to Ref.\cite{Gushchin:1982}.}
 \label{fig:extraction_vs_level} 
\end{figure}
    \item \textit{Amplification inside the LEMs.} The intrinsic LEM parameters that affect the amplification of the charge are the diameter of the holes,  the size of the dielectric rim (a small region at the periphery of the hole where copper is removed to reduce discharges), the geometrical arrangement of the holes and the thickness of the insulator. Their impact on the value and stability of the effective gain has been measured in \cite{Cantini:2014xza} and the parameters have been optimised accordingly. The thickness of the PCB dielectric has the largest impact, since it determines the distance over which the Townsend avalanche develops inside the hole. To preserve the uniformity of the gain, we require tolerances at the few percent level on the aforementioned parameters that are agreed with the supplier and verified upon reception of the LEMs. The manufacturing procedure of the LEMs as well as its quality control therefore constitutes a fundamental aspect of the detector installation. Both aspects are discussed in detail in \secref{sec_tpc}.
    
    \item \textit{Charge collection on the anode.} The electric field between the LEM and the anode, and hence their relative spacing, also plays a role on the value of the effective gain. As shown in \cite{shuoxing-thesis,filippo-thesis}, the effective gain exhibits a linear rise with the collection field mainly due the increase in focusing of the electrons on the anode. The distance between the LEM and the anode is maintained constant by precisely machined \SI{2}{\mm} thick pillars uniformly distributed over their area. The spacing uniformity is verified prior to installation to be within \SI{100}{\micro\meter} (see \secref{sec_tpc}), which would correspond to a less than 5\% variation of the gain.
\end{itemize}

\subsection{Technological milestones}

The design of the \three TPC is based on the outcome of an intensive R\&D: the optimisation and tests of the LEMs \cite{polina-thesis, filippo-thesis, Badertscher:2008rf,Badertscher:2010fi,Cantini:2014xza}, the simplification of the charge extraction scheme and design of low capacitance PCB based readouts \cite{Cantini:2014xza}, the development of the cryogenic front-end analogue electronics and the cost-effective scalable data acquisition system \cite{DeBonis:1692375, 1748-0221-6-01-C01005}, and the pursuit of high efficiency detection of the argon scintillation \cite{1475-7516-2017-03-003,Badertscher:2013ygt,1742-6596-39-1-028}. Last but not least the results reported in this paper also represent one of the first concrete outcomes of design studies performed with industry such as the use of a non-evacuable membrane cryostat and the operation of a cryogenic system capable of processing large volumes of ultra-pure liquid argon \cite{Stahl:2012exa}. The design and use of a \SI{300}{\kilo\volt}-rated high voltage feedthrough \cite{Cantini:2016tfx} as well as the investigation of a cost effective and scalable solution for the detector slow control and monitoring back-end are also important aspects of the project to benefit future detector development. As the first ever tonne scale dual-phase TPC to be constructed and operated, we list below some of its essential technological deliverables and operational milestones:
 \begin{itemize}
\item [i)] \textit{Performance of the cryogenic system and of the \SI{23}{\meter\cubed} membrane cryostat.} The  detector operation requires a thermodynamically stable cryogenic medium with a uniform gas density and a flat liquid argon surface. Furthermore, liquid argon impurities below 100\,ppt oxygen equivalent, or equivalently electron lifetimes of more than \SI{3}{\milli\second}, are required in order to efficiently transport the ionising charge over the $\mathcal{O}(10)$ meter drifts proposed in future large dual-phase TPCs \cite{Stahl:2012exa,Acciarri:2016ooe}.  This setup, although of a more modest drift distance, offers a unique test-bed to address the suitability of non-evacuable membrane cryostats for future large detectors.
     
\item[ii)] \textit{Extraction of the ionisation charge over an area of \SI{3}{\meter\squared}.} Never before has charge been extracted from liquid argon over such a large area. The stability of the liquid argon level and the high voltage operation of a \SI{3}{\meter\squared} extraction grid immersed only a few millimetres below the surface will be discussed. 

\item[iii)] \textit{Amplification in pure argon vapour by a combined operation of multiple \fifty LEMs.} Operating a TPC with \SI{12}{\meter} drift at an effective gain of 20 would provide a S/N of around 10 on both its collection views. The latter assumes an electron lifetime of \SI{3}{\milli\second} and an ENC of 1500 electrons at the input of the preamplifiers. A value of \Geffinfty$\approx 20$ therefore constitutes the baseline assumption for the DUNE Dual-phase Far Detector module \cite{Acciarri:2016ooe}. Even if stable effective gains of more than 20 were reported on smaller devices, the simultaneous operation of large area LEMs could represent new challenges. 

\item [iv)] \textit{Readout of the signal on two collection planes with strips of up to \SI{3}{\meter} length and performance of  analogue front-end electronics.} An important aspect of the dual-phase TPC with its vertical drift is the ability to position front-end electronics close to the readout strips, profiting from the cryogenic temperatures while at the same time ensuring its accessibility during detector operation. The charge readout scheme and its performance will be described in Sections \ref{sec_elec} and \ref{sec_operations}.

\item [v)] \textit{Detection of prompt and secondary scintillation}. The \three detector would allow to study the properties of LAr scintillation and the light propagation in \SI{3}{\meter\cubed} volume. The detection of the electroluminescence (secondary scintillation), produced by the ionisation charge in the argon vapour and never before measured in a LAr TPC with a \SI{1}{\meter} drift, could give additional handle on the amount of charge reaching the CRP.

\end{itemize}


\section{Cryostat and cryogenic system}
\label{sec_cryostat-cryo}
The cryostat functions as a totally sealed system containing the ultra pure liquid argon at its boiling point ($T_{LAr}\simeq\SI{87}{\kelvin}$), and slightly above the ambient atmospheric pressure ($P_{cryostat}\simeq\SI{1000}{mbar}$ compared to $P_{atm}$ of \SIrange{950}{980}{mbar} at CERN). The rate of evaporation, which depends mainly on the insulation quality of the cryostat, is controlled by the cryogenic system using a liquid nitrogen heat exchanger. The evaporated gas argon (the so-called \textit{boil-off}) and the liquid are constantly recirculated and purified in a closed loop during operation of the detector.

\subsection{The cryostat}\label{subsec_cryostat}

\begin{figure}[ht!]
 \centering
 \includegraphics[width=\textwidth]{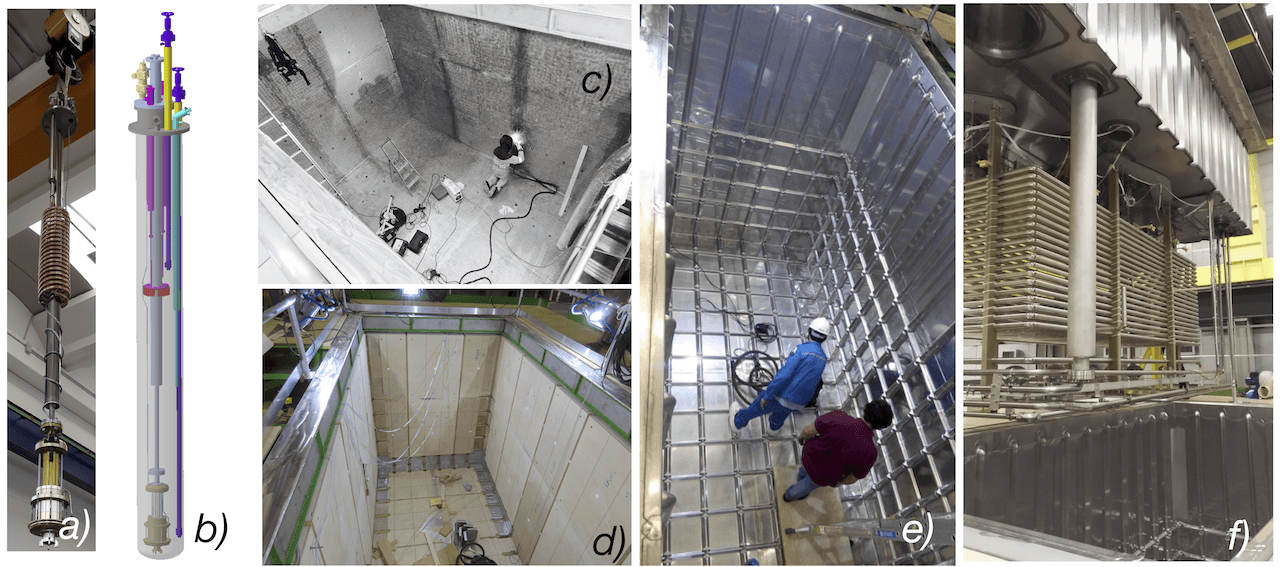}
 \caption{Illustrations and pictures of the extractable liquid pump and of the cryostat at various stages of construction. Picture of the pump (a) and CAD view with the pump tower in transparency (b). Pictures from inside the cryostat structure prior (c) and after the installation of the GRPF insulation blocks (d). The empty cryostat with the membrane is shown in (e) and the top-cap with the TPC suspended during installation in (f).}
 \label{fig:cryostat-assembly} 
\end{figure}

The cryostat is made from corrugated stainless steel \textit{membrane} panels that absorb the thermal stress and about 1 meter thick low density passive insulation. The LAGUNA-LBNO design studies  converged on the choice of the corrugated membrane technology licensed by GTT\footnote{Gaztransport et Technigaz www.gtt.fr} in France. A close partnership was put in place with GTT to construct the cryostat hosting the \three detector. The dimensions of the cryostat are provided in \tabref{tab:cryostat-cryo-spec} and pictures of its assembly are shown in \figref{fig:cryostat-assembly}. It comes in two separate elements: the main vessel and a thermally insulated lid called top-cap under which the detector is suspended. Their thermal insulation is based on glass reinforced polyurethane foam (GRPF) layers, interspersed with pressure distributing layers of plywood. The details of both elements are provided hereafter.

\begin{table}[ht!]
\begin{center}
\begin{tabular}{p{.2\textwidth}p{.4\textwidth}p{.2\textwidth}p{.1\textwidth}}
\hline
\hline
 &Component & value & unit\\
\hline
\textbf{\textit{cryostat}}& &\\
&outer dimensions (l$\times$w$\times$h)&$7.34\times4.88\times4.76$& m$^3$\\
&inner dimensions (l$\times$w$\times$h)&$4.76\times2.38\times2.03$& m$^3$\\
& main vessel (top-cap) insulation thickness & 1 (1.2)&m\\
&volume of LAr (GAr) & 18.01 (4.98)& m$^3$\\
&height of LAr (GAr)& 1590 (440)& mm\\
& main volume operated pressure & 1000 & mbar\\
& insulation volume operated pressure & 600 & mbar\\
& GAr temperature gradient 2 cm above the liquid& $\sim$2 & K/cm\\
& Insulation density & 70 & kg/m$^3$ \\
\textbf{\textit{liquid}}\\ \textbf{\textit{purification}} & &\\
&pump nominal (max)  flow rate & 22 (35) & Lpm\\
&pump efficiency at nominal flow rate & 40\% \\
&heat input to liquid argon at nominal flow rate & $\sim$300& W \\
&minimal height of liquid argon required & 5 & cm \\
&cartridge active volume (l$\times\oslash$) & 900$\times 306$& mm$^3$\\
&volume ratio copper:sieve & 5:1&\\
\textbf{\textit{boil-off}}\\ \textbf{\textit{compensation}}& &\\
& LN$_2$ temperature (pressure)& 85 (2.2) & K (bar)\\
& max. available cooling power during cooldown & 10.0& kW\\
& max. available cooling power during detector operation & 3.0 & kW\\
& measured cryostat heat input during detector operation& 0.8  & kW\\
\hline
\hline
\end{tabular}
\caption{List of relevant quantities from the cryostat and cryogenic system measured during operation of the detector. Pressures are quoted in absolute values.}
\label{tab:cryostat-cryo-spec}
\end{center} 
\end{table} 

The cryostat outer structure is formed by an exoskeleton made from carbon-steel I-beams covered with \SI{6}{\mm} thick stainless steel plates. Its role is to provide the mechanical support designed to sustain the forces of the over-pressure from the inner vessel. The passive insulation comes in prefabricated panels \SI{330}{\mm} thick made of GRPF sandwiched between two sheets of plywood. The blocks are superimposed in three layers to provide a uniform \SI{\sim1}{\meter} thick insulation. Forty-five temperature sensors are distributed in-between the insulation blocks to provide feedback on the gradient and quality of the insulation during cryogenic operation. Three millimetres thick corrugated stainless steel panels, so called membrane sheets, are then fixed on the insulation blocks and carefully welded together. The membrane sheets come in different dimensions with various shapes of corrugations to match the geometry and thermal shrinkage calculations of the cryostat. 

The top-cap is a \SI{1.2}{\meter} thick thermal insulating lid made from a stainless steel sheet on its upper surface and an INVAR\cite{INVAR_NIST} bottom plate. The sidewalls have vertical corrugations which are complementary to those from the main vessel in order to minimise the inter-space once the top-cap is inserted. The insulation is made from stacked sheets of GRPF with vertical plywood panels that are arranged to provide internal structural reinforcement.
Altogether twenty INVAR pipes of various diameters called \textit{chimneys} cross the top-cap in order to host the necessary feedthroughs as well the interfaces to the cryogenic system. Each pipe is extended to the exterior by about 30\,cm and terminated by a UHV flange so that the appropriate feedthrough can be fixed.  

Since the membrane from the main vessel and the top-cap ensure the tightness and liquid containment of the cryostat under normal operating condition, all the welds were systematically inspected before operation. Both the top-cap and the main vessel have a set of external ports communicating with their insulation volumes to allow for input of gas during leak checking and to regulate the insulation space pressure during operation. Upon delivery, both top-cap and the main vessel are checked separately by flushing gas inside their respective insulation volumes and by locally scanning the welds with a spectrometer. 
In the case of the main vessel, to increase the sensitivity of the test beyond the helium traces present in the atmosphere, custom designed vacuum plugs matching the shapes of the membrane corrugations have been developed. They allow to check for leaks on the membrane welds down to the sensitivity limit of the spectrometer of \SI{\sim1e-9}{mbar.\liter/\second}. The cryostat is thoroughly cleaned with a decreasing agent, rinsed and visually inspected. Once the top-cap with the detector are inserted in the cryostat, the top-cap is welded to the main vessel and their insulation volumes are linked together forming one single cryostat with a common insulation. The operating pressure of the cryostat's main volume ($P_{cryostat}$) is set to around \SI{1000}{mbar} (absolute pressure) and the insulation space is kept at a constant pressure of around \SI{600}{mbar}. Accidental over-pressures of the main volume are protected by a safety burst disk which is set to rupture if $P_{cryostat}\geq \SI{160}{mbar}$ over atmospheric pressure.

\subsection{The cryogenic and argon purification systems}

The principal tasks of the cryogenic and argon purification systems are:
\begin{enumerate}
    \item  displace the air from inside the cryostat by filling it with pure argon gas to the level that the main contaminants (oxygen, moisture and nitrogen) are reduced to the part-per-million (ppm) level and subsequently cool down and fill the cryostat with liquid argon in a uniform and controlled manner
    \item ensure a stable thermodynamic environment inside the cryostat to avoid variations in the LEM gain 
    \item keep the electronegative impurities in the liquid below the \SI{100}{ppt} level. 
\end{enumerate}

A detailed description of the system is provided in the next two sections. The performance of the cryogenic system and its impact on the detector operation and performance is instead given in \secref{sec_operations}.

\subsubsection{Gas argon piston purge, cooling down and filling}

In order to remove the air from inside the cryostat, argon gas is uniformly introduced at the bottom of the cryostat through a manifold of four pipes each containing three 12\,mm diameter holes and is exhausted through venting pipes placed at the upper most point of each penetration. This method, called ``piston purge", provides a uniform gas flow from the bottom to the top of the main volume and prevents the formation of any residual air pockets. The exhaust gas from the chimney vents can either be sent to the exterior through a control valve or recirculated through a purifying cartridge and returned to the cryostat. 

\begin{figure}[ht!]
 \centering
 \includegraphics[width=\textwidth]{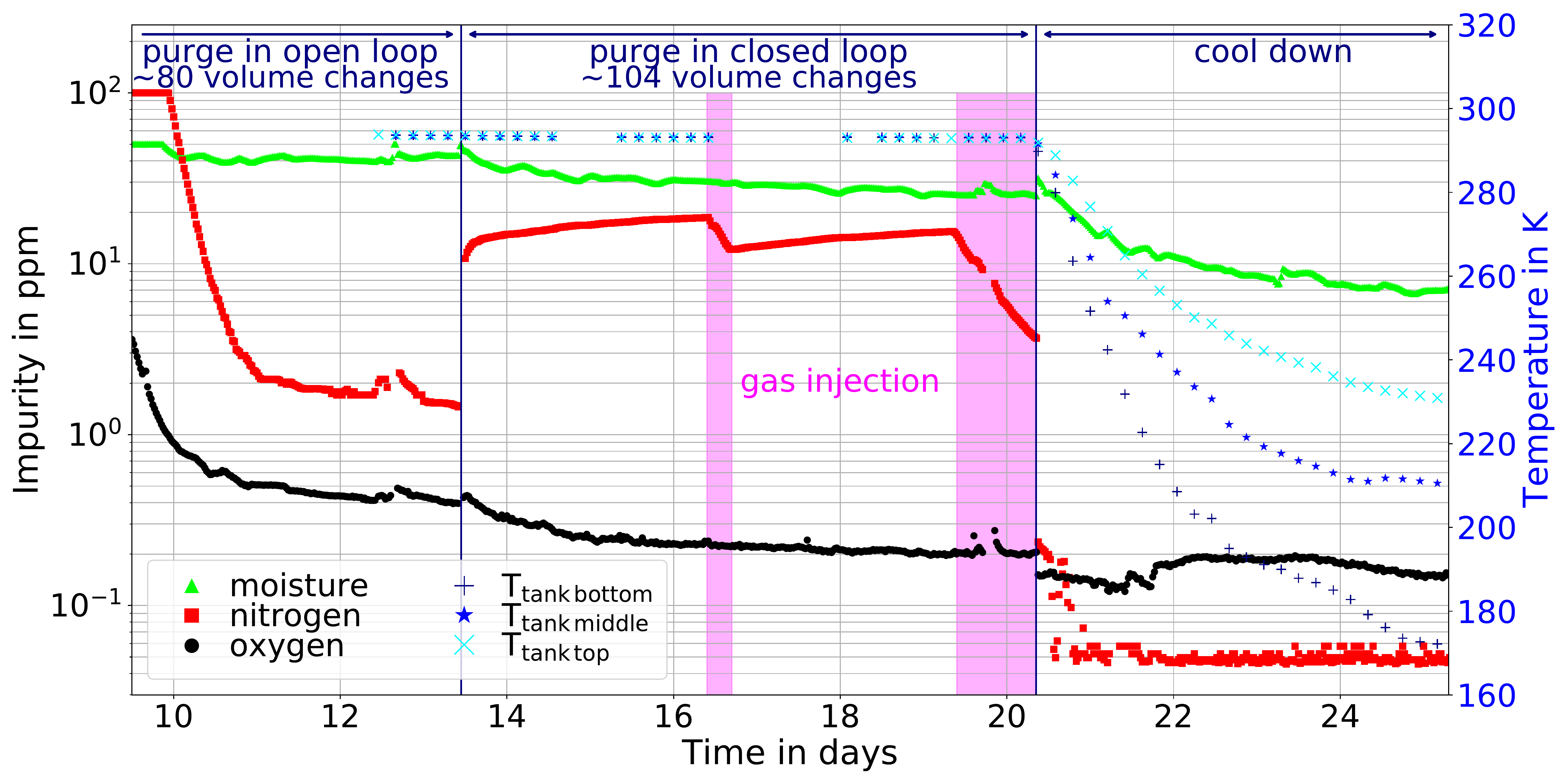}
 \caption{Evolution of the impurities measured in the gas during open, closed loop piston purge and cooling down. The measured temperatures inside the gas near the bottom, middle and top of the main cryostat volume are indicated in blue.}
 \label{fig:impurities_vs_time} 
\end{figure}

During the purge process, the gas from the main volume is constantly sampled and its impurities are continuously monitored with three residual gas trace analysers (RGTA) for oxygen, moisture and nitrogen. Their technical details are summarised in \tabref{tab:GTA}. In order to compensate for the gas taken by the sampling pump, a make-up gas line can inject pure argon gas through a commercial purifier\footnote{SAES MicroTorr MC400}. The evolution of the impurities during the piston purge is shown in \figref{fig:impurities_vs_time}.  The process is performed in two phases: first, in the so called open loop purge, the input gas is injected into the cryostat with a flow rate of about 2 L/s and vented to the exterior through a control valve. At the end of the open loop the impurities are measured to be 0.4~ppm, 1.7~ppm and 43~ppm for oxygen, nitrogen and moisture. The comparatively larger moisture content is interpreted as stemming from outgassing of various detector materials.
The gas is then recirculated in closed loop by a double diaphragm pump\footnote{KNF 0150.1.2 AN.12 E} at a flow rate of \SI{4}{\liter/\second} and a commercial gas purifier\footnote{SAES MicroTorr MC4500} which filters oxygen and moisture, but has no expected effect on nitrogen. Thus, during the closed loop stage only oxygen and moisture levels decrease, while nitrogen slightly rises, presumably due to outgassing of the detector components inside the cryostat. The sudden increase in the nitrogen level at the beginning of the closed loop is interpreted as coming from trapped volumes of air before the gas purification cartridge. Since nitrogen was not filtered, pure argon gas was injected through the makeup gas line to dilute its concentration eventually to below \SI{10}{ppm} level, as indicated by the pink areas labelled ``gas injection" in \figref{fig:impurities_vs_time}. At the end of the closed loop, the impurities are measured to be 0.2\,ppm, 3.5\,ppm and 25\,ppm for oxygen, nitrogen and moisture respectively.  

\begin{table}[ht!]
\begin{center} 
\begin{tabular}[\textwidth]{p{2cm} p{2.5cm} p{2.5cm} p{1cm} p{3cm} p{2.5cm}}
\hline
instrument & upper detection limit  & lower detection limit & number of ranges & precision at lowest range & provider \\
\hline 
Oxygen & 23\% & 50~ppb & 10 & $\pm$100~ppb & AMI (2001~R series) \\
Nitrogen & 200~ppm & 10~ppb & 2 & $\pm$ 20~ppb & Gow-mac (1200 series)\\
Moisture & 50~ppm & 10~ppb & 2 & $\pm$ 20~ppb & Gow-mac (1402 series)\\
\hline
\end{tabular}
\end{center} 
\caption{Parameters of the gas trace analysers. }
\label{tab:GTA}
\end{table}

The detector volume is cooled with a mixture of argon gas at room temperature and liquid argon at \SI{87}{\kelvin}, which are injected through four gas atomizing nozzles\footnote{SSCO-Spraying Systems 1/4J-SS+SUE18-SS} located at the bottom corner of the cryostat with flow rates of \SI{500}{\liter/\min} and \SI{21.1}{\liter/\hour}, respectively. The method provides a uniform and steady cooling by generating a flat pattern of atomised argon at temperature close to \SI{87}{\kelvin}. Such system was successfully used for the first time at FNAL in a 35 ton cryostat as reported in~\cite{dmonta:35ton}. Various temperature probes are present in the cryostat main volume either glued on the membrane surface or fixed to the TPC, and their readings provide feedback to the cryogenic system to adjust the input flow of LAr and hence control the cooling power. The temperature measured in gas at three different heights along the detector prior and during the cooldown is shown in \figref{fig:impurities_vs_time}.
The cryostat is cooled from room temperature to a minimum temperature measured in the gas of \SI{170}{\kelvin} in about five days at an initial rate of \SI{\sim2}{\kelvin} per hour. The cryostat is then filled with liquid argon at a flow rate of about \SI{12}{\liter/\min}. During the filling process the level is constantly monitored by a combination of vertical chain of temperature probes and liquid level meters (see \secref{sec_sc}). 

\subsubsection{Boil-off compensation and argon purification}

During detector operation the liquid is continuously recirculated and purified. A submerged centrifugal cryogenic pump 
operating at 22\,Lpm circulates the liquid through a custom built purification cartridge containing two separate volumes of molecular sieve\footnote{BASF 4A 8x12 mesh} and copper pellets\footnote{BASF CU 0226 S 8x14MESH} to remove moisture and oxygen respectively. About two cryostat volumes per day are filtered with the system. As shown in \figref{fig:311-full} and \figref{fig:cryostat-assembly}, the pump itself is confined at the bottom of a fixed vessel called \textit{pump tower} which has a \SI{350}{\mm} diameter and a \SI{3.5}{\meter} length. A liquid nitrogen heat exchanger with a cooling power of \SI{2}{\kilo\watt} is present inside the pump tower to compensate the heat generated from the pump. This unique design allows for the pump to be extracted for maintenance without polluting the argon of the main volume. Another attractive feature of the system is that possible waves and turbulence introduced by the pump are confined inside the tower and avoid perturbing the liquid level in the active region. The pump tower communicates with the main cryostat liquid and gas volumes via two \SI{25}{\mm} diameter ports. Their openings can be controlled from the exterior via two long stem cryogenics valves to regulate the flow of liquid and to equalise the pressures between the pump tower and the main volume.  

The boil-off gas argon in the main volume is continuously re-condensed by a liquid nitrogen heat exchanger designed to provide compensation for up to \SI{3}{\kilo\watt} of heat input. For a given heat input, the pressure inside the cryostat main volume is regulated by setting the flow and pressure of the liquid nitrogen. The condensed boil-off is re-injected into the pump tower to undergo the liquid recirculation cycle. A \SI{200}{\liter} liquid/gas phase separator is present on the liquid recirculation line in order to remove gas argon generated during the recirculation process. The produced gas argon is re-condensed and injected inside the pump tower. The recirculated and purified liquid argon is injected to the bottom of the main cryostat.



\section{Description of the detector}
\label{sec_tpc}

\begin{figure}[ht!]
 \centering
  \includegraphics[width=\textwidth]{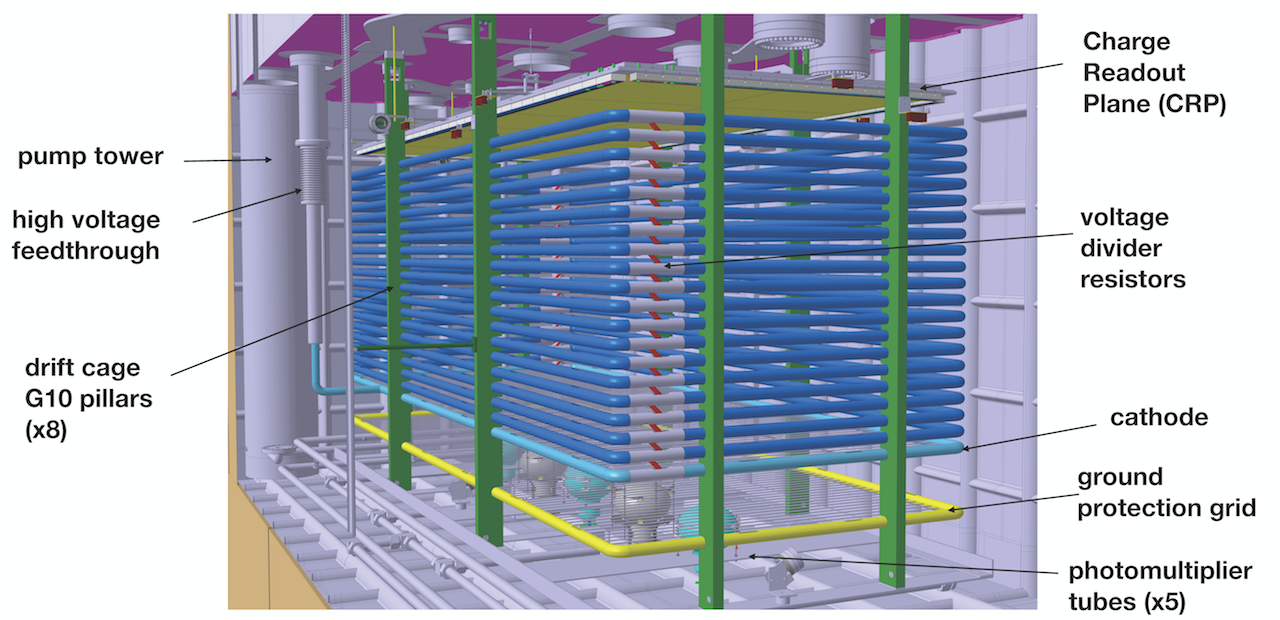}
  \caption{The \three TPC inside the cryostat.}
 \label{fig:311-TPC} 
\end{figure}

The \three TPC, illustrated in \figref{fig:311-TPC}, consists of three principal components: the drift cage, the CRP, and the photon detection system. Each component is described in detail in the next 3 subsections.

\subsection{TPC drift cage and drift field high voltage}

\label{sec_tpc_cdrift}
\begin{figure}[ht!]
 \centering
 \includegraphics[width=\textwidth]{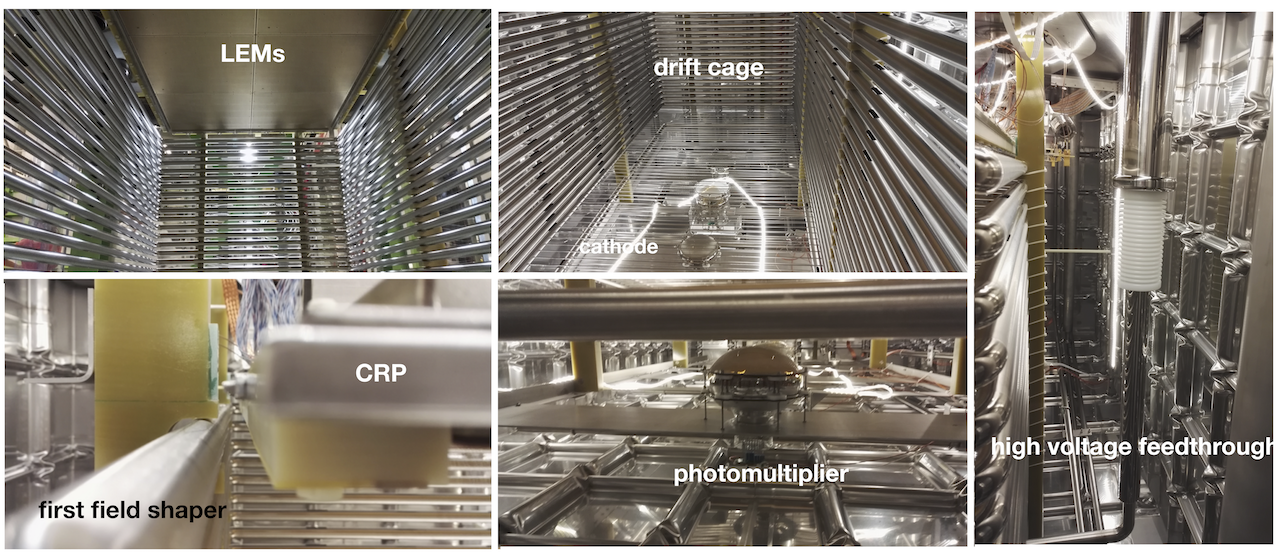}
 \caption{Pictures of the TPC inside the cryostat. Top-left picture from the drift cage interior looking up to the CRP LEM plane. Top-middle: from the top of the drift cage, the first field shaping ring. Bottom: picture of the CRP near the first field shaper (left) and one PMT shown below the ground grid (middle). The high voltage feedthrough connected to the cathode is shown in the right picture.}
 \label{fig:TPC-pictures} 
\end{figure}

Some pictures of the drift cage taken from inside the cryostat are shown in \figref{fig:TPC-pictures}. It is constructed from twenty identical field-shaping rings spaced vertically with a \SI{50}{\mm} pitch. The rings are made from \SI{2}{\mm} thick stainless steel tubes with a \SI{34}{\mm} external diameter. The tubes have regularly spaced \SI{2}{\cm} oval shaped openings in order to allow for the flow of gas during the purge process. The bottom-most field shaping ring, which serves as the cathode, is a grid made from \SI{4}{\mm} diameter stainless steel tubes point welded at a \SI{20}{\mm} pitch. The entire drift cage is interconnected with pairs of HV-rated resistors\footnote{Metallux 969.11} with a resistance at room temperature of  \SI{100}{\mega\ohm}, forming a voltage divider chain that ensures uniform drift field inside the TPC active volume. An average increase of about 7\% in the value of the resistance has been measured for these resistors at \SI{77}{\kelvin}. The cathode is electrically connected to the high voltage feedthrough and the top-most field shaping ring---referred to as the \textit{first field shaper} (FFS)---is terminated to ground outside of the cryostat via an interchangeable resistor. Depending on the high voltage applied to the cathode, this resistor fixes the appropriate voltage drop on the FFS to allow the electrons to drift towards the extraction grid. For the nominal drift field of \SI{0.5}{\kilo\volt/\cm}, the voltage on the FFS is \SI{-7.8}{\kilo\volt}. The entire drift cage assembly is suspended from eight G10 pillars fastened to the top-cap. The pillars also support the frame holding the five PMTs as well as a stainless steel grid (ground grid) that shields the photo-sensors from the cathode high voltage. 

The cathode high voltage is generated by a commercially available power supply\footnote{Heinzinger PNChp 300000-05-neg} and delivered to the cathode using a custom-made high voltage feedthrough. The feedthrough and the power supply were successfully operated up to \SI{-300}{\kilo\volt} in a dedicated set-up prior to their installation in the \three detector \cite{Cantini:2016tfx}. The feedthrough is approximately \SI{2}{\meter} long and has a coaxial structure consisting of a \SI{32}{\mm} diameter inner conductor, a \SI{36}{\mm} thick layer of High Molecular Density Polyethylene dielectric (HMDPE), and a \SI{2}{\mm} thick stainless steel ground shell. It is inserted through the top-cap and connected to the cathode via a spring-loaded contact. The designed length of the outer ground shield is such that it terminates below the liquid level, hereby avoiding the presence of a locally high electric field in pure argon vapour.

\subsection{Charge readout plane}\label{sec_tpc_crp}

\begin{figure}[th]
 \centering
 \includegraphics[width=.9\textwidth]{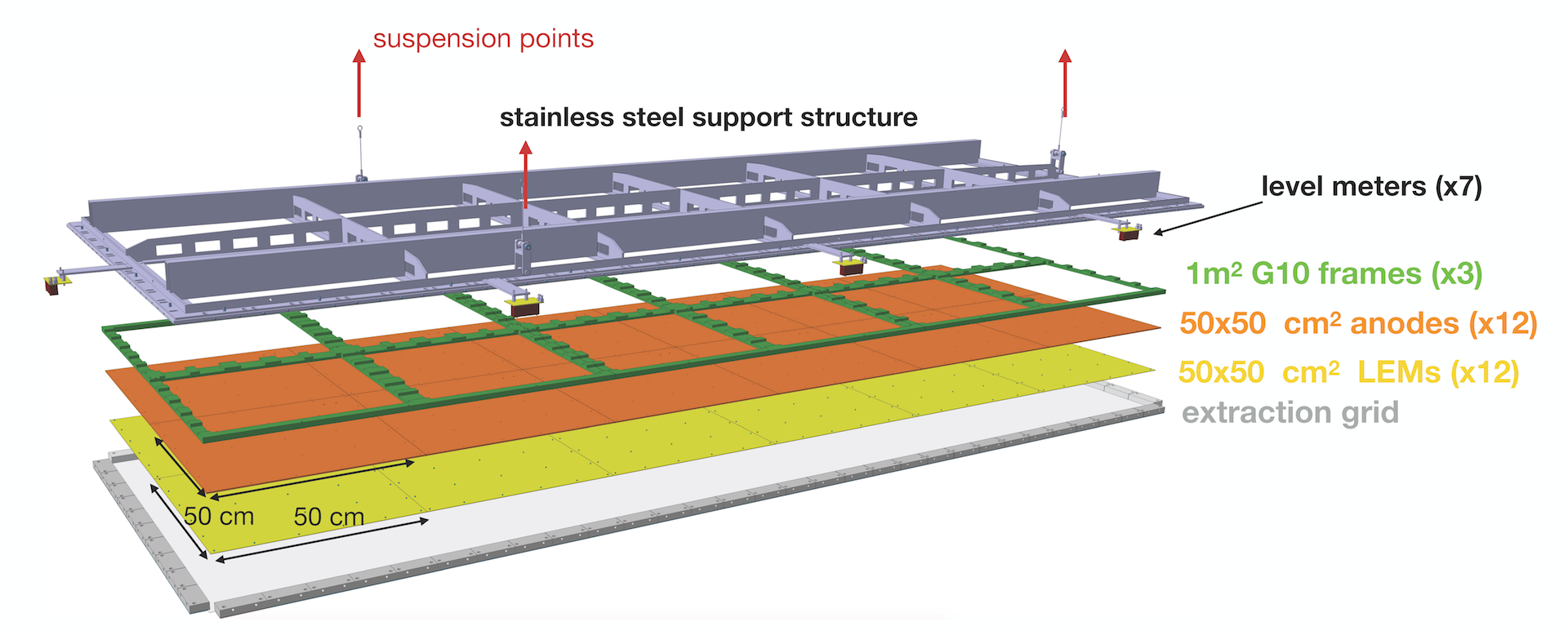}
  \caption{Exploded view of the \three CRP. The red arrows indicate the location of the three suspension cables.}
 \label{fig:CRP-exploded} 
\end{figure}

The concept of an independent CRP containing the charge extraction, amplification, and collection stages is a key element of the dual-phase design whose functionality is being tested for the first time in the \three. Since it is decoupled from the rest of the detector, the CRP can be entirely pre-assembled in a clean room and tested before installation in the cryostat. An exploded view of the CRP is shown in \figref{fig:CRP-exploded}. 
Its main elements are the \fifty LEM-anode modules (twelve units) and the extraction grid, which are all assembled with precisely defined inter-stage distances and alignment. They are fixed on three \SI{1}{\meter\squared} precisely machined G10 frames, which are in turn screwed on a stainless steel frame designed to provide the necessary mechanical rigidity in warm and cold conditions. The CRP is suspended from the top-cap by three cables. The cables are coupled via dedicated feedthroughs to remotely controlled precision step motors that allow for an alignment and positioning of the CRP with respect to the liquid level. The total range of the vertical movement is \SI{40}{\mm}.

\subsubsection{LEMs and anodes: design and quality assurance}
\begin{figure}[th]
 \centering
 \includegraphics[width=\textwidth]{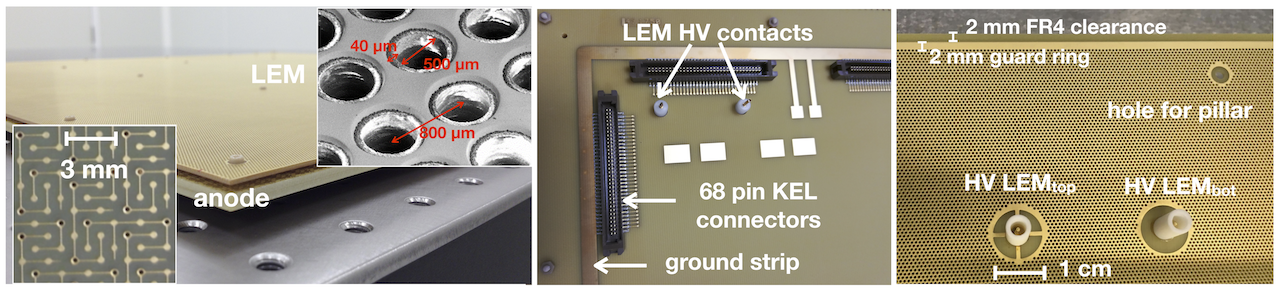}
  \caption{Left: one \fifty LEM and anode assembled as a "sandwich", with zooms on their patterns. Middle: top layer where the through going LEM HV connectors and the signal strips are visible. Right: picture of a LEM showing the details of the HV connections.}
 \label{fig:CRP-LEM-anode} 
\end{figure}

The \fifty LEMs and anodes are industrially manufactured in the PCB industry in large quantities and at affordable costs. Those installed in the \three detector are produced by ELTOS\footnote{www.eltos.it} in Italy. Close-up pictures of specific regions of the LEMs and anodes are shown in \figref{fig:CRP-LEM-anode}. 

The anode is a four-layer PCB having a set of orthogonal strips with a 3.125\,mm pitch that provide the two views of the event. The pattern of the readout strips, printed on the bottom PCB layer and used for charge collection, is optimised such that the charge is evenly split between both views~\cite{Cantini:2013yba}.
The readout strips are routed to the top layer towards 68-pin female connectors 
soldered on the anode periphery. Each connector reads 32 strips; its 36 remaining pins are connected to the detector ground via a copper strip which runs around the periphery of the top layer of the anode as shown in \figref{fig:CRP-LEM-anode}. The anode strips and the quality of the connector soldering are subject to an optical inspection by the company before delivery. Upon reception, further electrical tests to verify the electrical continuity of the strips and the absence of short circuits between neighbouring channels are performed. 

LEMs are built from \SI{1}{\mm} thick \fifty standard PCB epoxy plates. Holes of \SI{500}{\micro\meter} diameter are mechanically drilled in a honeycomb pattern with a pitch of \SI{800}{\micro\meter}, yielding about 180 holes per \units{cm}{2}. The copper surfaces around each hole is further removed producing a \SI{40}{\micro\meter} dielectric rim. The final copper thickness of the LEM electrode is about \SI{60}{\micro\meter}. To prevent high voltage discharges near or across the edges, the LEM has a \SI{2}{\mm} border free from metalisation and another \SI{2}{\mm} copper guard ring (see \figref{fig:CRP-LEM-anode}-right). High voltage contacts are made with \SI{1.2}{\mm} diameter male pins\footnote{Deutsch 6860-201-22278} that are soldered on specifically designed pads imprinted on the top electrode of the LEM. The pins are insulated with circular tubes made out of POM\footnote{Polyoxymethylene}. In addition, there is a \SI{10}{\mm} circular clearance around each pin. 

\begin{figure}[ht!]
 \centering
  \includegraphics[width=\textwidth]{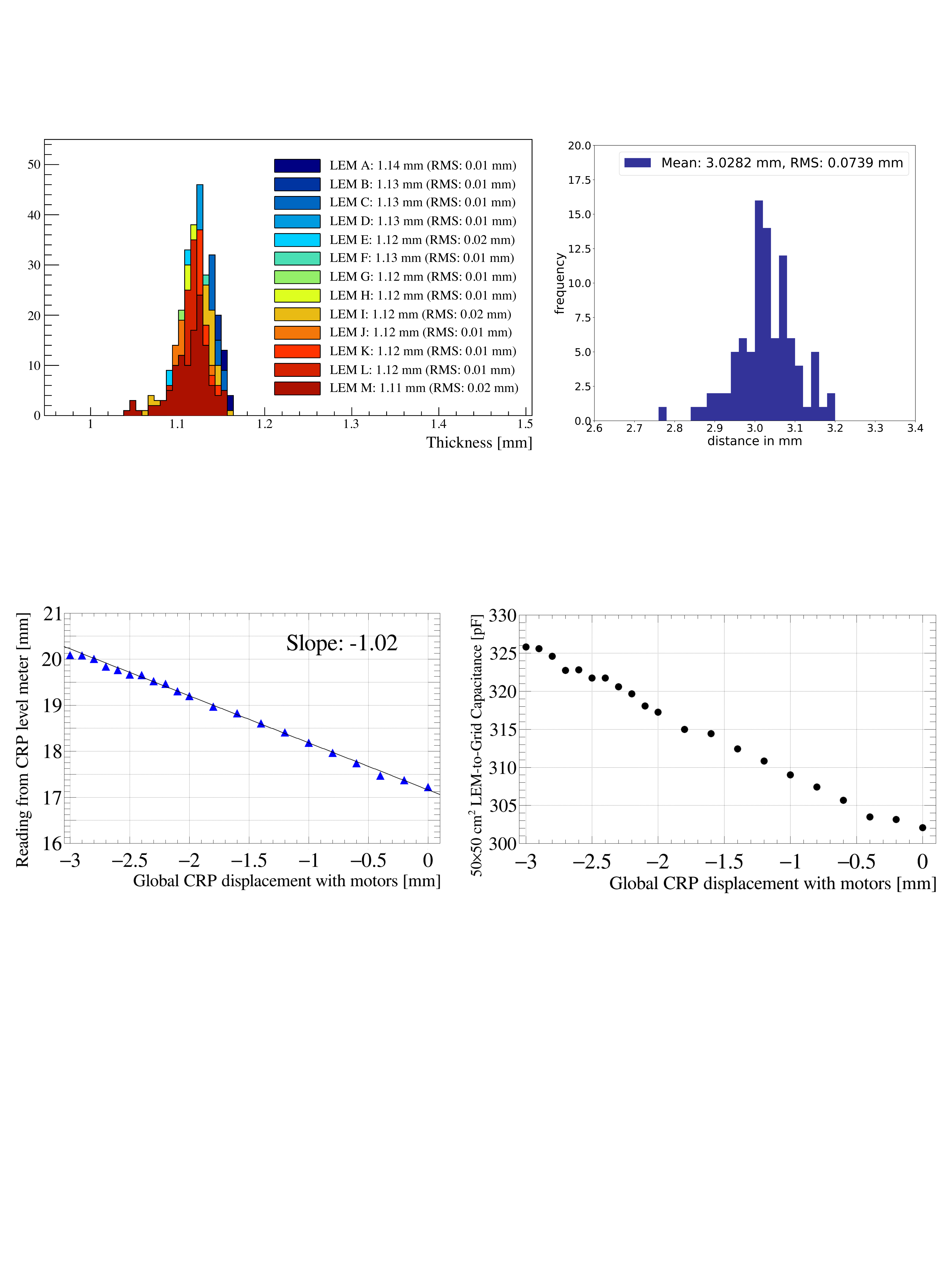}
  \caption{Left: the measured total thickness of 13 LEMs in approximately 100 points on their \fifty surface. The measurement includes the \SI{60}{\micro\meter}  of copper on both electrodes. Right: LEM anode inter-stage distance. The quoted errors represent the RMS of the distributions.}
 \label{fig:LEM-anode-survey} 
\end{figure}

The fabrication and acceptance of LEMs and anodes follow strict criteria agreed upon with the manufacturer. While the anodes are rather simple to produce and to certify, the LEMs require a more specific and time-consuming quality assurance procedure. The tolerances on the LEM thickness and dimensions of the holes and rims are typically required to be within a few per cent from the design values. To ensure maximal gain uniformity over the entire CRP the tolerance of \SI{30}{\micro\meter} on the plate thickness of the PCBs was agreed with the manufacturer. Since such tight tolerances are not common in the industry, the company had to carefully measure and select the raw epoxy plates to meet this requirement. The quality of the production is verified upon reception by inspecting a few samples under a scanning electron microscope and by measuring the rim size, hole diameter and pitch in different areas of the PCB. The thicknesses of all produced LEMs are also precisely measured at the CERN metrology laboratory. The measurements are taken over 100 different points for each LEM plate with a $\sim$\SI{2}{\micro\meter} precision and the results shown in \figref{fig:LEM-anode-survey} (left). The total thickness (insulation plus two copper layers) is uniform to within \SI{10}{\micro\meter} for the majority of the LEMs. The spread in the average values from one LEM to another is also within the \SI{30}{\micro\meter} requirement.

Each LEM is thoroughly cleaned in a 65$^\circ$C degreasing bath and then rinsed, with high-pressure water jet and with de-ionised water. The LEM is then baked for three hours at 180$^\circ$C in order to complete polymerisation of the epoxy plate above its T$_{g}$\footnote{Glass-transition temperature} point and to remove most of the water present inside the material. The performance of the LEM is then checked by applying a high voltage across its electrodes. The tests are done at room temperature and atmospheric pressure, first in a dry air environment and then in grade N6.0 (with \SI{1}{ppm} impurities) argon gas.
The LEMs are validated if they can sustain up to \SI{3.5}{\kilo\volt} in dry air and \SI{1.5}{\kilo\volt} in argon gas at room temperature and atmospheric pressure. Below those values, the monitored leakage currents should stay within a few \si{\nano\ampere}. Above those values, discharges can occur, their rate should not exceed about 1--2 per minute and should be randomly distributed over the LEM active area. If the above conditions are not satisfied, the cleaning procedure and the high voltage test are repeated. Out of an initial batch of 15 LEMs that underwent this procedure, all were accepted after only one cleaning iteration. The twelve LEMs selected for assembly on the CRP are those showing the best performance under the high voltage test and/or whose thickness was the most uniform. 

Once the quality of both LEMs and anodes is verified, the latter are assembled together as independent \textit{LEM-Anode Sandwiches} (LAS) as shown in \figref{fig:CRP-LEM-anode}-left. A constant \SI{2}{\mm} inter-stage distance between both \fifty plates is provided by 29 precisely machined pillars made of high density polyethylene\footnote{PE 1000}, uniformly located between both surfaces and held together by the means of M2 PEEK\footnote{Polyether ether ketone} screws. The uniformity of the spacing was measured on one LAS at the CERN metrology laboratory by a camera capable of visualising the anode plate through the \SI{500}{\micro\meter} diameter LEM holes. About one hundred points over the LAS surface were measured with an accuracy of about \SI{2}{\micro\meter}. The distribution presented in \figref{fig:LEM-anode-survey} shows an average distance of \SI{3}{\mm} with an RMS of around \SI{100}{\micro\meter}. This result is consistent with a \SI{2}{\mm} LEM-anode spacing taking into account the LEM thickness of about \SI{1}{\mm}.


\subsubsection{Mechanical frame assembly and cryogenic tests}
\begin{figure}[ht!]
 \centering
  \includegraphics[width=\textwidth]{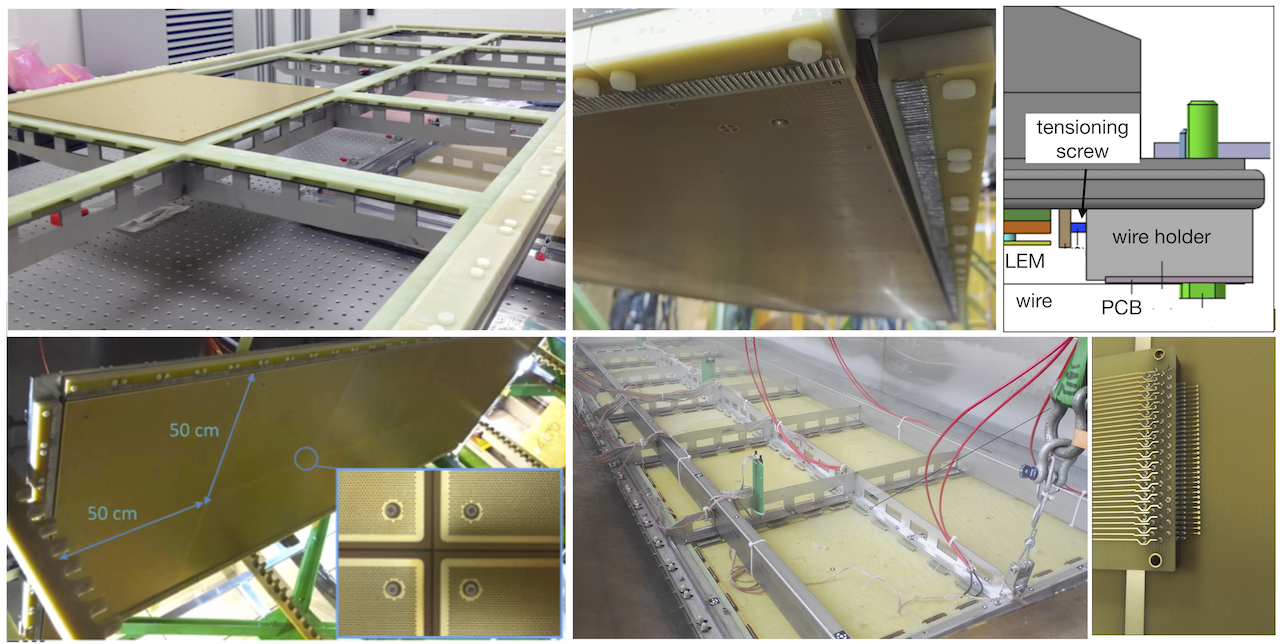}
  \caption{Pictures of the CRP during construction and installation. The top-left picture shows the frame upside-down with one LAS assembled on it, the extraction grid with the wire tensing pads being visible on the middle image and a CAD cross section of the CRP shows details on the wire tension mechanism. Bottom-left: picture of the fully assembled CRP under the top-cap before mounting of the drift cage. The space between four adjacent LEMs is shown in the zoom. Bottom-middle: picture of the CRP dipped in liquid nitrogen during a cryogenic open bath test. A closeup of one anode electrical jumper is shown on the right.}
 \label{fig:CRP-assembly-2} 
\end{figure}

Pictures of the CRP during assembly and installation are shown in \figref{fig:CRP-assembly-2}. The assembly takes place on an optical table. The LAS are precisely assembled on the G10 frames maintaining a \SI{0.5}{\mm} gap between adjacent modules. Since the LEM has a \SI{2}{\mm} copper clearance on its periphery, the distance between two neighbouring LEM copper electrodes is of \SI{4.5}{\mm}. The effective geometrical coverage of the LEM holes, taking into account the extra \SI{2}{\mm} of copper guard ring, the areas around the high voltage contacts and around the twenty-nine pillars, is about 96\% of the \SI[product-units=power]{3x1}{\meter} area.
The anodes are interconnected with jumpers made from \SI{\sim10}{\cm} long PCBs (\figref{fig:CRP-assembly-2}-bottom-right) to form the three meter long charge collection strips in one direction and the one meter long strips in the other. 

The extraction grid is constructed from \SI{100}{\micro\meter} diameter stainless steel wires\footnote{AISI 316L} tensed along both directions and aligned with respect to the anode inter-strips. The wire pitch of \SI{3.125}{\mm} matches that of the anode readout strips in order to provide a uniform extraction field and avoid charge shadowing by the grid wires. The wires are soldered in groups of \num{32} on a pair of independent tensing pads consisting of a PCB fixed on a mechanical wire holder as depicted in \figref{fig:CRP-assembly-2}-top-right. The PCB hosts the high voltage connection and \num{32} soldering pads with \SI{200}{\micro\meter} grooves for wire positioning. During soldering, the wires are tensed with \SI{150}{\gram} weights. The precision on the wire pitch after soldering is verified with a microscope to be better than \SI{50}{\micro\meter}. The mechanical wire holders housing the PCBs allow for adjustments of the tension for each group of \num{32} wires using a set of \textit{tensioning screws} (two per block). The wire holders are mounted along the outer perimeter of the CRP in an interlaced manner to partially constrain the wires in case of loss of tension. The tension of the \SI{3}{\meter} (\SI{1}{\meter}) long wires is adjusted to \SI{3}{\newton} (\SI{1}{\newton}), which is far below the \SI{15}{\newton} breaking force. Those tensions result in a sagging of around \SIrange{100}{200}{\micro\meter} at room temperature while distributing the weight uniformly around the frame. The PCBs with the soldered wires are electrically bridged together so that one contact is sufficient to provide the high voltage to the entire extraction grid.

Once fully assembled, the CRP is suspended above the optical table and the \SI{10}{\mm} LEM-extraction grid distance is verified in several points. The CRP is made only from materials that are carefully chosen for their cryogenic compatibility and the matching of their coefficients of thermal expansion.
Nonetheless, its behaviour under cryogenic conditions is validated prior to installation by shock cooling it in an open liquid nitrogen bath near \SI{77}{\kelvin}. The dimension of the frame, and especially its horizontal geometry, permits to perform such a test in relatively safe conditions with minimal infrastructure.
The main element that is verified is that the planarity of the frame, and hence of the gap between each stage is preserved at cryogenic temperature. We also visually inspect that the extraction grid remains under tension and verify after the cold test that no wires are ruptured. In addition, the electrical continuity of the anode strips is tested while the frame is near \SI{77}{\kelvin}.
It should be noted that dipping the entire CRP in an open liquid nitrogen bath exposes the CRP to a large temperature gradient and sudden stress, far more than what is observed during the controlled cooling down phase inside the cryostat (see \secref{sec_cryostat-cryo}). The planarity of the frame during the test is measured by photogrammetry. The method consists in taking multiple photographs of the frame on which fluorescent targets are previously glued (see \figref{fig:CRP-assembly-2}). The 3D coordinates of each target are then reconstructed with sub-mm accuracy, providing a high-resolution image of the entire frame and allowing to check for local deformations and potential differences between room and cryogenic temperature. It is the first time photogrammetry is successfully employed in cryogenic conditions at CERN and constitutes a promising method for similar measurements in the future. The structure exhibits a deformation not exceeding \SI{1}{\mm} along its shortest side; a sagitta of about \SI{2}{\mm} and \SI{3}{\mm} in warm and cold conditions, respectively, is measured for the \SI{3}{\meter} long side. The possible impact of this deformation on the operation of the detector is discussed in \secref{sec_operations}. The extraction grid is visually inspected during and after the cooling shock to ensure that the wires are intact. Furthermore, the electrical continuity of the anode strips is not affected  when the frame is immersed in liquid nitrogen. From the latter observations, we conclude that the CRP is compatible with cryogenic operation.

\subsubsection{Electrical connections and properties, charge injection}\label{subsec_CRP-electrical}

Once the CRP is suspended under the top-cap, each element is connected to its respective feedthrough. In terms of high voltage, the CRP requires twenty-four channels to bias both electrodes of each LEM (referred to as the LEM-top and LEM-bottom electrodes) and one channel to provide the high voltage to the extraction grid. Two high voltage channels and contacts are however present on the extraction grid for redundancy. Those high voltages are supplied externally and brought to the CRP with the system described in \secref{sec_sc}.  The remaining instrumentation present on the CRP, such as the LEDs, temperature probes or level meters, are connected to double sided SMA or SUB-D feedthroughs. All the necessary precautions are taken during cabling in terms of stress release and cable length to allow for the 40 mm vertical movement of the frame.

\begin{figure}[ht!]
  \centering
  \includegraphics[width=\textwidth]{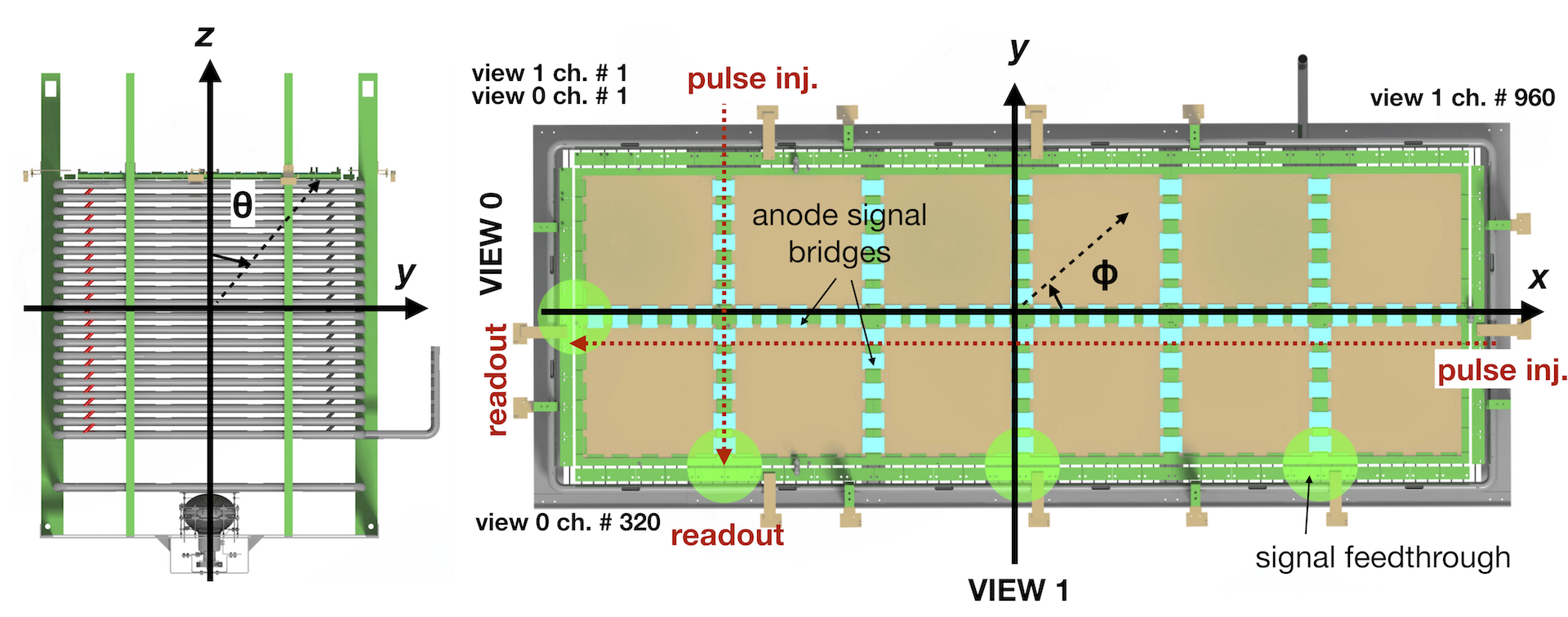}
  \caption{Side view (left) and top view (right) of the TPC illustrating the coordinate system and naming conventions of both orthogonal views from the charge readout. The location of the signal feedthroughs where the charge pre-amplifiers are located are also schematically indicated. The dotted red arrows represent one channel of each view with the readout and pulse injection sides indicated.}
     \label{fig:311-coordinates} 
\end{figure}

The anodes are connected to the charge pre-amplifiers with a multiplicity of thirty-two channels through twisted pair ribbon cables. The pre-amplifiers are located inside custom-designed \textit{signal feedthroughs} described in \secref{sec_elec}. The charge readout coordinate system and channel mapping convention of the TPC are shown in \figref{fig:311-coordinates}. 
View 0 (view 1) consists of 320 three meter-long (960 one meter-long) strips running parallel to the \textit{x} (\textit{y}) coordinate axis. The pre-amplifiers read signals from one end of the strips, while, on the opposite side, the strips are connected to a charge injection system consisting of a \SI{1}{\pico\farad} test capacitor driven by an external pulse generator. The system can inject charge into groups of 32 neighbouring channels to check electrical continuity, monitor potential dead or problematic channels, and verify the channel mapping assignment in the data acquisition system. 

\begin{figure}[ht!]
  \centering
  \includegraphics[width=\textwidth]{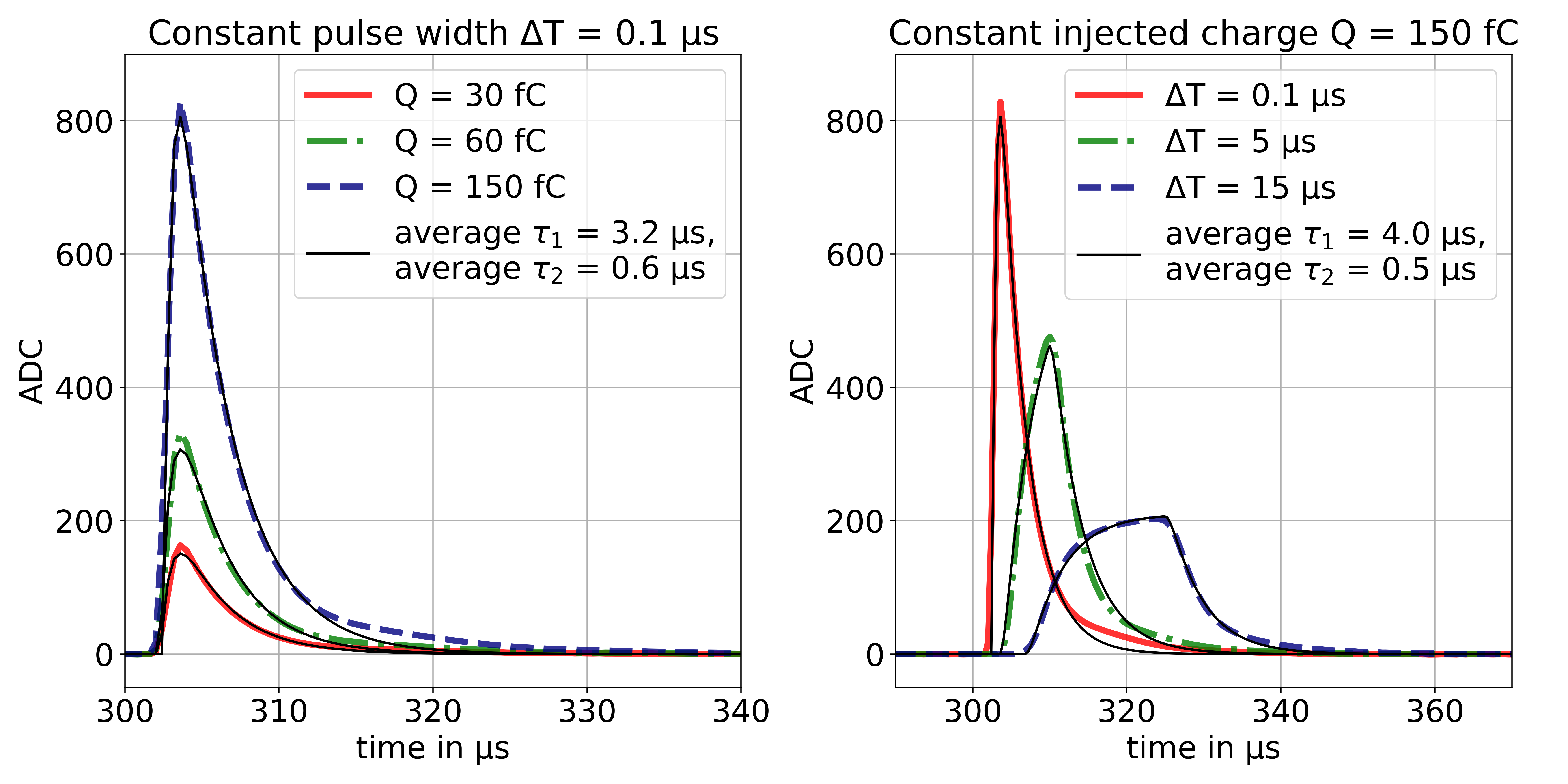}
  \caption{Response of one channel from view 0 (three meter length) to the anode pulsing system described in the text.
 Left: increasing pulse amplitude. Right: increasing pulse duration.}
     \label{fig:pulse-response} 
\end{figure}

The response of one channel to various pulse profiles is shown in \figref{fig:pulse-response}. The data is collected when the entire detector is at room temperature. The waveforms in the left plot are obtained by varying the amount of the injected charge. Those shown in the right plot are made by injecting a constant charge over different time periods $\Delta t$. The waveforms are fitted with the preamplifier response function (see \secref{sec_elec}):
\begin{equation}
    V_{response}(t)= \frac{\tau_1}{\tau_1 -\tau_2}\cdot \left( e^{-t/\tau_1}-e^{-t/\tau_2} \right)
\end{equation}
where $\tau_2$  and $\tau_1$ are the rise and fall time of the waveform. For the right plot the response function is convoluted with a constant current over the corresponding time $\Delta t$. The average fitted values of $\tau_1, \tau_2$ correspond to a $V_{response}$ peaking time of $\sim$\SI{1.2}{\micro s} which is in relative good agreement with the expected value from \tabref{tab:preamp-para}.

In \tabref{tab:CRP-capa} we list the capacitive couplings measured between the various TPC electrodes. The measurements are systematically performed between both mentioned electrodes while setting all others to ground. We note that these inter-electrode couplings may represent a nuisance in terms of high voltage operation of the entire CRP with potentially one element inducing currents on neighbours in the case of a discharge. On the other hand, the fact that most of the electrodes are coupled with measurable capacitance is useful for practical aspects. The measurement of the LEM-grid capacitance provides information on the position of the liquid level between both planes, with a \fifty granularity (see \secref{sec_sc}). In addition dead or problematic channels on the anode can also be identified by injecting a low amplitude pulse on either the grid or the LEMs. 
\begin{table}[ht!]
\begin{center}
\begin{tabular}{p{.5\textwidth}p{.2\textwidth}p{.2\textwidth}}
\hline
\hline
 Electrode & distance (mm) & capacitance\\
           &               & in GAr (pF)\\
\hline
3 m (1 m) anode strip to GND&-&480 (160)\\
\fifty LEM-top to anode plane&2&1000\\
\fifty LEM-top to LEM-bot &1&7000\\
\fifty LEM-top to neighbour LEM-top&4.5&$\leq$1 \\
\fifty LEM-bot to extraction grid&10&150\\
3$\times$1 m$^2$ cathode to GND &200& 4000\\
\hline
\hline
\end{tabular}
\caption{Approximate inter-electrode capacitance measured in the \three (see text for nomenclature).}
\label{tab:CRP-capa}
\end{center} 
\end{table} 

\subsection{Photon detection system} \label{sec_tpc_lrp}

The detection of the VUV scintillation from particles interacting in the gas and liquid argon is performed by an array of five cryogenic Hamamatsu R5912-02MOD photomultiplier tubes (PMT) with an 8-inch diameter borosilicate window. 
The PMTs have a bialkali photo-cathode with platinum underlay to preserve its electrical conductivity at cryogenic temperatures. Each PMT features 14 dynode stages to operate at high gain (up to $10^{9}$) to compensate for the loss in dynode performance at cryogenic temperatures. The photo-cathode quantum efficiency for \SI{420}{\nano\meter} photons measured at room temperature is around 17\% as specified by the manufacturer.  

\begin{figure}[!hpb]
\includegraphics[width=6.6cm]{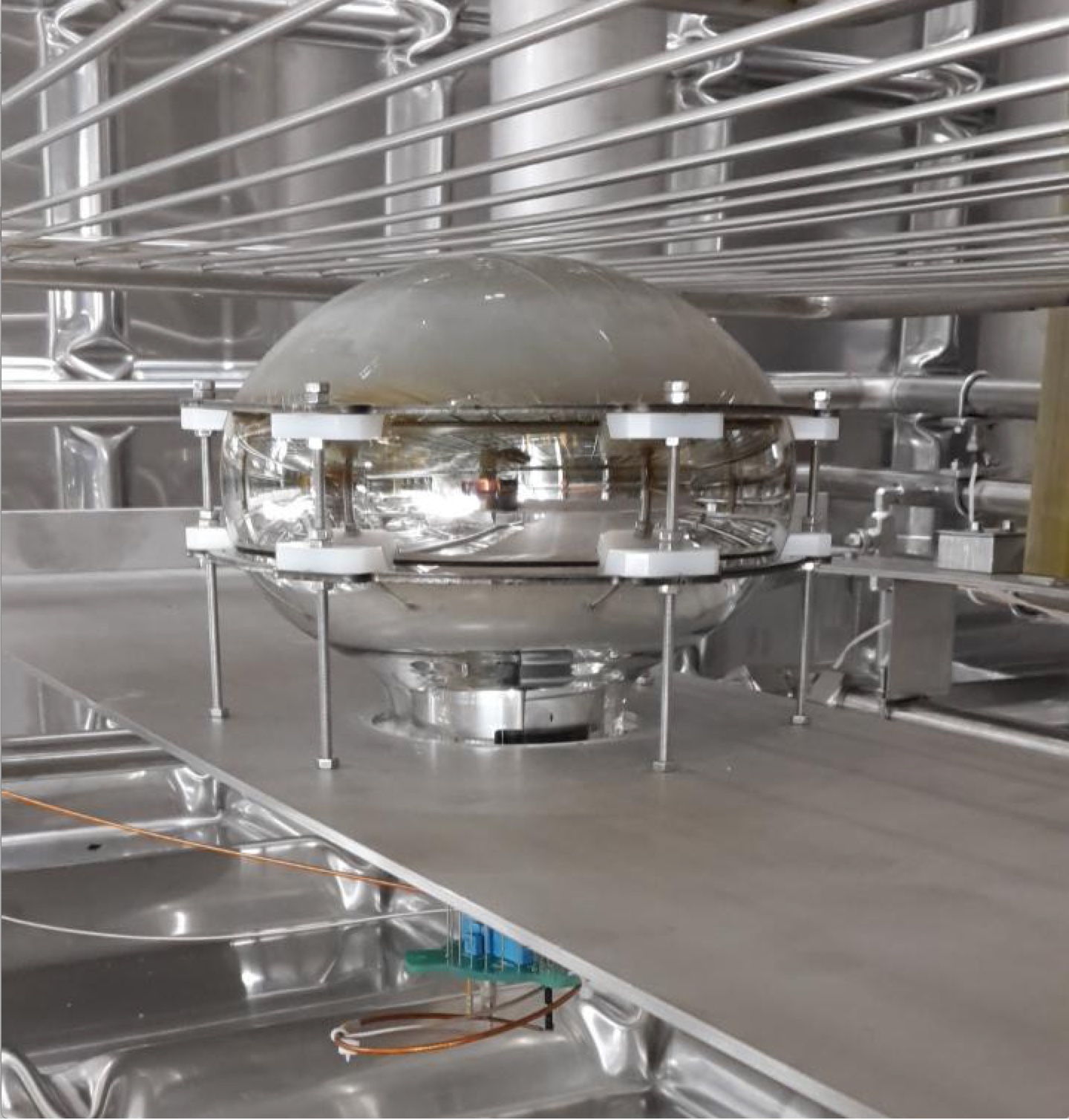}
\includegraphics[width=5.0cm]{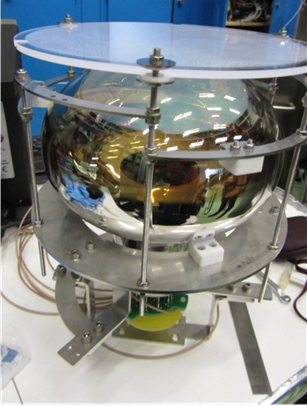}
\centering \caption{Left: picture of one PMT coated with TPB. Right: a PMT unit with the coated plate mounted in front of the photo-cathode and ready for its installation.} 
\label{fig:PMTConfigurations}
\end{figure}

A thin layer of TPB is used to convert the argon scintillation light centred at \SI{128}{\nano\meter}~\cite{Lippincott} to the wavelengths in the range of the PMT photo-cathode sensitivity. TPB is particularly well-suited for the detection of VUV light in cryogenic environment and commonly used in experiments using liquid argon~\cite{Spanu:2018xwr,Pate:2017cux}. Three of the five PMTs are coated with TPB by evaporating it directly on the glass window while for the other two, the coating is applied on a transparent \SI{4}{\mm} thick acrylic\footnote{PLEXIGAS GS233/0F00} plate mounted in front of the photo-detectors. A better performance is expected when the TPB is directly deposited on the PMT. However, the plates are easier to coat, handle, store, and install. The two configurations are shown in \figref{fig:PMTConfigurations}. The TPB was deposited under vacuum using an evaporator system at the CERN thin film laboratory. A layer with TPB density of about \SI{0.2}{\mg \per\square\cm} was chosen based on the previous experience~\cite{1748-0221-4-06-P06001,boccone-thesis}. The quantum efficiency after the TPB coating was measured to be about 60\% at \SI{200}{\nano\meter} in air. The large increase in the efficiency, compared to the intrinsic one of the photo-cathode, is due to the re-emission of more than one photon from the TPB layer.

Two polarity configurations of power supply are employed for the PMT bases in order to evaluate the optimal solution for future detectors. The \textit{negative base} operates with a negative bias voltage applied at the photo-cathode, while the anode is grounded, requiring one cable for the HV supply and another for the signal.  The \textit{positive base} functions with positive biasing applied at the anode and carries the HV and signal on a single cable. The signal is then decoupled from the HV externally. The latter solution represents a more optimal choice in view of minimising the number of feedthroughs and cables as well as offers noise reduction. 

Table~\ref{tab:PMT_configuration} gives a summary for the configurations of the five installed PMTs and shows the nominal operating voltages with the corresponding gain and the observed noise levels. 

\begin{table}[ht!]
\begin{center}
\begin{tabular}{cccccc}
\hline
\hline
PMT & Base & Coating & Voltage (kV) & Gain (${10}^{6}$) & Noise (ADC)\\
\hline
1 & - (2 cables) & Coated & 1.2 & $0.92\pm0.13$ & 0.7\\
2 & - (2 cables) & Plate & 1.2 & $1.01\pm0.12$ & 0.7\\
3 & + (1 cable) & Coated & 1.1 & $0.95\pm0.11$ & 0.4\\
4 & + (1 cable) & Plate & 1.1 & $1.26\pm0.15$ & 0.4\\
5 & - (2 cables) & Coated & 1.2 & $1.33\pm0.15$ & 0.8\\
\arrayrulecolor{black}
\hline
\hline
\end{tabular}
\caption{Details of the 5 PMTs installed in the TPC. The gain value for each PMT at the given operating voltage was obtained from the extrapolation of calibration curves in liquid argon (see \figref{fig:single_pe_charge}(right)).}
\label{tab:PMT_configuration}
\end{center} 
\end{table} 

\begin{figure}[h!]
\includegraphics[width=\textwidth]{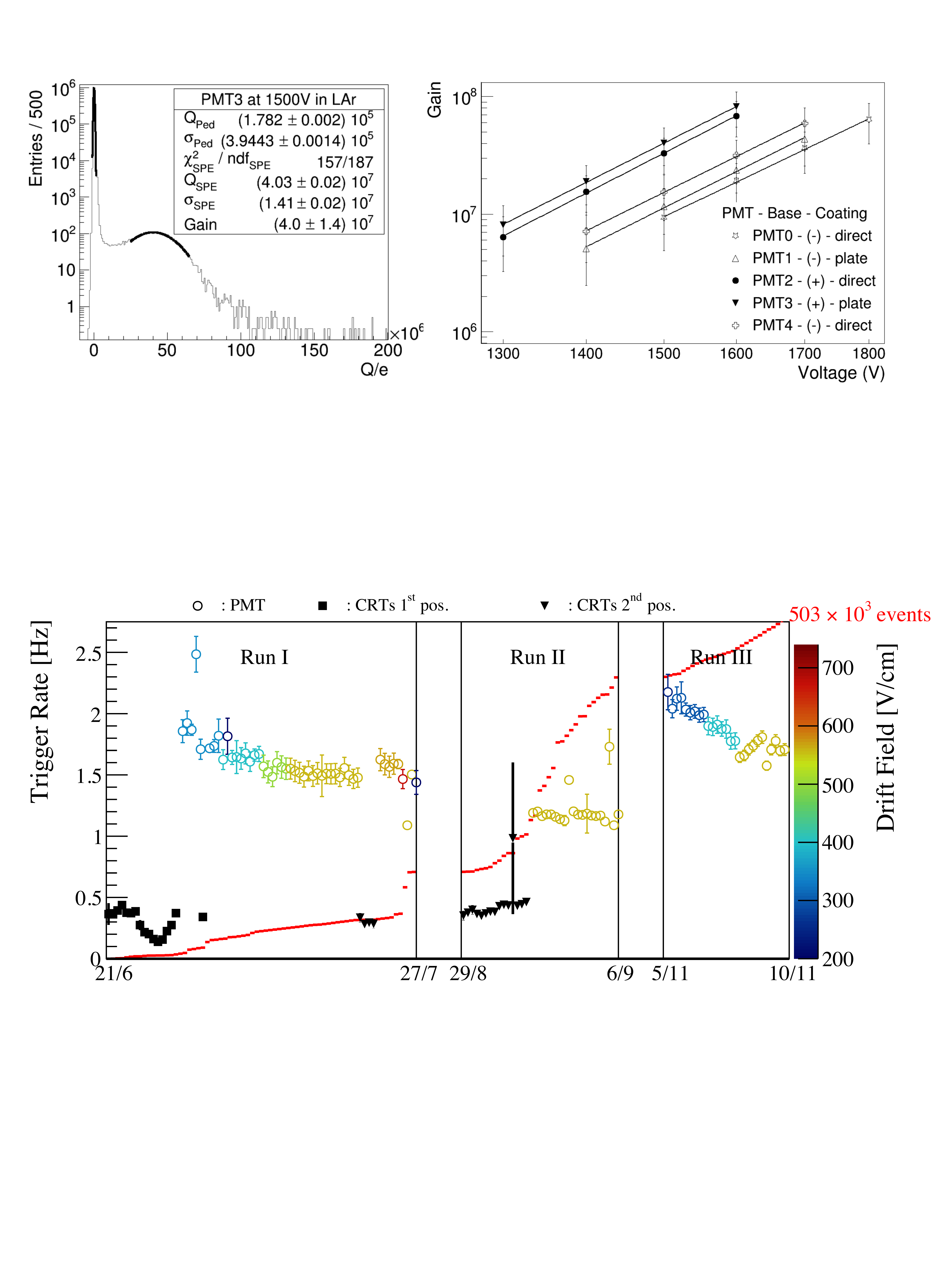}
\centering \caption{Left: distribution of the integrated charge for one PMT in the single PE analysis. Right: PMT gain in liquid argon for different operating voltages. Fits with the power law parametrisation are superimposed. }
\label{fig:single_pe_charge}
\end{figure}

Prior to installation, the gains of the five PMTs are measured in air at room temperature by studying the response to a single photoelectron (PE) produced at the photo-cathode.
Since at cryogenic temperatures a decrease of the gain is expected, the latter must be remeasured once the detector is filled in order to equalise the PMT responses in liquid argon \cite{Belver:2018erf}.
Dedicated runs were taken using a pulse generator as a trigger running at a frequency of \SI{100}{\Hz}. Data were digitised with the sampling of \SI{250}{\MHz} and the gain was measured in an offline analysis of the collected waveforms. Every fluctuation from the baseline along the waveform is integrated to get the corresponding charge. Since a random trigger is used, the most common signal should be dark current at the level of a single PE, as it is shown in the resulting charge histogram in \figref{fig:single_pe_charge}-left. This spectrum consists of two Gaussian distributions, one centred at zero, corresponding to the pedestal, and the other representing the PMT response to a single PE. The PMT gain is then directly calculated from the distance between the means of the two gaussians, expressed in units of the electron charge. A voltage scan is performed to obtain the gain calibration curve for every PMT, as is shown in \figref{fig:single_pe_charge}-right. The PMT's equipped with positive base provides larger gain for equal applied voltage.
The dependence of the gain (G) on the voltage (V) applied to the PMT is fit to a power law $G=AV^{B}$, where A and B are constants dependent on the number, structure, and material of the dynodes. Finally, the PMT gain at the operation voltage is obtained by extrapolation from the corresponding calibration curve.
As shown in \tabref{tab:PMT_configuration}, the PMT gain during most of the data taking is set to about $10^{6}$ in average.


\section{Ancillary instrumentation and detector control system}
\label{sec_sc}
\begin{figure}[ht!]
  \centering
  \includegraphics[width=.9\textwidth]{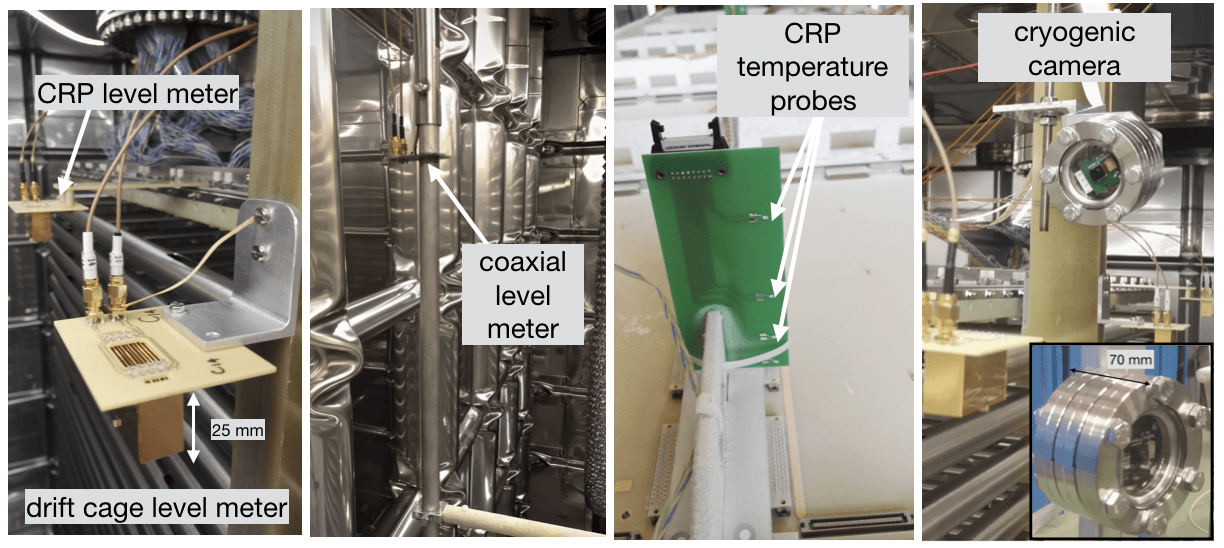}
  \caption{Pictures of some of the instrumentation in the \three. From left to right: the high resolution capacitive level meters (a, b), a PCB with temperature probes to measure the gas temperature gradient near the LEMs (c) and one cryogenic camera (d).}
     \label{fig:SC-LM-T} 
   \end{figure}

The slow control system of the \three detector is the result of a continued effort aimed at developing a system dedicated to multi-ktonne liquid argon dual-phase TPCs. Its role is to provide a monitoring and control of the parameters that affect the operation of the detector such as for instance temperatures, pressures, high voltages, liquid argon level, and the CRP position. The list of the equipment and sensors managed by the system is given in \tabref{tab:SC-para}. Pictures of some of those sensors are shown in \figref{fig:SC-LM-T}. 

\begin{table}[ht!]
\begin{center}
\begin{tabular}{>{\raggedright}p{.25\textwidth}>{\raggedright}p{.15\textwidth}p{.25\textwidth}p{.25\textwidth}}
\hline
\hline
Name & number of channels & resolution & range\\
\hline
HV LEM & 24 & \SI{50}{\pico\ampere}/\SI{10}{\milli\volt}  & \SIrange{0}{20}{\micro\ampere}/\SIrange{0}{8}{\kilo\volt} \\
HV Grid & 2 & \SI{10}{\nano\ampere}/\SI{1}{\volt} & \SIrange{0}{100}{\micro\ampere}/\SIrange{0}{12}{\kilo\volt} \\
HV PMT & 5  & \SI{50}{\nano\ampere}/\SI{100}{\milli\volt}  & \SIrange{0}{1}{\milli\ampere}/\SIrange{0}{3}{\kilo\volt}\\
HV cathode & 1 & 0.05\%(I)/0.001\%(V) & \SIrange{0}{0.5}{\milli\ampere}/\SIrange{0}{300}{\kilo\volt}\\
Coaxial level meters & 2  & \SI{1}{\mm} & \SI{470}{\mm}, \SI{1200}{\mm}    \\
Plate level meters   & 13 & \SI{100}{\micro\metre} & \SIrange{0}{25}{\mm}        \\
CRP step motors & 3 & \SI{100}{\micro\metre} & \SI{40}{\mm} \\
temperature sensors& 153 & \SI{\leq0.5}{\kelvin} & \SIrange{50}{350}{\kelvin}\\
LED strips & 5 &- & -\\
Heaters& 4 & - & \SI{100}{\watt}\\
$P_{\mbox{abs}}$ gauges & 4 & 1\% full range & \SIrange{900}{1100}{mbar}\\
$P_{\mbox{diff}}$ gauges & 5 & 1\% full range & \SI{\pm50}{mbar}\\
Cryogenic cameras& 4 & 5 Mpixels sensor& -\\
\hline
\hline
\end{tabular}
\caption{List of devices which are monitored or operated by the detector control system. All high voltages are negative apart from two of the HV PMT channels.}
\label{tab:SC-para}
\end{center} 
\end{table} 

\subsection{Pressure and temperature monitoring}
\label{sec_sc_tp}
 The atmospheric pressure as well as the values of pressure inside the cryostat main volume and insulation space are constantly monitored by a set of 9 transducers providing either absolute or relative readings. They have the primary role of verifying that the entire cryostat is operated safely within the specified pressure ranges (see \tabref{tab:cryostat-cryo-spec}).
 
All temperatures inside the cryostat are measured with platinum resistance thermometers (PT1000) which provide a precision of better than 0.5 K over the \SIrange{50}{350}{\kelvin} range. The sensors are positioned in various locations throughout the cryostat, on the detector, and inside the insulation space. A sub-set of the temperature probes is placed at a regular \SI{4}{\cm} interval over the entire height of the cryostat inner volume to provide the information on the temperature gradients. The sensors register an abrupt temperature drop of a few kelvin as soon as they become submerged in the liquid. This effect, along with the knowledge of the probe positions, can be also used to infer the liquid argon level during filling. Groups of probes are also placed between \SI{2}{\cm} to \SI{7}{\cm} above the CRP frame to monitor the temperature gradient in the vapour. During the cool down phase, the temperature measurements are used to adjust the cooling power in order to minimise the thermal stress on the membrane and on the detector. 

\subsection{Liquid argon level monitoring and CRP motorisation systems}
\label{sec_sc_lm}

The system to measure the liquid argon level inside of the cryostat consists of a set of coaxial and parallel plate capacitors (see \figref{fig:SC-LM-T}). Due to the difference in the relative permittivity of the liquid and gas argon the capacitance of these level meters changes as the function of the depth to which they are submerged. Custom electronics located outside of the cryostat measures the capacitance of each level meter probe. There are two coaxial level meters: one in the main cryostat volume near the TPC with an operating range of 470 mm and the other inside the pump tower with a range of 1200 mm. Both have a sensitivity of around 1 mm and are used to follow the evolution of the liquid level during filling and emptying of the cryostat.

\begin{figure}[ht!]
  \centering
 \includegraphics[width=.8\textwidth]{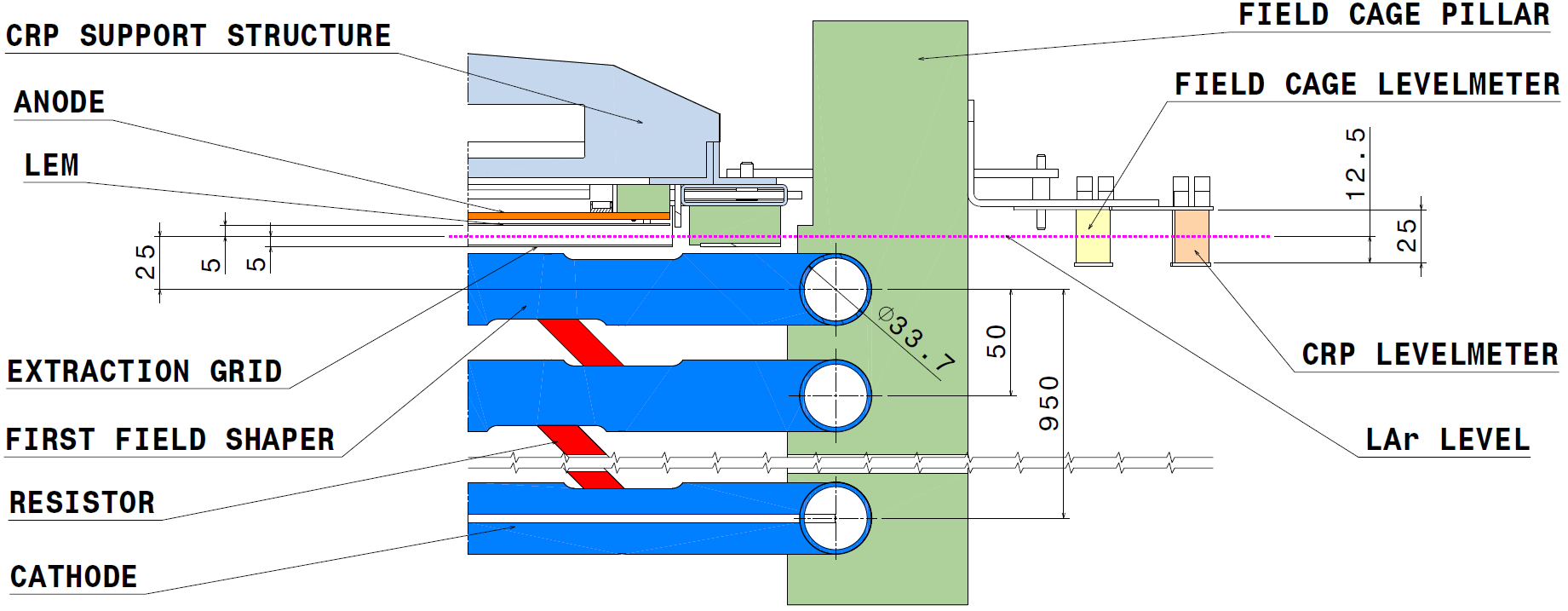}
  \caption{Cut view of the TPC near the liquid level.}
     \label{fig:SC-levelmeters} 
\end{figure}

The parallel plate capacitor level meters have a narrow operating range of 25 mm and a sensitivity of better than \SI{100}{\micro\meter}. They are intended to provide the  monitoring of the level once the liquid reaches the top of the TPC field cage. There are 13 of these devices installed on the TPC: six (DC-LM) are fixed on the G10 pillars that support the field cage and seven (CRP-LM) are attached along the perimeter of the CRP frame. The function of DC-LMs is to ensure that the field cage is completely submersed. On the other hand, the CRP-LMs are used for the adjustment of the CRP with respect to the LAr surface and monitoring the level between the extraction grid and the bottom electrode of the LEMs. \figref{fig:SC-levelmeters} shows a cut view of the TPC indicating the design positions of two level meters (DC-LM and CRP-LM) when the liquid level is at its nominal height with the field cage completely submerged and the extraction grid covered by \SI{5}{\mm} of LAr.

The leveling of the CRP with respect to the liquid surface is performed by adjusting the height of each of the three suspension points based on the feedback provided by the CRP-LMs.

\begin{figure}[ht!]
  \centering
 \includegraphics[width=\textwidth]{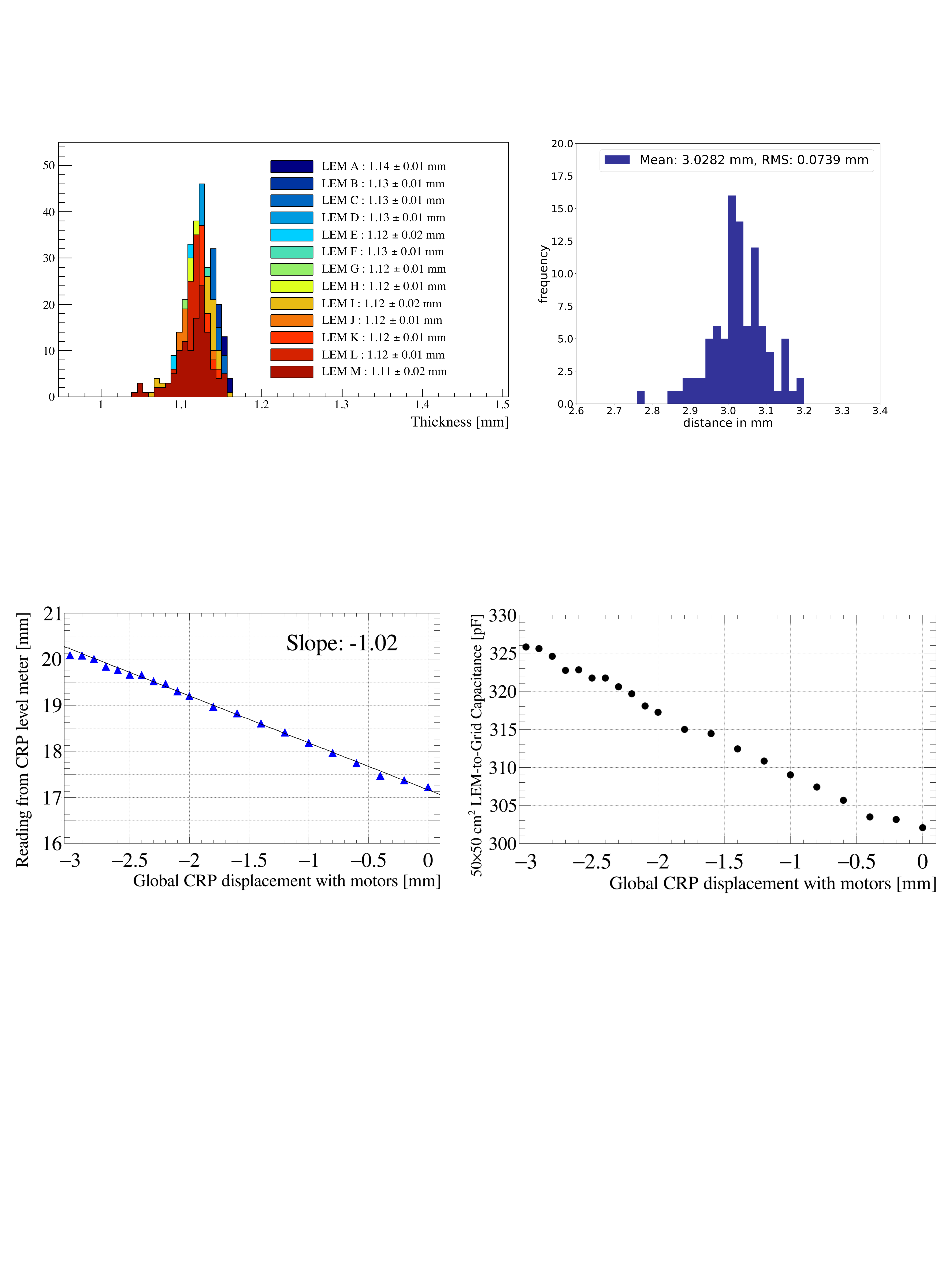}
  \caption{Reading from one CRP level meter (left) and \fifty LEM to extraction grid capacitance (right) as a function of the vertical position of the CRP given by the motorised system.}
     \label{fig:level-vs-motor} 
\end{figure}

The sensitivity of the CRP-LMs to the variations in the liquid level is illustrated in \figref{fig:level-vs-motor} (left), which shows the response of one of the CRP-LM to the CRP movement. The liquid level is measured as the CRP is lowered using the motors by \SI{3}{\mm} in steps of \SI{0.1}{\mm}.

In addition to the installed level meters, the liquid height in the extraction region of the CRP can be inferred by measuring the capacitance between the grid and the bottom electrode of each LEM. Averaging over all 12 LEMs the measured values of this capacitance typically range from \SI{150}{\pico\farad} with the liquid below the grid to around \SI{350}{\pico\farad} when the LEMs are submerged (see \tabref{tab:CRP-capa}). The sensitivity of this technique is illustrated in \figref{fig:level-vs-motor} (right), which shows the measured capacitance between one  LEM and the grid as a function of the CRP displacement. This method offers the potential advantage of monitoring the liquid level in the CRP extraction region with a \fifty granularity and could be used for the CRP level adjustment in the future large-scale detectors where, due to the space constraints, placement of the level meters along the CRP perimeter would not be possible.

Prior to the leveling of the CRP with respect to the liquid argon surface, the readings of the CRP-LMs were calibrated to better than $15\%$ and their pedestal offsets were adjusted within \SI{1}{\mm} accuracy.  The deformation of the CRP frame mentioned does not affect our capacity to immerse the entire grid surface without the liquid touching the LEM plane. This is verified by measuring the capacitance between each \fifty LEM and the grid and further confirmed during high voltage operation of the CRP as explained in \secref{sec_operations}.

 \subsection{Cryogenic cameras}

Four identical digital video cameras capable of operating in liquid argon are installed throughout the cryostat. They provide the visual feedback for the monitoring of the detector filling and inspection of the stability of the liquid argon surface. In addition, they can also be used to detect potential electric discharges. Five LED strips, each approximately \SI{5}{\meter} long, provide the necessary illumination inside the cryostat. The cameras and LED lights are turned off and disconnected during data taking, since they generate significant noise on the charge readout electronics. They also generate light that could damage the PMTs in operation.

The principal parameters of the camera system are summarised in \tabref{tab:camera-para}. The cameras are based on the commercially available Raspberry Pi V1 digital camera module. The main selection criterion for the camera model was its demonstrated ability to undergo long-term operations at cryogenic temperatures in absence of a local heat source (which could induce formation of bubbles) without exhibiting any image degradation. Low power consumption, cost-effectiveness, compact size, and the capability of reading the sensor remotely over a distance of a few meters were also important aspects.

\begin{table}[htb]
\begin{center}
\begin{tabular}{l|c} 
\hline
\hline
Size & \SI[product-units=power]{25x24x9}{\mm}\\
Weight & \SI{3}{\gram}\\
Sensor & OmniVision OV5647\\
Sensor resolution & 2592 $\times$ 1944 pixels\\
Focal length & \SI{3.6}{\mm}\\
Fixed focus & \SI{1}{\meter} to $\infty$\\
Focal ratio & 2.9 \\
Max frame rate & 120 fps\\
Connection to Raspberry Pi & 15 wire FFC\\
\hline
\hline
\end{tabular}
\caption{Specifications of the Raspberry Pi digital camera module.}
\label{tab:camera-para} 
\end{center} 
\end{table}
 The cameras are contained in a custom-made case composed of a DN40 CF Quartz window, a 20\,mm thick spacer flange and a 15 pin SUB-D flange at the back (see \figref{fig:SC-LM-T}). The system is assembled in an argon atmosphere to avoid development of condensation on the lens once it is cooled to cryogenic temperatures. Each camera is connected via a 15 wire flat flexible cable (FFC) to its own Raspberry Pi computer for image acquisition. Cable lengths of up \SI{8}{\meter}, tested at room and liquid argon temperatures, showed no perceivable image distortions. 
 
 \begin{figure}[ht!]
 \centering
 \includegraphics[width=\textwidth]{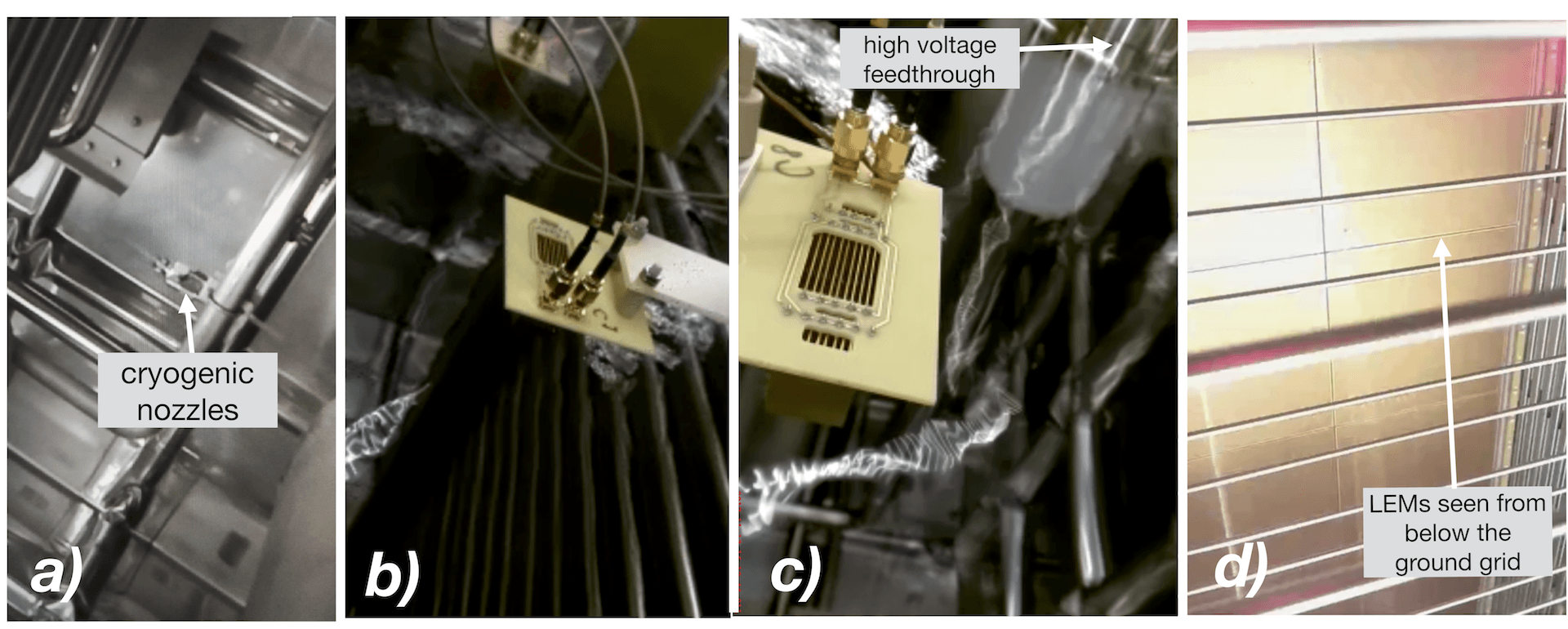}
 \caption{Images of inside the cryostat provided by the four cryogenic cameras. Camera a) monitors the functioning of the cryogenic nozzles during cool down. The liquid level during operation can clearly bee seen by cameras b) and c). Camera d) is placed under the ground grid and cathode (clearly visible on the picture) and observes the \fifty LEMs.}
 \label{fig:camera-views} 
\end{figure}

\figref{fig:camera-views} shows some of the recorded images and also illustrates the parts of the detector visible to each camera. The three cameras, providing pictures \textit{a}, \textit{b} and \textit{c}, are fixed about 30\,cm above the CRP frame in the gas and orientated downwards to monitor the state of the liquid argon surface in the vicinity of the CRP level meters. As can be seen in picture \textit{c}, the entire length of the high voltage feedthrough is also visible, particularly the ground termination ring where the electric field is expected to be the highest \cite{Cantini:2016tfx}. The fourth camera (picture \textit{d}) is located in the liquid argon below the ground protection grid of the PMTs. It faces the CRP, capturing most of the LEMs.

\subsection{CRP high voltage system}

The nominal operation of the CRP requires biasing the extraction grid to \SI{-6.8}{\kilo\volt} and the LEM bottom (top) electrode to \SI{-4.3}{\kilo\volt} (\SI{-1.0}{\kilo\volt}). The two LEM electrodes are polarised independently. In addition, there are two contacts for the extraction grid for redundancy. The CRP operation therefore necessitates altogether 26 independent high voltage channels, which are conservatively required to support voltages of up to \SI{10}{\kilo\volt}. Due to substantially lower dielectric strength of gas argon relative to air, the transport of such high voltage inside the cryostat is a delicate matter requiring a careful selection of cables, HV contacts, and feedthrough interfaces.  

\begin{figure}[ht!]
  \centering
 \includegraphics[width=\textwidth]{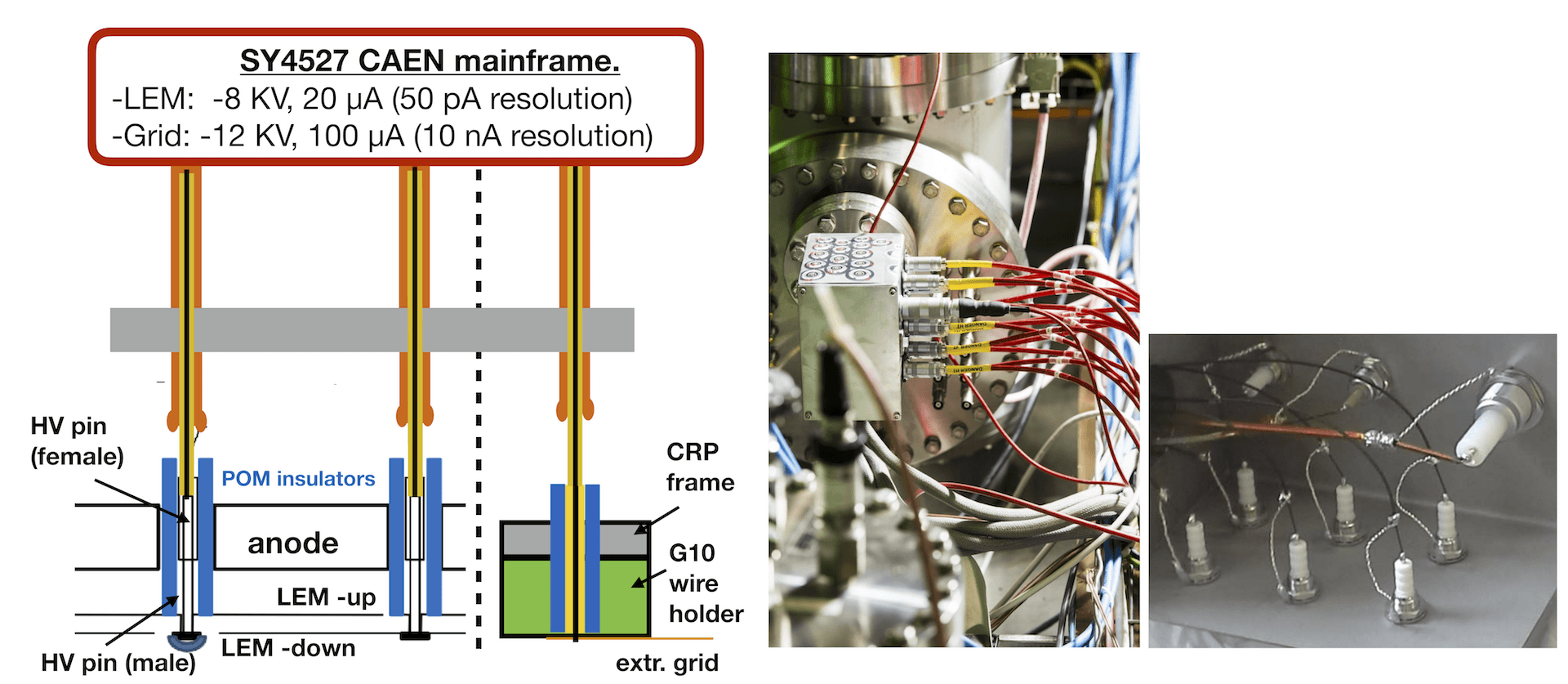}
  \caption{Illustration and pictures of the CRP high voltage distribution. Left: a schematic for the biasing of two LEM electrodes and the extraction grid. Centre: external picture of the CRP high voltage feedthrough during operation. Right: internal view of the feedthrough before casting with the polymer.}
     \label{fig:CRP-HV} 
\end{figure}
A schematic of the high voltage distribution for two LEM electrodes along with pictures of the feedthrough are shown in \figref{fig:CRP-HV}. The high voltage to the CRP elements is transported internally by coaxial cables\footnote{KAPW-50ohm www.lewvac.co.uk} rated for 10 kV operation in high vacuum applications. At the LEM and grid contacts on the CRP, the cable ground shield is stripped $10-15$ cm away from the internal conductor and protected with an insulating retractable sheath. The contacts for both LEM electrodes are made with the high voltage pins described in \secref{sec_tpc_crp}. These pins traverse the anode PCB and are protected by circular POM insulators. For the extraction grid, the inner conductor is soldered directly on the PCB that supports the grid wires. 

The high voltage for the CRP elements (as well as the PMTs) is delivered inside the cryostat via two feedthroughs, each containing 17 channels qualified for up to \SI{10}{\kilo\volt} operation (see \figref{fig:CRP-HV}). The feedthroughs are custom-built, since no commercial solution capable of withstanding the required voltages in pure argon gas was found. The feedthroughs are made from a stainless steel box within which the conductor and shield of the coaxial cable are soldered on a SHV connector for each channel. The entire inner volume of the box is subsequently filled with an electrically insulating polymer. The feedthroughs are tested prior to installation to ensure that they can withstand up to \SI{-10}{\kilo\volt} in argon gas at room temperature and are leak-tight at the level of \SI{\leq1e-9}{mbar.\liter/\second} (limit of the spectrometer sensitivity). 

The high voltage for the extraction grid and LEMs is provided by the SY4527 CAEN mainframe. The system allows to monitor the LEM (grid) currents with a resolution of about \SI{50}{\pico\ampere} (\SI{10}{\nano\ampere}). The values of the measured current and voltage for each channel are read at a frequency of \SI{2}{\hertz}, the maximal rate supported by the power supply unit, by the slow control system and stored in a database.

\subsection{Detector slow control back-end}
\label{sec_sc_backend}
The detector slow control system manages the supervision of all the quantities of interest listed in \tabref{tab:SC-para}. A supervisory system based on the Siemens WinCC Open Architecture software continuously reads the values, reacts to changes, compares them to settings issuing warnings and alarms when necessary.
All the data are recorded in a mySQL database and can be retrieved later for offline analysis. The system is built on robust and versatile National Instruments 9082 compact RIO controllers\footnote{http://www.ni.com/compactrio}, which run on FPGAs and hardware links guaranteeing fast and fail-proof operation with minimal human intervention. Hardware links are present between sub-systems that have safety implications on the detector operation, for instance between high voltage power supplies and readings related to the cryogenic status of the cryostat. A loss of vacuum insulation in a cryogenic pipe, unforeseen variation on pressure or any indication of a liquid argon spill from Oxygen Deficiency Hazard sensors would automatically send an interlock signal to all detector high voltage channels.

\begin{figure}[ht!]
  \centering
 \includegraphics[width=\textwidth]{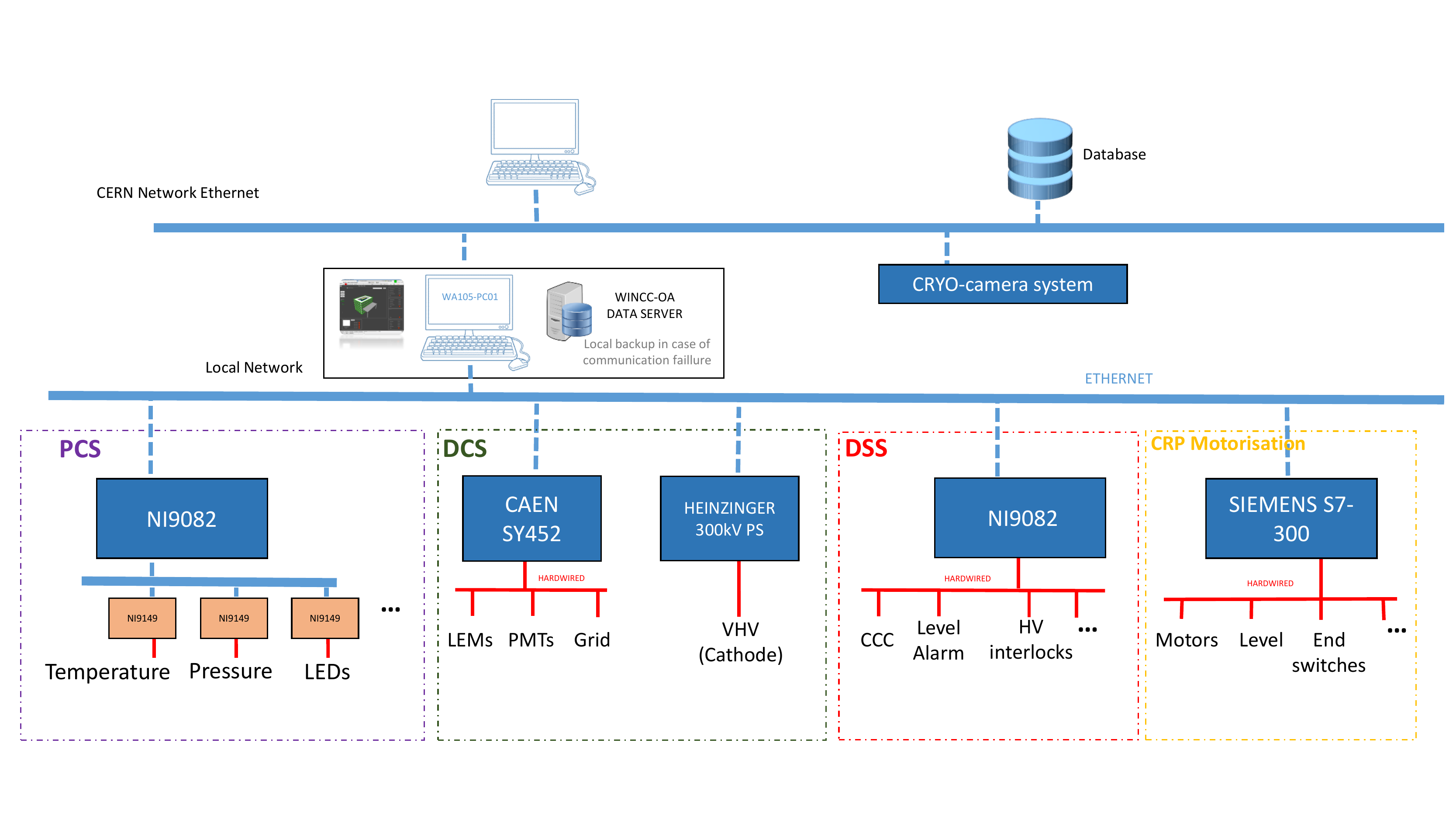}
  \caption{Schematic of the detector slow control system.}
     \label{fig:SC-backend} 
\end{figure}

The overall architecture of the system is represented in \figref{fig:SC-backend}. It can be divided in four sub-systems according to their functionality:
\begin{itemize}
\item \textbf{Process Control System (PCS).} The central control and acquisition system is based on fast (40 MHz) FPGA cards contained inside a module that communicates using protocols developed according to CERN standards. As an example for temperature measurements, we use the modules NI9219 which can acquire data at a rate of up to 100 Hz with 24 bits precision. The central module is linked through Ethernet to multiple other sub-modules each containing their own FPGA, that process information such as temperature and pressure. The system can thus be easily extended by adding sub-modules according to the required number and type of sensors.
\item \textbf{Detector safety system (DSS)}. The safety system has its own FPGA module. The sensors whose readings are subject to issue major alarms are hard-wired directly to the module. This is the case for instance for the level meters and some pressure sensors. The clear advantage of using FPGAs is their fast processing and reaction time, but also the simplicity of using simple logic binary gates instead of microprocessors, making the system more robust. The DSS therefore acts directly on some chosen subsystems through hardware links without human intervention. To be noted that less critical alarms or warnings can be set and modified at the level of the supervision software (WINCC-OA) at any time independently of the DSS. 
\item \textbf{Detector control system (DCS)}. The DCS system controls and acquires the values of all high voltage channels with their respective currents. Some of the actions can be interlocked by the DSS for security reasons
\item{\textbf{CRP motorisation.} A dedicated PLC connected to the global slow control system provides the remote control and monitoring of the CRP motors.}
\end{itemize}


\section{Detector readout, data acquisition and processing}
\label{sec_elec}
In this section we provide a detailed description of the detector  readout system which includes the amplification, digitisation and storage of the charge and scintillation signals. In the case of the charge readout, some processing of the events is also performed online in order to provide rapid feedback on the performance of the detector.
The scintillation signals from the PMTs, in addition to serving as inputs to the event trigger, are also digitised and stored for analysis. An event by event matching of scintillation and charge data is performed offline. For future detectors \cite{DeBonis:1692375,Acciarri:2016ooe}, both systems will be fully synchronised by using similar electronics to that of the charge readout adapted to the PMT signals. 

\subsection{Charge readout overview}
The analogue and digital charge readout system utilised in the \three detector is the result of more than 10 years of developments aiming at finding an integrated and cost effective solution for the readout of very large dual liquid argon TPCs. The R\&D focused on three main aspects:
\begin{itemize}
    \item{Development of multi-channel cryogenic low-noise charge amplifier ASIC.}
    \item{Development of low-cost digital electronics supporting high data throughput without zero suppression.}
    \item{Usage of synchronous Ethernet with Precision Time Protocol (PTP) exchanges for time synchronisation and event timing as well as triggering.}
\end{itemize}
The analogue and digital electronics and the DAQ components are part of the anticipated production and procurement for the charge readout system of the ProtoDUNE-DP detector.

One of the key features of the dual-phase TPC is the possibility to place the analogue front end (FE) cards near the readout thereby profiting from the cryogenic environment while at the same time having the possibility to extract the cards for maintenance during operation of the detector. The FE cards amplifying the charge from the anode strips are mounted inside dedicated feedthroughs called \textit{Signal Feedthrough} (SFT) that traverse the entire cryostat walls (see \figref{fig:SFT}). The cards are fixed at the bottom of the SFT close to the CRP. The differential analogue signals are then guided up towards the exterior through twisted pair ribbon cables where they are digitised with \SI{12}{bit} resolution at a frequency of \SI{2.5}{\MHz} in Advanced Mezzanine Cards (AMC) located in commercial uTCA crates \cite{uTCA}.  Each crate services one SFT processing the signals from 320 channels. The digitised signals are subsequently sent to an event-building machine in the global DAQ via a \SI{10}{\giga bit \per\second} optical fiber link. The back-end of DAQ includes an online storage and processing facility that is designed to cope with high data throughput. The time synchronisation of different AMC cards as well as the trigger distribution scheme utilises a dedicated White Rabbit\footnote{https://www.ohwr.org/projects/white-rabbit} network.

As shown in \figref{fig:311-coordinates} the \num{960} one meter long anode strips are connected to three SFTs and the 320 three meter strips to one SFT. The charge readout therefore has to deal with 1280 channels organised in four uTCA crates. The full one meter drift of the ionising charge in \SI{\sim500}{\volt/\cm} electric drift field corresponds to about \SI{700}{\micro\second} time window. The entire time window is digitised at \SI{2.5}{\mega\hertz} corresponding to \num{1667} 12-bit samples providing an event size from all of the \num{1280} channels of about \SI{3}{MB}. The events are triggered by a logic signal (TTL) which can be provided by either a pulse generator (for testing and debugging purposes), the Cosmic Ray Taggers (CRTs) or PMTs during cosmic ray data taking.

\subsection{Charge signal feedthroughs}
\begin{figure}[ht!]
 \centering
 \includegraphics[width=\textwidth]{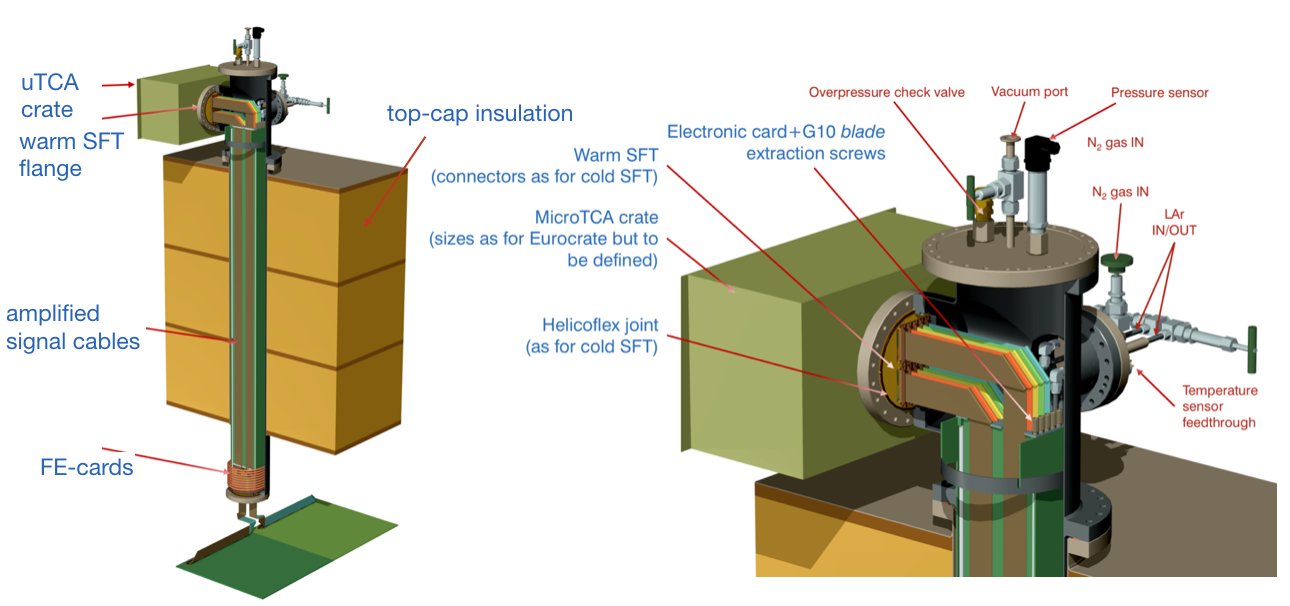}
 \caption{Details of one charge readout signal feedthrough.}
 \label{fig:SFT} 
\end{figure}

\begin{figure}[ht!]
 \centering
 \includegraphics[width=\textwidth]{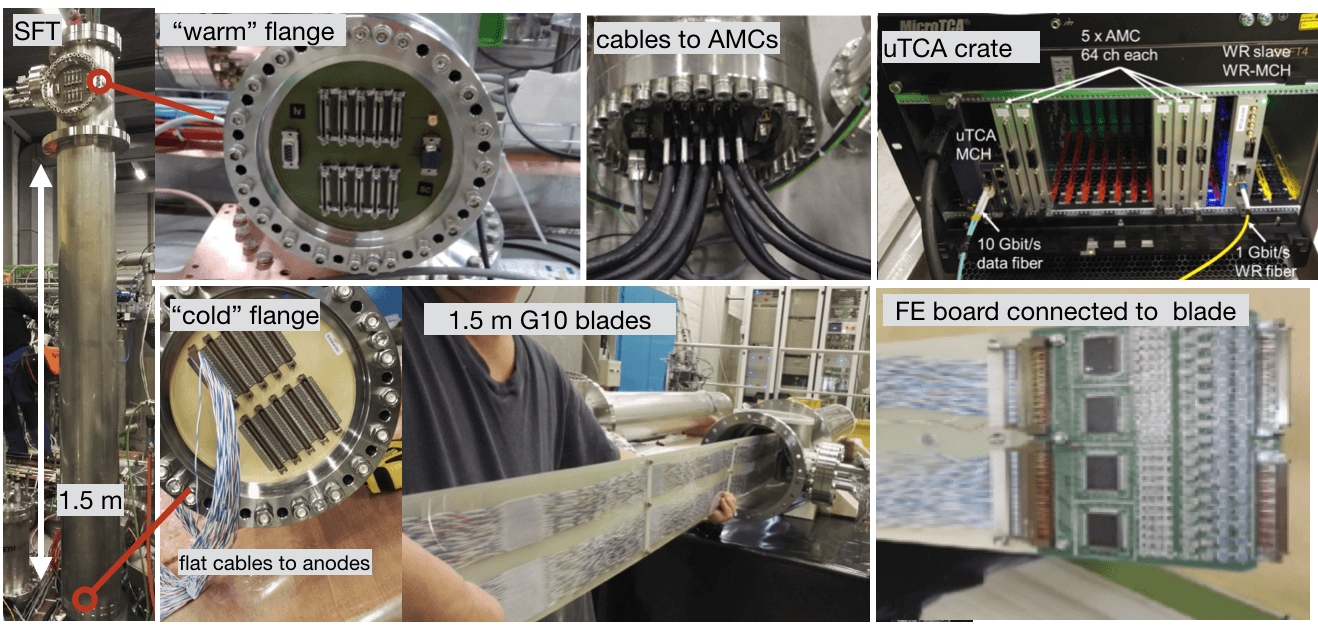}
 \caption{Pictures with details of a signal feedthrough and of a uTCA crate.}
 \label{fig:CRO-overview} 
\end{figure}

The SFTs consist of a \SI{1.5}{\meter} stainless vacuum tight steel tube terminated with the appropriate UHV flanges that provide the interface for routing the signals. They are designed to enable the access of the FE analogue electronics for possible repair or exchange without contaminating the ultra-pure argon in the cryostat. In addition, their metallic structure acts as a Faraday cage completely shielding the FE cards from the external environment and potential noise induced by the digital electronics. Details of the design of an SFT are shown in \figref{fig:SFT} and some pictures are provided in \figref{fig:CRO-overview}.

The SFTs  are closed at their bottom and top with vacuum tight flanges made from thick multitlayer PCBs. The flange at the bottom (\textit{cold flange}) interconnects the signals from the anodes to the analogue FE electronics cards located inside the tube and directly connected to the cold flange. The flange at the top (\textit{warm flange}) acts as interface to route the analogue signals towards the digitisers and also feeds the required low voltage and control lines to the FE cards. If required the temperature inside the SFT may be adjusted using a LAr-cooled heat exchanger coil surrounding the FE cards.

The FE cards are mounted on \SI{\sim1.5}{\meter} long blades made from G10, which allow for the insertion or extraction of the analogue electronics from the top of the SFT. The blades also support the flat cables that transmit the signals, low voltage, and other control lines. The blades are guided by rails precisely positioned inside and over the entire height of the tube to ensure that the FE card is guided and correctly plugged into its respective connectors on the cold flange. 

Prior to cooling the detector for operation, the SFTs are evacuated and then filled with nitrogen gas. In order to access the electronics the stainless steel top flange must be removed. Opening the SFT once the detector is in operation requires continuous flushing of nitrogen gas at slight over-pressure with respect to the atmospheric so as to prevent the humidity from entering. Once the top flange is removed, the blades with the FE cards can be extracted after unplugging the flat cables (two per blade-card assembly) connected on the inner side of the warm flange (see \figref{fig:SFT}). The procedure to access the electronics at cold was successfully tested at one point during detector operation. The top of the SFT remains at room temperature allowing for easy handling of the cables connected to the inner side of the warm flange. The movement of the blades on the rails and insertion/extraction of the FE cards on the cold flange exhibited similar behaviour as observed during the same operations at warm. In addition, while unplugging/plugging the cards, the electronics was left on; no adverse changes were subsequently seen in its performance. The access during cryogenic operation lasted approximately one hour during which time the SFT was continuously open and the temperature at the bottom increased by about \SI{30}{\kelvin}.

\subsection{Analogue front-end cards}
\begin{figure}[ht!]
 \centering
 \includegraphics[width=.8\textwidth]{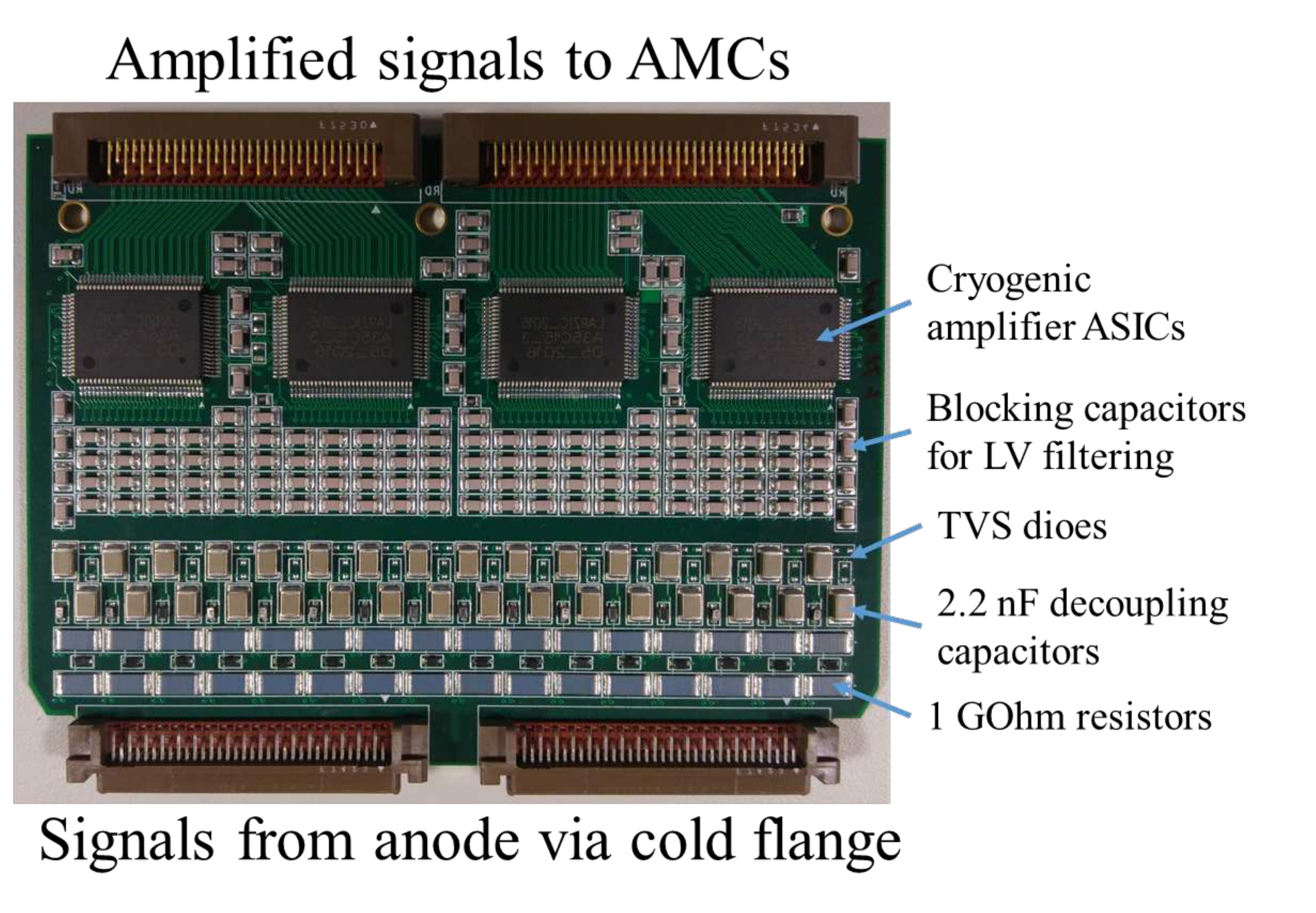}
 \caption{Details of an analogue FE card.}
 \label{fig:FE_card} 
\end{figure}

Each analogue FE card, shown in \figref{fig:FE_card}, hosts four ASIC charge amplifier chips and a few passive discrete components. Some of the principal properties and parameters of the front-end electronics are summarised in \tabref{tab:preamp-para}. The input stage of every channel has a decoupling capacitor of \SI{2.2}{\nano\farad} and a \SI{1}{\giga\ohm} resistor that connects the anode strips to ground. An ESD device\footnote{TVS diodes Bourns CDSOD323-T08LC} is also included into each input stage and is used to protect the amplifiers against discharges coming from the detector.
The particular device was selected after a long campaign aimed at checking the performance of different ESD components subjected to many discharges of a few kV and a stored energy similar to the one of the LEMs. In addition, blocking capacitors to filter the low voltage power lines are included in the card.

\begin{table}[ht!]
\begin{center}
\begin{tabular}{p{.5\textwidth}p{.2\textwidth}}
\hline
\hline
 Parameter & Value \\
\hline
Number of channels per ASIC & 16 \\
Number of channels per FE card & 64\\
CSA dynamic range & \SI{1200}{\femto\coulomb} (40 MIP) \\
Peaking time & \SI{1}{\micro\second} \\
Noise at \SI{110}{\kelvin} for \SI{160}{\pico\farad} (\SI{480}{\pico\farad}) $C_{det}$   
 &  950 (1980) $\mbox{e}^{-}$ \\
Power consumption per channel & \SI{<18}{\milli\watt} \\
\hline
\hline
\end{tabular}
\caption{ASIC and FE card parameters and specifications.}
\label{tab:preamp-para}
\end{center} 
\end{table}

\begin{figure}[ht!]
\centering
\includegraphics[width=.49\textwidth]{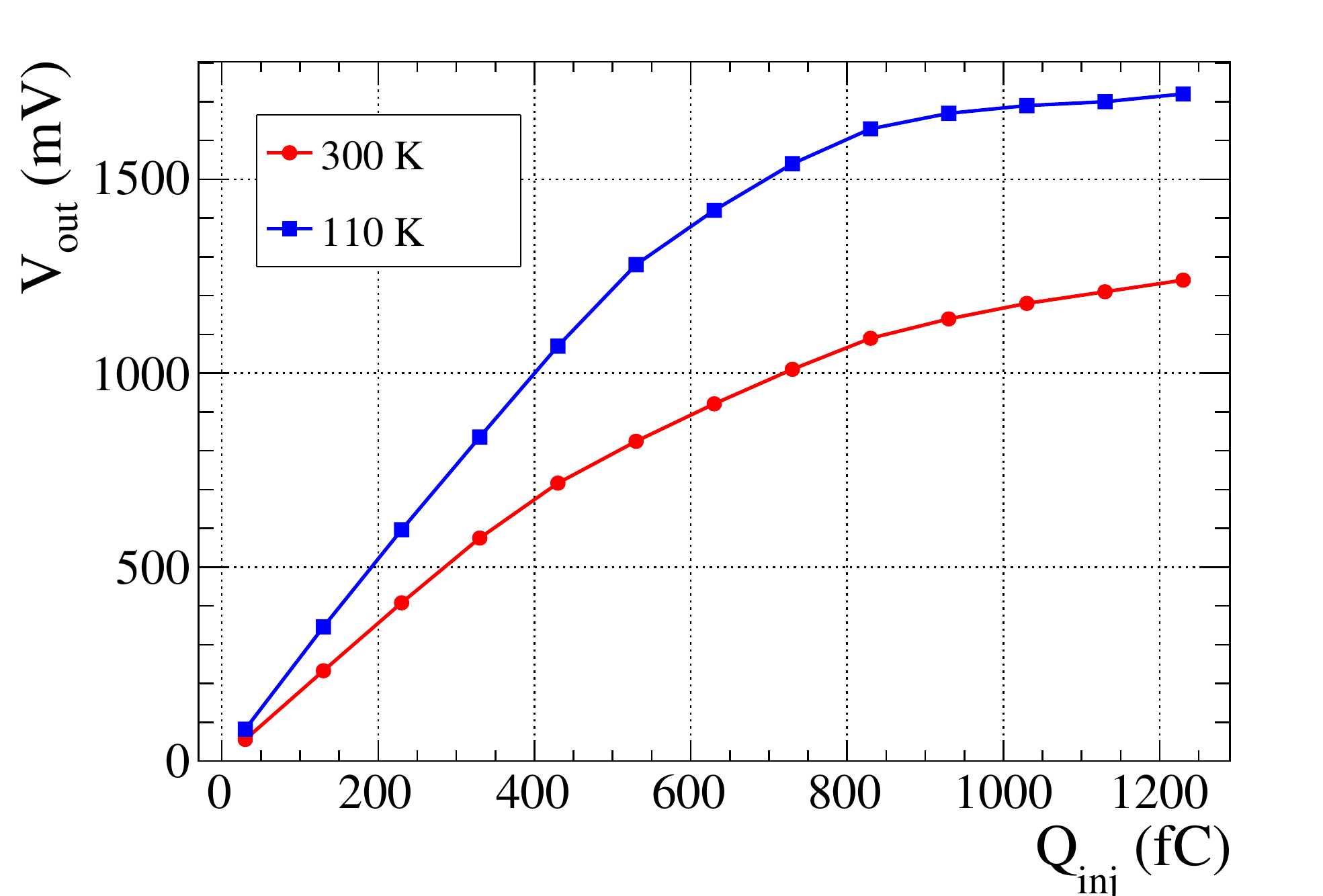}
\includegraphics[width=.49\textwidth]{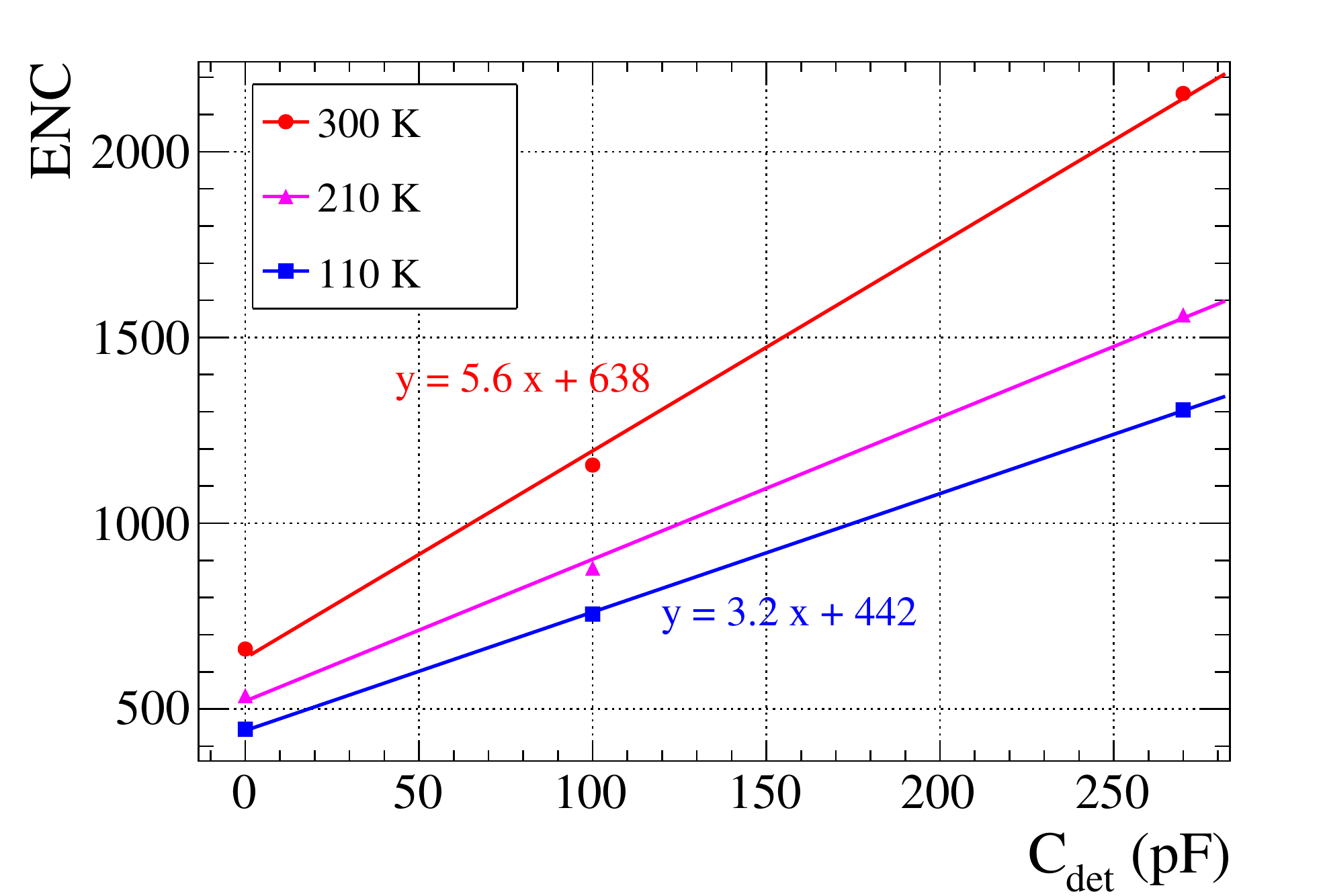}
 \caption{Cryogenic FE ASIC properties: amplifier response (left) and noise (right) at different
temperatures measured at the output of differential buffer. ENC is expressed in terms of the number of electrons.}
 \label{fig:fe-asic-response} 
\end{figure}

The principal component of the FE analogue cards is the cryogenic charge amplifier ASIC based on CMOS \SI{0.35}{\micro\meter} technology with a large dynamic range (\SI{1200}{\femto\coulomb}) to cope with charge amplification in the CRP. The chip features \num{16} input channels each consisting of a charge sensitive amplifier (CSA), differential output buffer stage acting as low pass filter, and \SI{1}{\pico\farad} test capacitor. The CSA has a linear gain for input charges of up to \SI{400}{\femto\coulomb} and a logarithmic response in the \SIrange{400}{1200}{\femto\coulomb} range. This double-slope behaviour is obtained by using a MOSCAP capacitor in the amplifier feedback loop that changes its capacitance above a certain signal threshold of injected charge. The ASIC power consumption is less than \SI{18}{\milli\watt} per channel. \figref{fig:fe-asic-response} (left) shows the response of the amplifier measured at the output of the differential buffer stage at room and cryogenic temperatures for different values of the injected charge. The input charge per channel for a MIP is of around \SI{1.5}{\femto\coulomb} (\SI{30}{\femto\coulomb}) for the CRP operated at $\Geff=1$ ($\Geff=20$). The measured noise in multiples of the amplifier input or Equivalent Noise Charge (ENC) as a function of the input detector capacitance $C_{\mbox{det}}$ at different temperatures is shown in \figref{fig:fe-asic-response} (right).

\subsection{Digital electronics, timing, and data acquisition}

\begin{figure}[ht!]
 \centering
 \includegraphics[width=.87\textwidth]{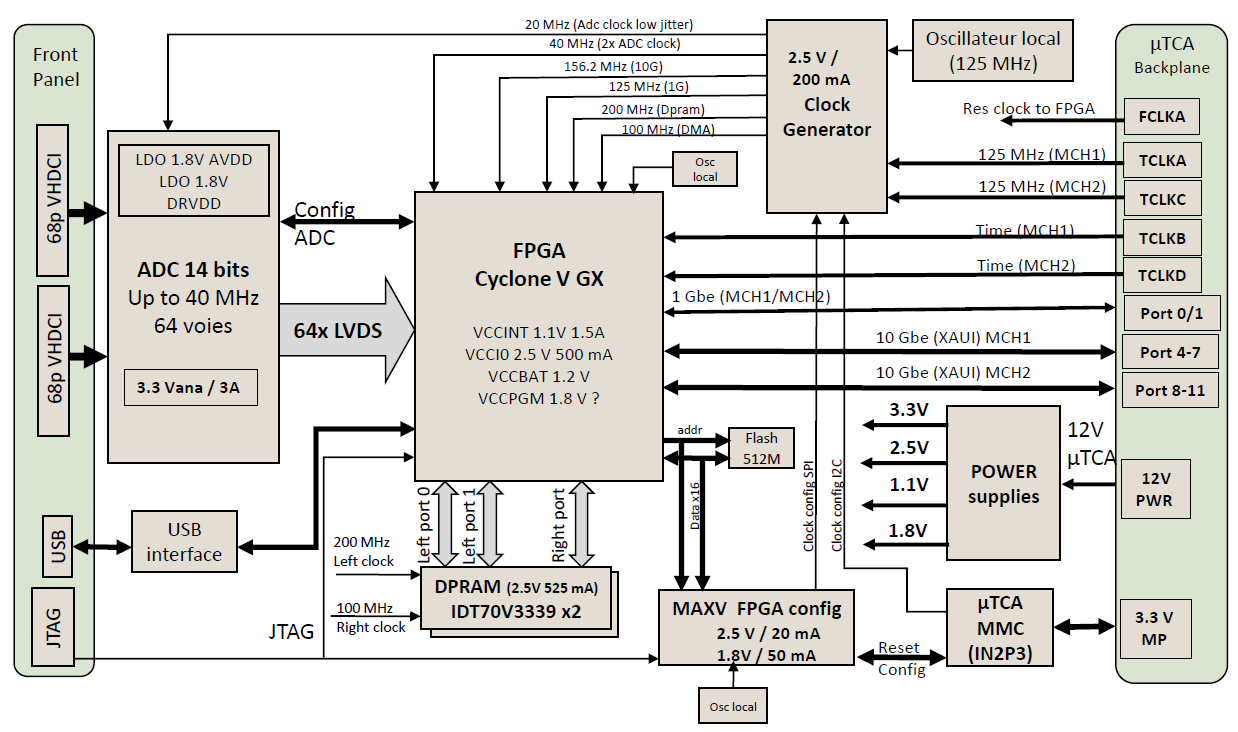}
 \caption{The synoptic of the AMC digitisation card.}
 \label{fig:AMC_scheme} 
\end{figure}

The AMCs read and digitise the data from the FE amplifiers and transmit them over the DAQ network to the event-builder station. Each card has eight ADC chips\footnote{AD9257} two dual-port memories\footnote{IDT70T3339} and an FPGA\footnote{ALTERA Cyclone V} on board. The FPGA provides a virtual processor, NIOS\footnote{https://www.altera.com/products/processors/overview.html},  which handles the readout as well as the data formatting  and transmission. \figref{fig:AMC_scheme} shows the general synoptic of the actual card. Each AMC has 64 channels and reads one analogue FE cards.

The data is continuously sampled at \SI{2.5}{\MHz}, digitised with \SI{14}{bit} accuracy, and recorded into a local memory buffer. The \SI{14}{bit} ADC was chosen to ensure the linearity over the \SI{12}{bit} range required for the TPC signals, since the least significant bit (LSB) is typically affected by noise. The two LSBs of each sample are suppressed prior to transmitting the data to the DAQ back-end. The data is acquired with a common timestamp generated by an external trigger, which defines the start time for the data sequence to be retrieved from the memory buffer. The length of the sequence covers the entire drift window for the ionising charge to arrive from the cathode to the CRP. 
For a time window of around \SI{667}{\micro\second}, \num{1667} of \SI{2.5}{\MHz} samples per channel are needed. The AMCs organise the collected data in packets for transmission without applying any zero-suppression. These are collected, via the network lines implemented on the uTCA crate back-plane, by the MicroTCA Carrier Hub (MCH) switch present in each crate. Digitised data is eventually sent from the uTCA crate to an event-building machine via a 10 Gbit/s optical fiber link.    

\begin{figure}[ht!]
 \centering
 \includegraphics[width=.89\textwidth]{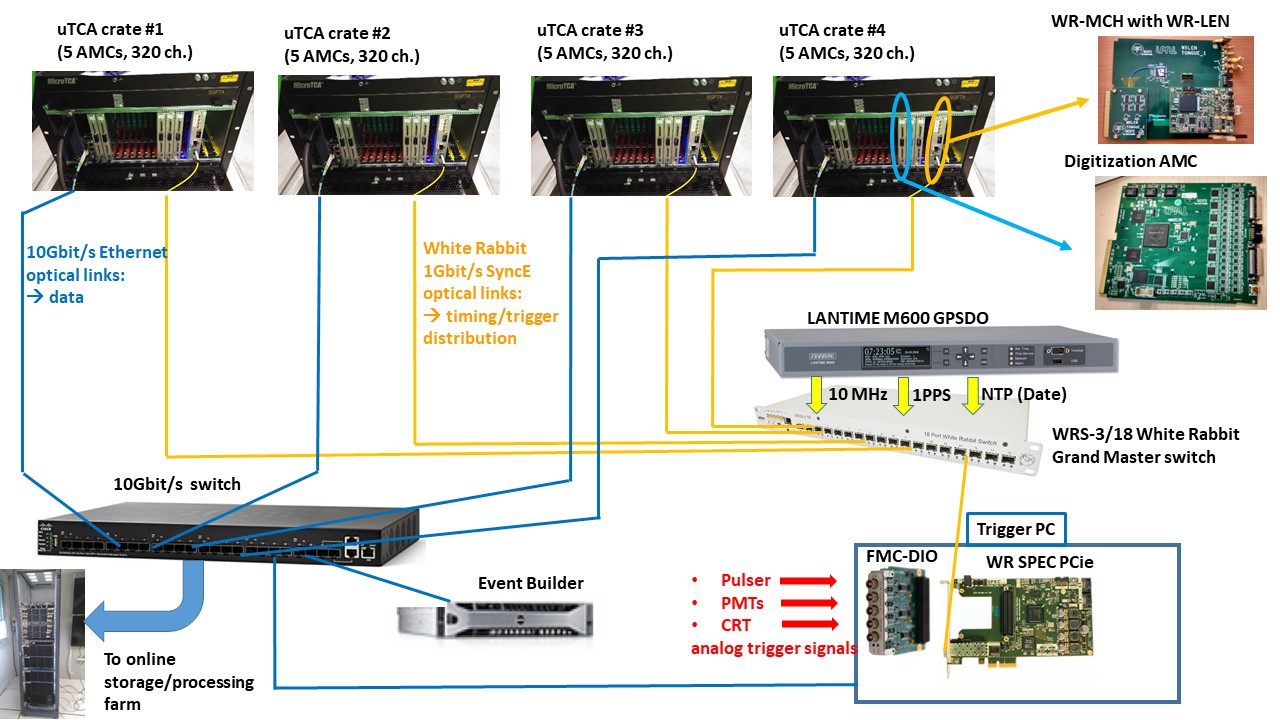}
 \caption{Scheme of the  DAQ/timing system implemented for the \three detector.}
 \label{fig:DAQ_scheme} 
\end{figure}
The overall scheme of the DAQ system and the layouts of the data and timing/trigger networks are illustrated in \figref{fig:DAQ_scheme}. The time synchronisation of the AMC units is facilitated by a White Rabbit (WR) system that operates on a dedicated synchronous \SI{1}{\giga bit \per \second} Ethernet (SyncE) network. A GPS disciplined clock unit\footnote{Meinberg LANTIME M600} feeds 10 MHz and 1 pulse per second (PPS) reference signals to a commercial White Rabbit switch\footnote{Seven Solutions WRS V3.4}. The switch acts as Grand Master for all the connected nodes synchronising them to the common reference time. It also communicates through a standard Ethernet port with the GPS clock unit for its date and time synchronisation via NTP. The WR system automatically performs periodic self-calibrations to account for propagation delays achieving sub-ns accuracy on the clock synchronisation between distant nodes. Through a dedicated R\&D the WR network has been also adapted for transmission of the trigger timestamps generated from external sources (e.g., cosmic ray taggers, PMTs, pulse generator).

Each uTCA crate is equipped with a specially designed slave node, White Rabbit MicroTCA Carrier Hub (WR-MCH), which distributes the timing and trigger information to the AMCs via the back-plane of the uTCA crate using dedicated lines and data-frame protocol. The card contains a WR slave node card, the White Rabbit Lite Embedded Node\footnote{Seven Solutions OEM WR-LEN}, as a mezzanine that runs on a customised firmware enabling it to decode the trigger timestamp data packets.

The trigger timestamps are generated by a White Rabbit TimeStamping Node (WR-TSN) card hosted in a dedicated computer (\textit{trigger server}). The WR-TSN module is built from commercial components: an FMC-DIO mezzanine card\footnote{Seven Solutions FMC-DIO 5CH} mounted on a Simple PCIe FMC Carrier (SPEC) board\footnote{Seven Solutions  White Rabbit SPEC FMC PCIe Carrier V4}. SPEC operates with a specialised firmware and has an SFP optical transceiver for the connection to the WR network for synchronisation and distribution of the trigger timestamp information. The FMC-DIO mezzanine accepts TTL-level signals from external trigger systems. Once a trigger signal is received, FMC-DIO generates its timestamp. The trigger server, in communication with the event builder station, monitors if the latter is able to accept new events. If the event builder is ready when a trigger signal is received, a dedicated timestamp data packet is sent over the WR network (via the SPEC card) for the connected WR-MCH nodes triggering the detector readout. The event would be lost if, in the case, very rare for this configuration, that the event builder is not ready.

The light readout signals are acquired using commercial electronics. The analogue signals from each PMT are digitised over a \SI{1}{\milli\second} window, which corresponds approximately to the maximum drift time of the electrons. The digitisation is performed with a resolution of 12 bit sampled at \SI{250}{\mega\hertz} using a CAEN v1720 board. 
To limit the data volume in some runs, a \SI{4}{\micro\second} event time window was acquired  to record only the S1 signal. The ADC has a total dynamic range of \SI{2}{\volt} limiting the PMT gain. The board is read out via an optical link to a PC equipped with a CAEN A2818 PCI CARD. The software for the display and acquisition is based on the MIDAS framework \cite{MIDAS} and runs on the same PC. The board also allows to program a simple majority coincidence trigger. 

\subsection{Event triggering}\label{sec:trigger_conditions}
\label{sec:CRT}
\label{PMTDAQ}

The charge readout system is triggered via TTL-level pulses sent to the WR-TSN module from different trigger signals depending on the desired rate and functionality. Triggers can be generated by a pulse generator for test purpose and anode-strip charge injection measurements. During data taking events can be triggered by either the prompt scintillation in liquid argon, referred to as \textit{PMT trigger}, or on crossing muons by using the CRT panels (\textit{CRT trigger}). Each selects a distinct event topology at different acquisition rates. Quasi-horizontal muons crossing the entire fiducial volume are triggered by the CRTs at a rather low rate of around \SI{0.3}{\hertz}. The PMTs produce a trigger if the scintillation signal amplitude exceeds a certain threshold. The threshold is set so that the charge readout DAQ reads events at 3 Hz (this value was chosen because of network limitations of the experimental hall). Events triggered with PMTs will therefore consist of a broader set of events including for instance showering events. 



The Cosmic Ray Tagger (CRT) system \cite{Auger:2016tjc} detects cosmic ray muons and measures their crossing time and coordinates relative to events internal to the TPC.  It consists of two \SI[product-units=power]{1.8x1.8}{\meter} planes mounted on either side of the cryostat external structure walls (see \figref{fig:311-full}) providing an azimuthal angular acceptance of $\pm13.5^\circ$. Each CRT plane is composed of two modules providing the 2D hit coordinates with an accuracy of \SI{\sim1.8}{\cm}. Each module contains 16 scintillator strips equipped with wavelength-shifting fibers that are read by two Silicon Photomultipliers (SiPMs). The SiPM signals are amplified, processed, and digitised on a dedicated Front-End Board (FEB)~\cite{FEB}. The back-end of the FEB is connected to the CRT DAQ computer via an Ethernet daisy-chain. When hit, each CRT module provides a \SI{160}{\nano\second} CMOS pulse. The pulses from the four modules are processed and combined into a \SI{500}{\nano\second} output trigger that is sent to the charge and light data acquisition systems. Both NIM and \SI{3.3}{\volt} LVCMOS signals can be provided. For each trigger, the CRT provides information on the SiPM signal amplitudes as well as the absolute timing, accurate to within \SI{\sim3}{\nano\second}, of the hits with respect to a PPS signal originating from the GPS clock unit used by the White Rabbit system. 

To trigger on the scintillator light, a simple majority coincidence logic is implemented on the CAEN v1720 board in order to send a TTL trigger to the charge readout DAQ when eac hof the five PMTs sees simultaneously a signal greater than \SI{250}\it{ADC} in amplitude (roughly \SI{125}{\milli\volt}) within an \SI{80}{\nano\second} coincidence time window. With this condition, the trigger rate is around \SI{3}{Hz}. A veto window of \SI{200}{\milli\second} is opened once a trigger is generated in order to avoid losses of data packets on the network observed during the data transmission for events too close in time. 

The event builder runs a dedicated multi-threaded application that handles data packets coming from the AMCs. Once the trigger timestamp is generated, the event builder sends a query packet to each card (each is seen as a unique node on the network) to initiate the data transmission. The cards respond by sending their respective data in UDP (User Datagram Protocol) packets, which the event builder assembles into the full event record writing it in the end in a binary format to a file. Each event contains an event header with the event number and trigger information (timestamp, trigger input type) along with some data quality flags (e.g., a code for number of missing UDP packets) followed by the event data, which are arranged in a sequential list of samples from each channel.  

\subsection{Online data processing and storage}
\label{sec:online}

Future large LAr TPCs need to cope with a very large data flow which necessitates a high performance distributed storage system at the PB scale supporting \SI{20}{GB \per\second} bandwidth. A processing farm coupled to the storage servers is also required in order to perform online reconstruction and analysis for checking data quality and monitoring the detector characteristics e.g., liquid argon purity and CRP gain. 

The online analysis relies on reconstructing and selecting some of the cosmic ray muon tracks overlapping within a given readout window of the detector volume. The raw data, written by the event builders, are clustered in files and channeled by the DAQ back-end to the online storage facility hosted at the experimental site. All of the events are processed by running a set of fast reconstruction tools. The reconstructed cosmic ray tracks are then used for the data quality monitoring as well as the online purity and CRP gain measurements. 

A reduced scale of the online storage and processing system was set up for the \three detector. It employs four dedicated storage servers for a total of \SI{192}{TB} of disk space in conjunction with one meta-data server machine. The distributed file system deployed on the storage servers is EOS\footnote{eos.web.cern.ch}. It was chosen after comparative performance studies of the currently available distributed storage systems were carried out on a test-bench setup. The online-processing system is built from a set of batch workers (each containing 7 CPU units for a total of 112 available processors), a batch manager, and a configuration server. TORQUE\footnote{http://www.adaptivecomputing.com/products/open-source/torque/} is used as the resource manager of the system. 

\begin{figure}[ht!]
 \centering
 \includegraphics[width=.49\textwidth]{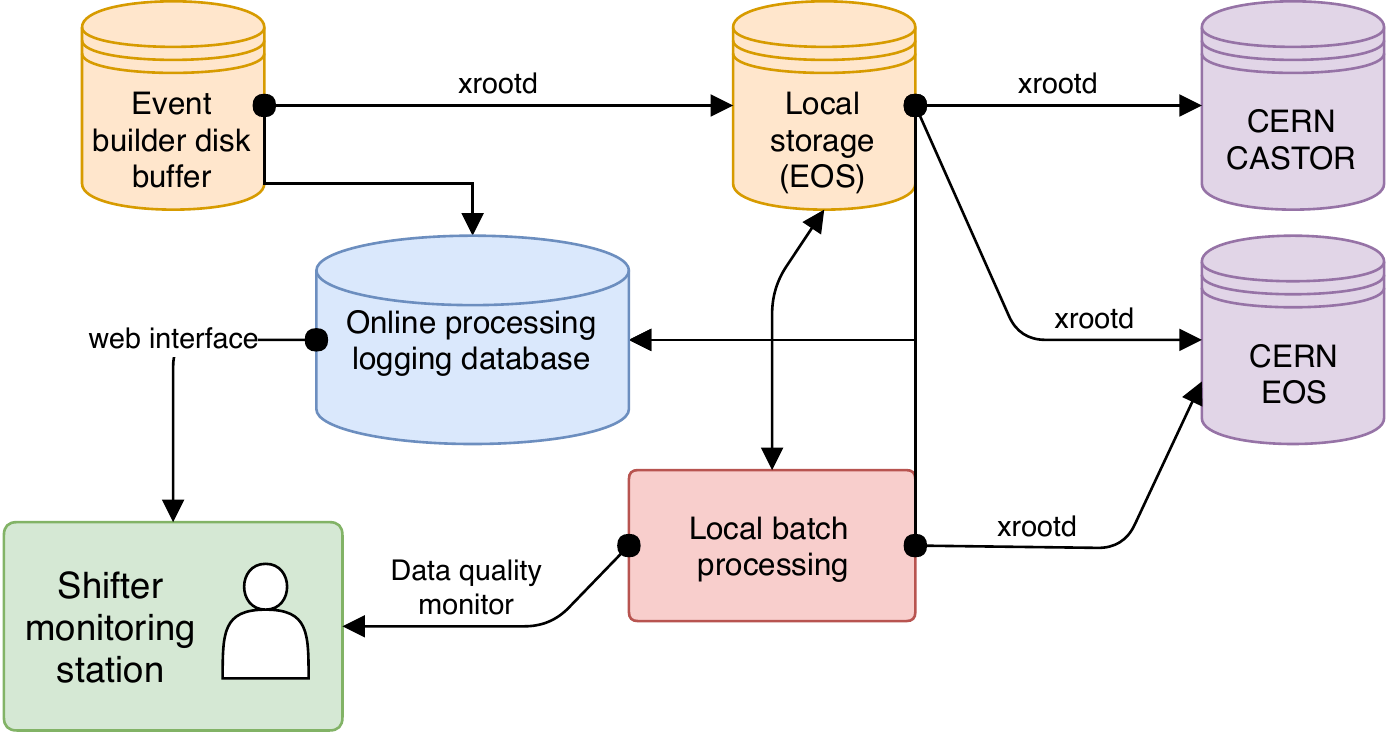}
 \includegraphics[width=.49\textwidth]{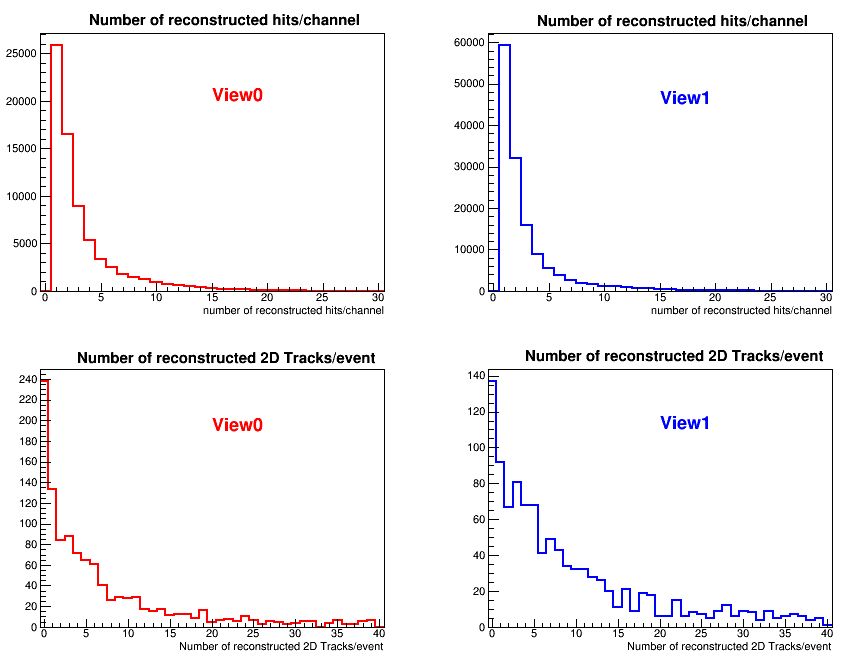}
 \caption{Left: data flow on the online storage and processing farm. Right: example of the online quality monitoring that shows the total number of reconstructed hits per channel and total number of reconstructed 2D tracks per event for each collection view.}
 \label{fig:farm_flow_dqm} 
\end{figure}

An overview of the complete data processing work flow on the online storage and processing farm is shown in \figref{fig:farm_flow_dqm} (left). A data file produced by the DAQ (with a size of about \SI{1}{GB} corresponding to \num{335} events) is immediately copied to the local EOS storage area and scheduled for the transfer to the CERN central storage (EOS and CASTOR). The file transfer protocol is based on xrootd with an addition of the check-sum calculation to strengthen the reliability. To reconstruct all the events in the file, a script is automatically generated and submitted to the batch farm. Reconstruction and analysis results are consequently stored on local EOS and as well scheduled for transfer to the central CERN EOS. Depending on the raw data type (cosmic, pedestals or calibration events) different processing paths are followed.  Reconstruction results are used to perform data quality monitoring (see an example in \figref{fig:farm_flow_dqm}, right). All steps of the online processing are automatically logged in the database, where each processing step has a dedicated table to avoid deadlocks.  

No major problems appeared throughout the operation period: the whole data processing chain after its installation in December 2016 has worked smoothly, without manual interventions. About \num{7500} (\SI{3}{TB}) files have been pushed from the local storage to CERN EOS and CASTOR and around \num{3500} batch jobs have been executed on the online farm.

\subsection{Grounding and noise in charge readout chain}

Although the FE amplifier ASICs are in the shielded environment provided by the SFT metallic structure, interference from other equipment via ground loops or a noisy ground reference could significantly worsen their noise performance from the design target. The overall grounding scheme of the \three relies on having the cryostat as the ground reference. The low voltage power supplies for the FE analogue electronics are powered via an insulating transformer ensuring that they see no other ground reference. The low voltages are delivered to the SFTs via a distribution box that filters the residual power supply ripple. The box is connected to each SFT with shielded cables thus taking the ground reference of the cryostat. The uTCA crates are put on a separate insulation transformer in order to have the cleanest possible conditions for the low voltage power supplies and avoid any noise injected from the crates. Unfortunately, a similar insulation scheme was not possible for the slow control and HV power supply connections at the time. These equipment see the ground from the building electrical network and thus could interfere with the cryostat ground. 

\begin{figure}[ht!]
 \centering
 \includegraphics[width=.49\textwidth]{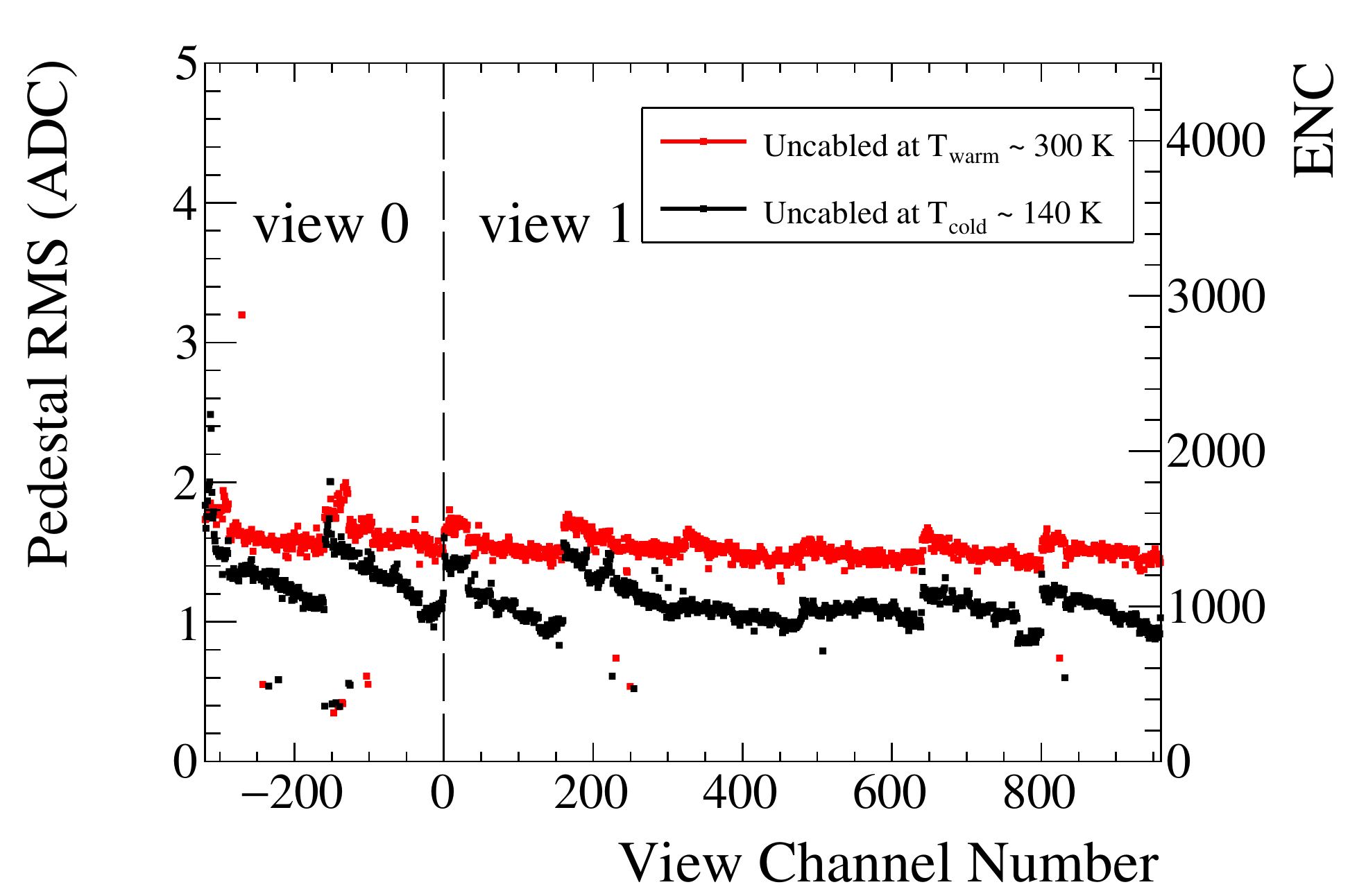}
 \includegraphics[width=.49\textwidth]{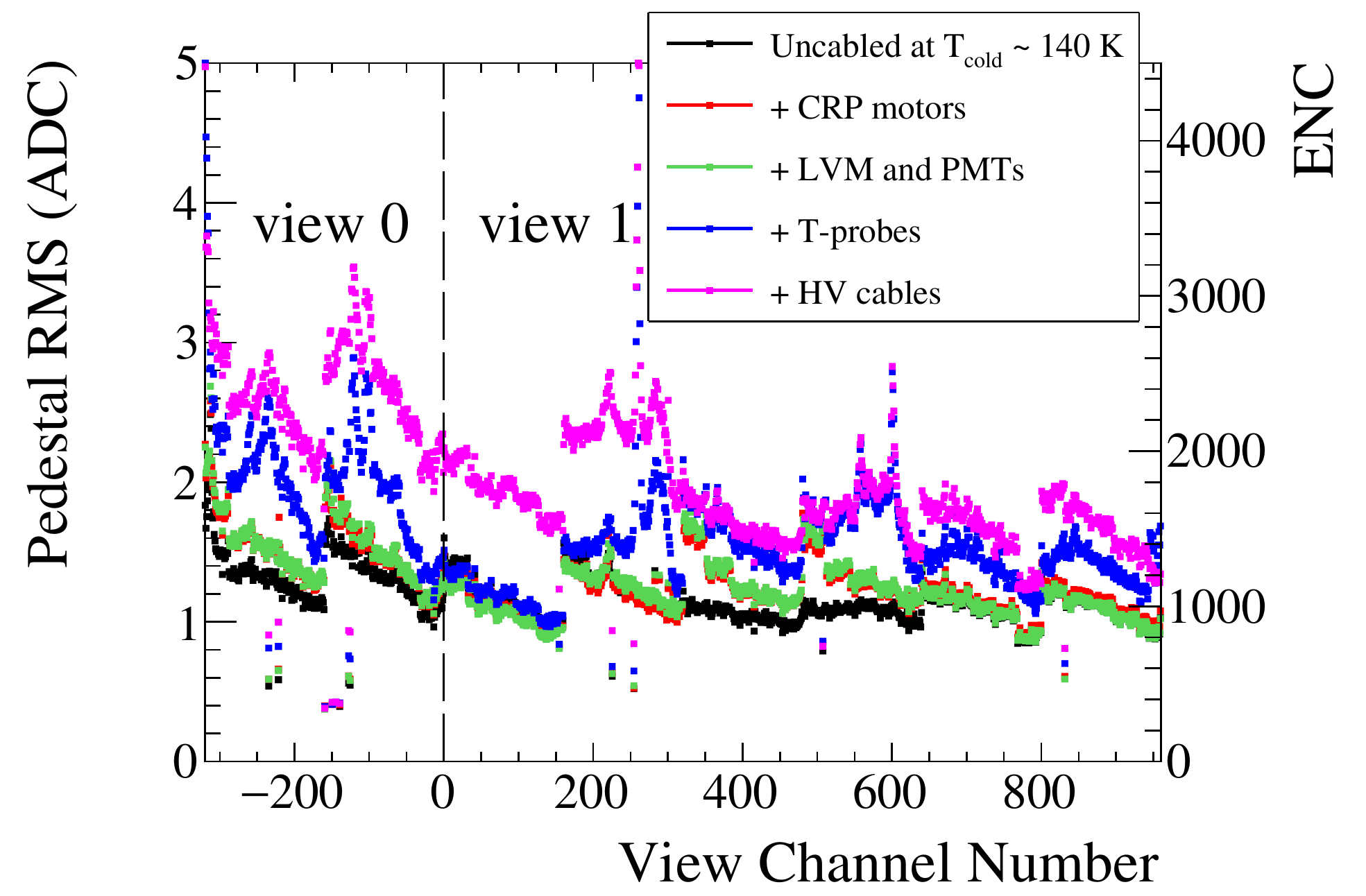}
 \caption{Left: comparison of noise measurements at warm (red points) and cold (black points) with the slow control cables disconnected. 
Right: comparison of noise measurements taken  at cold by progressively reconnecting various equipment. }
 \label{fig:noise_warm_cold} 
\end{figure}

A particular care was taken during the detector commissioning phase to investigate ground loops and to quantify the contribution of different noise sources. We made a complete set of noise measurements by systematically disconnecting the various electrical connections to the cryostat (temperature probes, level meters, cameras and LED, high voltage cables,..) referred to as \textit{slow control} connections and/or by switching devices on and off. The data, taken both at room temperature and with the FE at around \SI{140}{\kelvin} (referred hereafter as at \textit{cold}) after the detector was filled with the liquid argon, indicate that the noise was dominated by the contribution coming from the ground loops in the slow control connections. By disconnecting all of the slow control connections, the intrinsic noise of the analogue electronics could be measured at room temperature. At cold, measurements in exactly the same conditions could not be made as the LAr re-circulation pump was active and the cathode HV connection could not be removed. 

\figref{fig:noise_warm_cold} (left) shows a comparison of the measured RMS fluctuations (noise) in the baseline (pedestal) of each readout channel for data taken at warm (red) and cold (black). These measurements were collected with the full readout chain in place (analogue plus digital part). The channels reading \SI{3}{\meter} (\SI{1}{\meter}) long strip correspond to negative (positive) channel numbering in the plot and the \SI{1}{ADC} count is equivalent to about \SI{900}{e^-} ENC (a MIP signal with $\Geff=1$ is around \SI{9000}{e^-} corresponding to about \SI{10}{ADC}). The pedestal RMS, averaged over all channels, decreases by about 25\% from roughly \SI{1.5}{ADC} (about \SI{1350}{ENC}) to \SI{1.1}{ADC} (about \SI{1000}{ENC}) when the FE analogue cards are operated at cold. The visible step pattern delineates the boundaries between neighbouring anodes; its underlying cause is under investigation. Another interesting feature is that the noise is similar for the channels connected to the \SI{1}{\meter} and \SI{3}{\metre} long strips. Given that the long strips have three times the input capacitance than the shorter ones, the expected noise for these should be larger by a factor of \num{2.1} at warm (\figref{fig:fe-asic-response}). In addition, the noise on the short strips is also lower (\SI{1.5}{ADC} or \SI{1350}{ENC}) than expected with the \SI{160}{\pico\farad} input capacitance (\SI{1.7}{ADC} or \SI{1530}{ENC}). The reason for such behaviour of the noise is under investigation.

The effects of the slow control connections on the noise in the analogue FE electronics from the measurements taken at cold are illustrated in \figref{fig:noise_warm_cold} (right). The plot shows the noise evolution as different slow control sub-systems are progressively connected. The magenta data points in the figure reflect the actual noise when the detector is in operation. On average, they correspond to about \SI{2}{ADC} (\SI{1800}{ENC}) pedestal RMS. The camera system and the cables for the cryostat illumination (LED power) were always kept disconnected throughout the noise measurement campaign as well as during the data taking, since they were found to generate a substantial noise at the level of around \SI{20}{ADC} pedestal RMS.


\section{Detector performance}
\label{sec_operations}

The cryostat is purged, cooled down and filled with liquid argon until the entire field cage is fully immersed as explained in \secref{sec_cryostat-cryo}.
Once the filling is complete the argon recirculation is switched on and maintained throughout the operation. At this point in time the CRP is adjusted parallel to the liquid argon level as described in \secref{sec_sc} and the detector can be operated. The entire operation period ranges from roughly the end of June until mid November 2017, it includes high voltage trials, technical investigations and data taking in various detector configurations. A total of about $5\times10^5$ events were collected with simultaneous charge and light readout. In the first subsection we describe the performance of the cryogenic system. First feedback and results from the detector operation are then discussed.

\subsection{Cryogenic system performance and stability of the liquid argon surface}

The active liquid nitrogen cooling is continuously functional in order to compensate for the heat load. The main source of heat input to the liquid argon bulk comes from the cryostat insulation, the feedthroughs, and liquid recirculation pump. The total heat load (including \SI{300}{\watt} from the pump), estimated from the consumption of liquid nitrogen, is around \SI{1.3}{\kilo\watt}.

\begin{figure}[ht!]
 \centering
 \includegraphics[width=.9\textwidth]{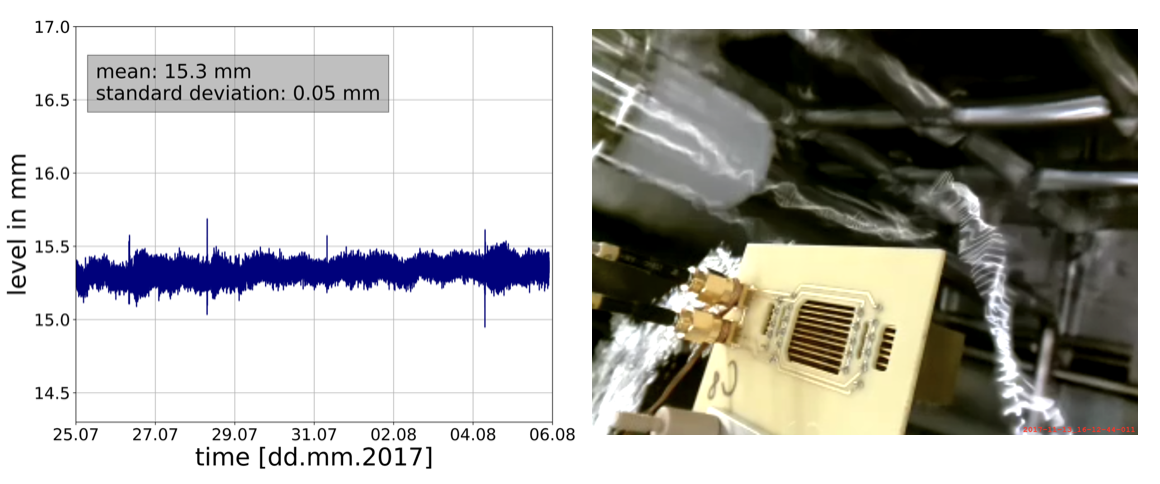}
  \caption{The image on the left shows the liquid argon level recorded by a drift cage level meter during a 10 day period. The image on the right, taken with one of the cryogenic cameras, shows the level meter and the high voltage feedthrough; the liquid level is visible in the picture.}
 \label{fig:level-stab} 
\end{figure}

The cryogenic PLC continuously checks the measurements from one CRP-LM (see \secref{sec_sc_lm}) in order to regulate the flow from the liquid recirculation to maintain a constant liquid level inside the cryostat. \figref{fig:level-stab}-left shows the measurements from one of the DC-LMs taken over a one week period during data taking and illustrates the stability of the liquid level within the \SI{100}{\micro\meter} intrinsic precision of the instrument. The observation of the level is complemented by live feeds from the custom built cryogenic cameras (see \secref{sec_sc}), hereby providing qualitative feedback on the position and flatness of the surface. An example of an image from one of the recordings is shown in \figref{fig:level-stab}-right. 

\begin{figure}[ht!]
 \centering
 \includegraphics[width=\textwidth]{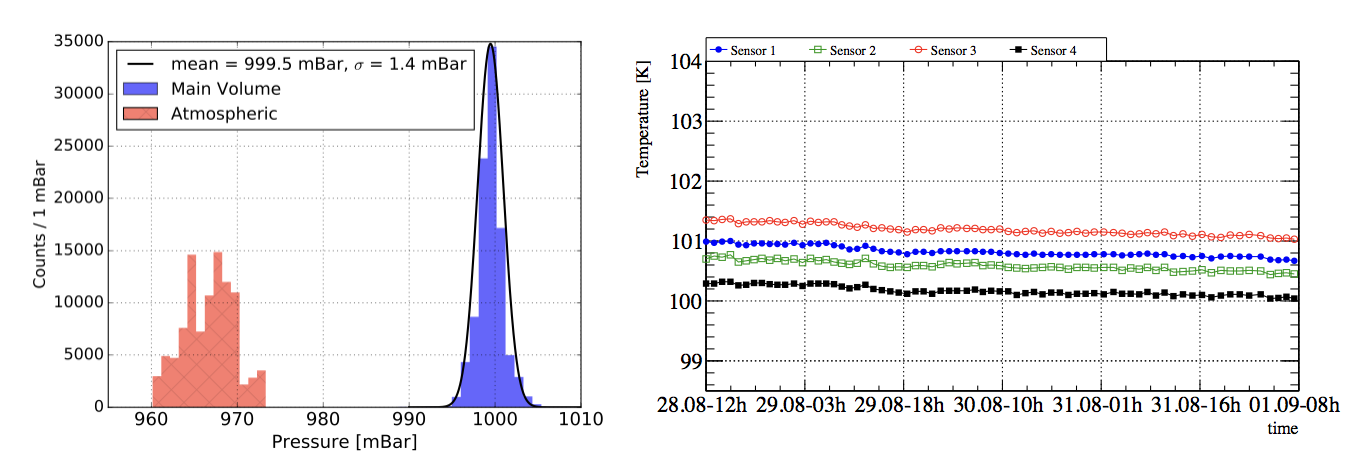}
 \caption{Pressure and temperatures measured over a one week period during data taking. Left: the pressure inside the cryostat main volume compared to that of the insulation and atmospheric. Right: the temperature measured in 4 different points on the CRP at a distance of \SI{2.6}{\cm} above the LEMs.}
 \label{fig:pressure-temperature-cryostat} 
\end{figure}

The pressure inside the cryostat and the temperatures measured \SI{2.6}{\cm} above the LEMs in various points over a week of data taking are shown in \figref{fig:pressure-temperature-cryostat}. The measurements under stable conditions indicate a stable $P_{cryostat}=\SI{999.5 \pm 1.4}{mbar}$ and a uniform gas argon temperature of around \SI{101}{\kelvin} with fluctuations of less than a degree.

Based on the above mentioned performance, this membrane cryostat satisfies a minimal set of fundamental criteria for operating a LAr TPC in dual-phase conditions. Namely: i) the heat input is of $\mathcal{O}(1)$\,kW and can be adequately compensated by active cooling using liquid nitrogen ii) the liquid level remains flat within our requirements for long duration, and iii) the thermodynamic conditions of the gas phase are stable and compatible with the dual-phase operation in pure argon vapour near \SI{87}{\kelvin}.

\subsection{Observation of prompt scintillation and electroluminescence} 
\begin{figure}[ht!]
  \centering 
  \includegraphics[width=\textwidth]{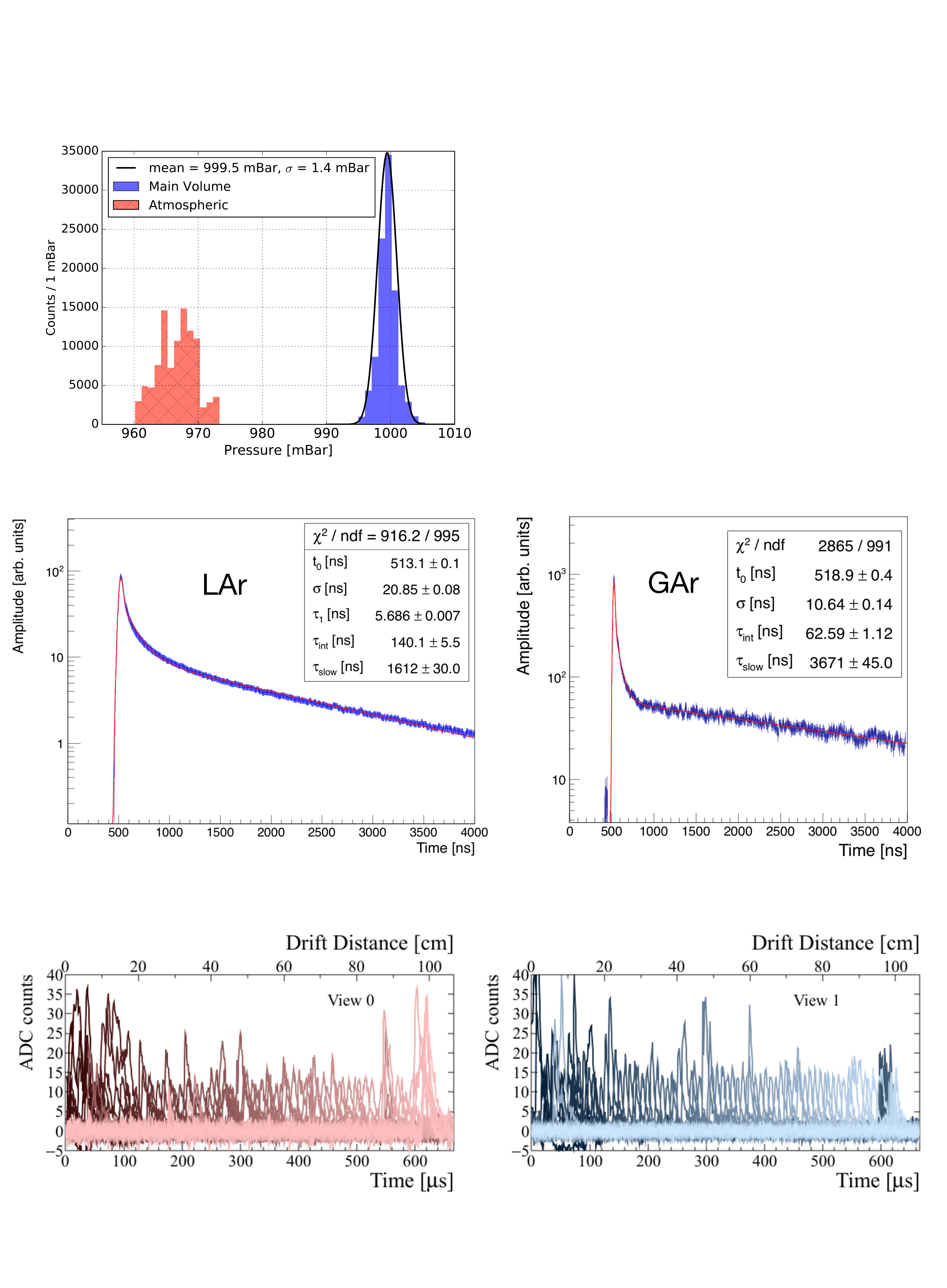}
  \caption{Average digitised signals from one PMT fitted with the function described in the text. The lifetime of the fast component is fixed to \SI{7}{\nano\second} in the fit to the gas data. Vertical scales on both plots are arbitrary and not related.}
     \label{fig:scint_prompt} 
   \end{figure}

We first study the prompt liquid argon scintillation by not applying any drift electric field on the TPC. The average of multiple digitised waveforms from  one photomultiplier (PMT-5, see \tabref{tab:PMT_configuration}) for cosmic ray data is shown in \figref{fig:scint_prompt} as an example. It is fitted with a superposition of three exponential functions convoluted with a Gaussian in order to include both the scintillation time structure and any experimental effects such as the jitter of the trigger device (\SI{\sim5}{\nano\second}), the time of flight of the primary ionising particle and the light propagation time (\SI{\sim10}{\nano\second}). 
The scintillation time profile of liquid argon \cite{0022-3719-11-12-024} consists of a fast de-excitation of both a singlet state (fast component, $\tau_1$) and a triplet state (slow component, $\tau_{slow}$), whose value depends on the purity of the liquid. A third,\textit{intermediate} , component ($\tau_{int}$) is needed in order to improve the agreement with the data.
The intermediate component has been observed at different setups ranging from tenths of \si{\nano\second} \cite{PhysRevB.27.5279,ACCIARRI2009169,Acciarri:2008kv} to values above \SI{100}{\nano\second} \cite{Whittington:2015gov} in good agreement with present results. The origin and the $\tau_{int}$ value of the intermediate component is still under discussion \cite{Acciarri:2008kv,Segreto:2014aia}. For comparison, in \figref{fig:scint_prompt}-right we show the waveform  from the same PMT acquired in \SI{215}{\kelvin} gas argon prior to filling. The preliminary values obtained for the triplet lifetimes are \SI{1.61\pm0.03}{\micro\second} in liquid and  \SI{3.67\pm0.05}{\micro\second} in gas. 

\begin{figure}[ht!]
  \centering
  \includegraphics[width=.8\textwidth]{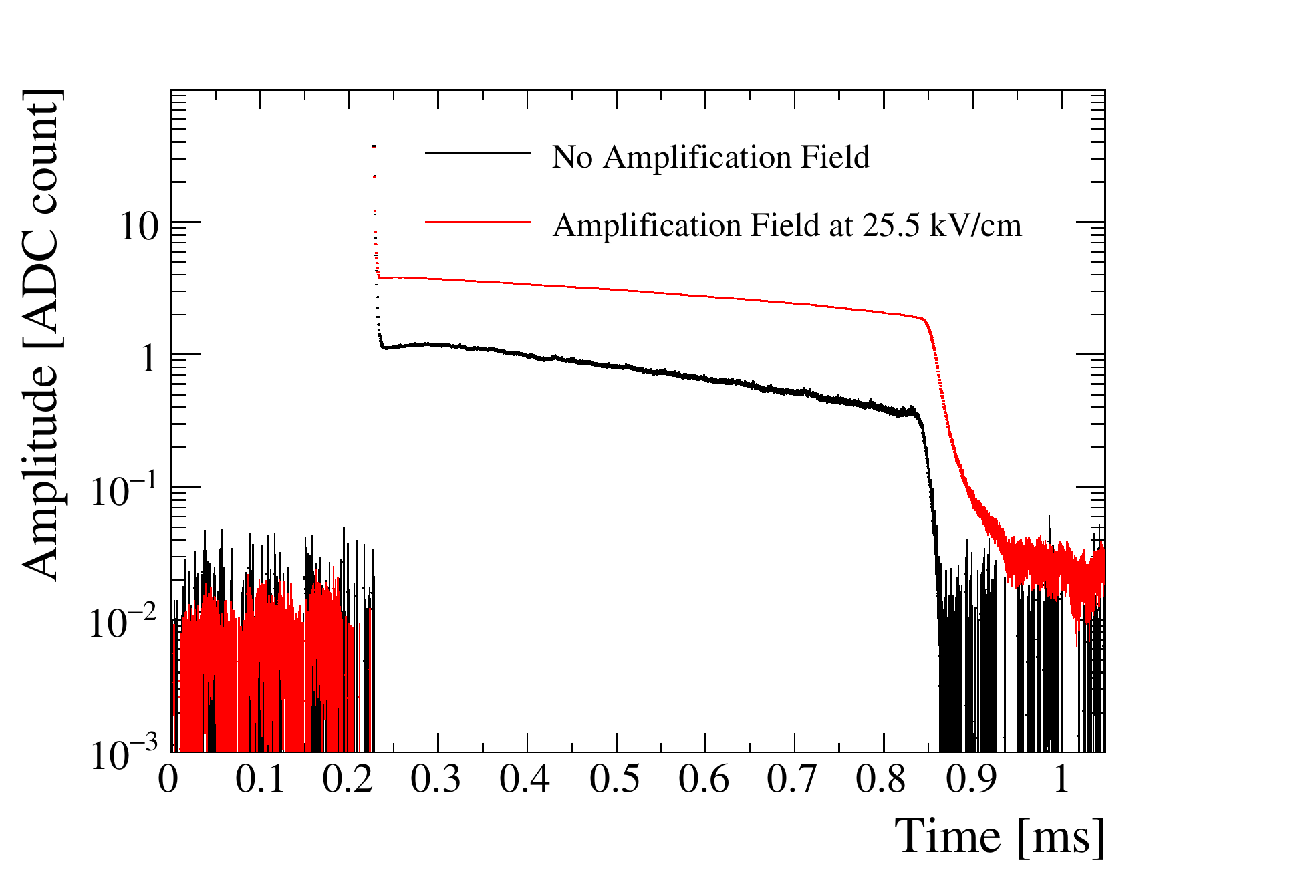}
  \caption{Sum of averaged waveforms from the 3 negative based PMTs showing the prompt and proportional scintillation components. The prominent peaks centred at around \SI{200}{\micro\second} correspond to the prompt scintillation in liquid (S1) and the continua extending for \SI{\sim600}{\micro\second} after are from the proportional scintillation in gas (S2). The higher S2 yield is clearly visible when the LEMs are polarised (red curve).}
  \label{fig:s1_s2_wf} 
\end{figure}

By operating the TPC with a drift and extraction field, we are able to observe the proportional scintillation of the ionising charges in the argon vapour. The proportional scintillation, also referred to as electroluminescence or S2, is expected to take place in the gas in regions where the electric field is above \SI{\sim2}{\kilo\volt/\cm} \cite{Monteiro:2008zz}, which is the case in the extraction gap but also inside the LEMs and in the induction gap with electric fields of the order of \SI{30}{\kilo\volt/\cm} and \SI{5}{\kilo\volt/\cm}.
\figref{fig:s1_s2_wf} shows the sum averaged waveforms from \num{5} PMTs acquired with the TPC operated at a drift field of \SI{0.5}{\kilo\volt/\cm} (i.e., a drift velocity of \SI{\sim1.6}{\mm/\micro\second} \cite{Gonzalez-Diaz:2017gxo}) and extraction fields in liquid above \SI{2}{\kilo\volt/\cm} (corresponding to \SI{\geq3}{\kilo\volt/\cm} in GAr with the liquid in the middle of the extraction gap). The black curve shows the measurements taken without LEM field, while the red curve corresponds to those acquired with the LEM field of \SI{26}{\kilo\volt/\cm}.
The first peaks centred at around \SI{200}{\micro\second} are from the primary scintillation in liquid and the flat continuum is from the proportional scintillation in the gas. The proportional scintillation signal lasts until the farthest electrons reach the extraction region. The maximum time extension of the secondary light signal is therefore comparable to the maximum electron drift time. As can be seen from the figure, the secondary light contribution extends for about 
\SI{600}{\micro\second}, which is the expected time for an electron drifting over \SI{1}{\metre} in a field of \SI{0.5}{\kilo\volt/\cm}. The almost flat S2 continuum qualitatively indicates that the liquid argon purity is sufficiently high so that there is no substantial attenuation in the extracted ionising charge over the entire drift distance. The effect of the larger electric field inside the LEM is clearly visible and contributes as an enhancement of the S2 signal, as well as to an increase of the lifetime. Nevertheless, at those electric field settings, the amplitude of the S2 continuum remains several orders of magnitude lower than that of the primary scintillation peak. 
Since oxygen contamination would reduce the light signal because of quenching, the liquid argon purity can be verified by studying the lifetime of the triplet state in liquid. In the liquid phase, the measured value of 1.6 $\mu$s is compatible with oxygen equivalent impurities better than \SI{\sim1}{ppm} according to estimations based on previous experimental data \cite{Acciarri:2008kx}.

\subsection{TPC charge readout and operational feedback}\label{subsec_op_feedback}
Charge readout events can be triggered by either the prompt scintillation in liquid argon, referred to as \textit{PMT trigger}, or on crossing muons by using the CRT panels (\textit{CRT trigger}) as described in \secref{sec:trigger_conditions}. From the $5\times10^5$ events collected by the charge readout DAQ, 15\%  are CRT triggered events and the remaining 85\% are PMT triggered events.

The nominal settings for the electric fields and electrode potentials shown in \figref{fig:DP-principle}  correspond to the operation of the TPC with an effective gain \Geffinfty$\approx 20$ after complete charging up of the LEMs (\secref{ssec_overview_dpconcept}) according to the measurements with the small dual-phase TPCs\cite{Badertscher:2008rf, Badertscher:2009av, Badertscher:2010zg,Badertscher:2013wm,devis-thesis,filippo-thesis}. The operation of the \three focuses essentially on two goals: i) explore the maximally achievable electric fields for each stage (drift, extraction, amplification and induction) by disregarding the TPC imaging performance as a whole and ii) collect data with the high voltage configuration that maximises the stability, extraction and collection efficiencies, and LEM gain in order to characterise the detector response. A first feedback from both above cited points is provided in the next two sections.

\subsubsection{Electric field settings}
The electric fields in the drift, extraction, amplification, and induction regions are scanned individually by reducing the other field values in order to be compatible with the overall limitation due to the grid HV breakdown. The method consists in increasing step by step the electric field in a specific region up to a limit imposed by a high voltage breakdown. At each step, the setting is kept during about half an hour in average and charge and scintillation data are acquired. While doing so the electric fields in the other regions are reduced to a minimal value, generally below their nominal settings, meaning that the overall electrical transparency and hence the charge imaging performance of the TPC is not optimised. The main reason for doing so arises from a limitation on the potential at which the extraction grid can be operated stably, with details explained below.  The range of scanned electric fields are reported in the left column of \tabref{tab:HV_configurations}. In this column the maximal reported value corresponds to the setting at which data are stably acquired for the duration of the run without any discharges. The details on the scans in each specific region are discussed below.
\begin{table}[ht!]
\begin{flushleft}
\begin{tabular}{p{.25\textwidth}p{.2\textwidth}p{.25\textwidth}p{.2\textwidth}}
\hline
\hline
 Electric field (kV/cm) & Scanned range & Data taking settings & \textit{Nominal} \\
\hline
Induction  & $1-5$ & 1.5 & \textit{5}\\
Amplification\tablefootnote{The four corner LEMs in the \three are operated at 24 kV/cm}  & $23-31$ & 28 & \textit{33}\\
Extraction (in liquid) & $0.6-2.5$ &1.7&\textit{2}\\
Drift & $0.18-0.7$ & 0.5 & \textit{0.5}\\
\arrayrulecolor{black}
\hline
\hline
\end{tabular}
\caption{Electric field settings explored during and data taking with the \three TPC (see text for more details). The values quoted as nominal correspond to \Geffinfty of 20 as achieved in the dual-phase TPC with the \SI[product-units=power]{10x10}{\cm} readout area \cite{Cantini:2013yba, Cantini:2014xza}.}
\label{tab:HV_configurations}
\end{flushleft}
\end{table} 

\begin{itemize}

\item Drift field: a stable drift field of at least \SI{500}{\volt/\cm},  corresponding to a cathode voltage of \SI{-56.5}{\kilo\volt} (see \figref{fig:DP-principle}), is sustained during the $\sim$4 months of the \three operating period. No high voltage discharges are observed. This demonstrates the robustness of the entire drift cage up to its design operating high voltage and also validates the quality of the feedthrough and of the Heinzinger power supply for long duration operations at \SI{-56.5}{\kilo\volt}. At one point in time the high voltage on the cathode was lowered further to \SI{-75}{\kilo\volt} which corresponds to drift field of \SI{\sim700}{\volt/\cm}. Data was collected at this increased drift field for about half an hour until a high voltage discharge was recorded on the power supply. The cathode was then polarised back to \SI{-56.5}{\kilo\volt} and we did not perform a second attempt at increasing the cathode high voltage above this value.

\item Extraction field: as specified in \figref{fig:DP-principle} operating the TPC at an extraction field in
the liquid of \SI{2}{\kilo\volt/\cm} while simultaneously maintaining the amplification and induction fields at \SI{33}{\kilo\volt/\cm} and \SI{5}{\kilo\volt/\cm} respectively requires biasing the extraction grid at a nominal voltage of \SI{-6.8}{\kilo\volt}. However, due to a technical issue, the grid could only be operated with stable high voltages down to \SI{-5}{\kilo\volt}: below this value high
voltage discharges occurred. Investigations point to a combination of a faulty high
voltage contact on the extraction grid and an un-tensed group of wires. The latter
is supported by the presence of a short circuit between the grid and two of the twelve
bottom LEM electrodes observed when the liquid level is increased. The visual inspection inside the cryostat after the end of data-taking  (at room temperature) confirmed an issue on the tensioning system holding a group of 32 one meter long wires located under those specific LEMs. The \SI{-5}{\kilo\volt} limit on the grid high voltage
compromises the ability to operate the \three TPC with induction, amplification
and extraction at their nominal values simultaneously\footnote{During the electric field scans when one field is increased the others must be adjusted to ensure the extraction grid remains at a potential above \SI{-5}{\kilo\volt}}. Despite this limitation, the extraction field is nonetheless scanned up to a field of \SI{2.5}{\kilo\volt/\cm} in liquid, where the extraction efficiency starts to plateau to a maximal value \cite{Gushchin:1982}. We verified that when not immersed in liquid the extraction grid can only sustain a voltage down to \SI{-2}{\kilo\volt}, below this value discharges appear regularly. The fact that it can be operated at \SI{-5}{\kilo\volt} during data taking is a clear indication that the extraction grid is completely submerged. The $\mathcal{O}$(\si{\mm}) sagitta in the CRP frame, measured during the cryogenic test (\secref{sec_tpc_crp}), is not expected to lead to substantial variations in the effective gain given the small impact that the changes in the grid-LEM distance have on the extraction efficiency (see \secref{sec_overview}). The uniformity of the gain over the entire CRP area is nevertheless an aspect that will be verified during a detailed analysis of the data.

\item Amplification field: 
a maximal field applied across the LEM electrodes of \SI{31}{\kilo\volt/\cm} is achieved by setting the LEM top and bottom electrode at \SI{-0.2}{\kilo\volt} and \SI{-3.3}{\kilo\volt} respectively while keeping the extraction grid at \SI{-5}{\kilo\volt}. The four \fifty LEMs located in the corners must however be operated at  electric fields below \SI{27}{\kilo\volt/\cm}. The reason for a drop in performance of those corner LEMs needs to be investigated. When a discharge occurs across one \fifty LEM a peak in current lasting 1 or 2 seconds and about \SI{100}{\pico\ampere} in amplitude is occasionally observed on the HV channels of neighbouring LEMs. We interpret this as a current induced from  capacitive couplings between the LEM plane and the extraction grid (see \tabref{tab:CRP-capa}). 

Being all LEM plates polarised, discharges occurring in a LEM module do not propagate to the neighbours, so high voltage discharges between \fifty LEMs plates have never been observed.
Before decommissioning the detector, a set of tests are carried out to evaluate the performance of each LEM powered individually without any extraction field. The extraction grid is left at a non-defined floating potential by disconnecting the high voltage supply cable. During these tests we observe that six LEMs are able to sustain \SI{32}{\kilo\volt/\cm} field during one hour.

\item Induction field: 
the anode-LEM field is scanned from \num{1} to \SI{5}{\kilo\volt/\cm} on the entire surface of the readout without any discharges occurring. The induction field is further explored up to \SI{10}{\kilo\volt/\cm} with the grid high voltage contact disconnected and without applying any amplification field.
\end{itemize} 

We do not observe any increase of electromagnetic pickup noise on the charge readout when all the above-cited high voltage are applied on the electrodes. The noise is instead dominated by the physical connections of the various cables as detailed in \secref{sec_elec}.

In summary the main limitations on the high voltage settings that prevent us from operating the entire dual-phase TPC at its nominal field settings and quantify the charging up of the LEM dielectric are i) a technical issue that limits the operating potential of the extraction grid to \SI{-5.0}{\kilo\volt} and ii) the fact that the \fifty LEMs themselves cannot be operated stably at their nominal field of \SI{33}{\kilo\volt/\cm}. Based on the above-cited limitations a compromise is found in order to operate the TPC with the best possible settings that provide a large extraction efficiency and some amplification inside the LEMs. Those settings are reported in the centre column of \tabref{tab:HV_configurations}. The top, bottom LEM electrodes and the extraction grid are biased to \num{-0.3}, \num{-3.1} and \SI{-5}{\kilo\volt} providing an induction, amplification and extraction field in the liquid of \num{1.5}, \num{28} and \SI{1.7}{\kilo\volt/\cm}, respectively. Data are acquired for about four hours at those settings collecting more than \num{30000} cosmic events. The run was interrupted by a high voltage discharge of the extraction grid. The first evaluation of the performance based on this data and events are discussed in the next section.

\subsubsection{First look at data: cosmic muons with gain}
\begin{figure}[ht!]
\centering
\includegraphics[width=.7\textwidth]{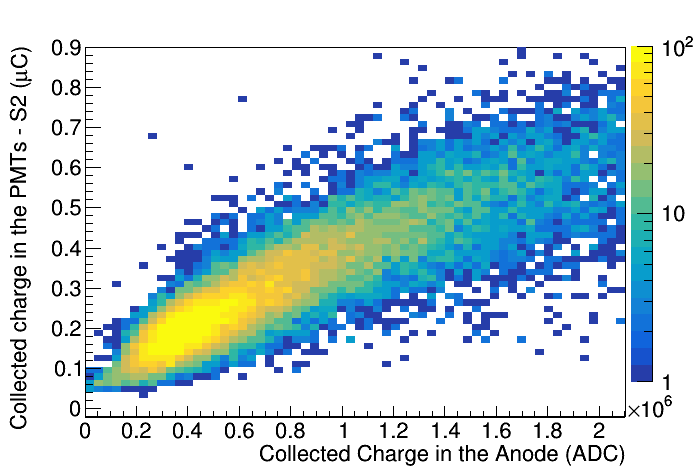}
\caption{The S2 light collected by the PMT array as function of the integrated charge collected on the anode. The data are collected with the field settings described in the centre column of \tabref{tab:HV_configurations}.}
\label{fig:charge_light_corr}
\end{figure}

\figref{fig:charge_light_corr} shows the correlation between the amount of collected charge (abscissa) versus the amount of detected S2 light (ordinate) with the PMT array after the contribution of the prompt S1 signal is subtracted in the analysis. These data were acquired with the TPC HV settings described in the centre column of \tabref{tab:HV_configurations}. The evident correlation between two signals demonstrates the sensitivity of the photon detection system to the electroluminescence in the vapour phase and the integrated amount of the S2 light could serve to provide additional information on the total amount of the deposited charge.

\begin{figure}[ht!] 
\centering
  \includegraphics[width=0.9\textwidth]
  {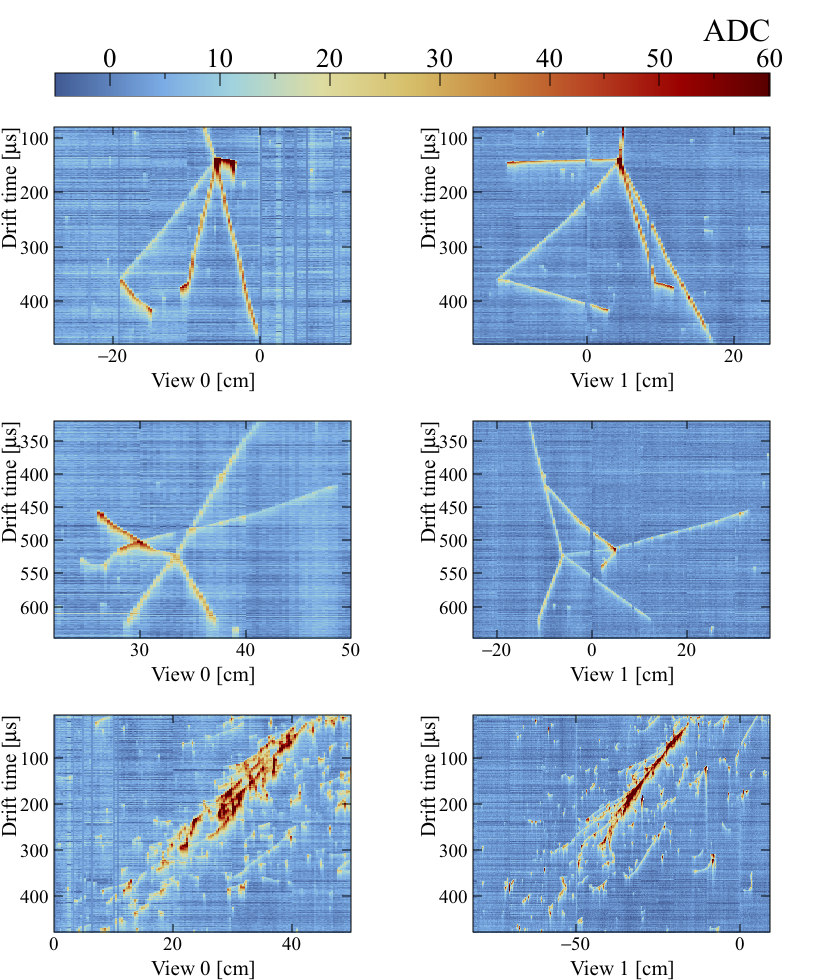}
  \caption{Cosmic ray events recorded in the \three operated at the electric fields settings explained in the text. The windows are cropped to focus on the area of interest. The zoom on view 0 (left) and view 1 (right) of the same event are shown on each row. The three events show from top to bottom: two hadronic showers and an electromagnetic shower candidate.}\label{fig:ED-interactions-zoomed} 
\end{figure}

In \figref{fig:ED-interactions-zoomed} we show a set of raw events collected with the TPC electric field settings described in the centre column of \tabref{tab:HV_configurations} and triggered with the PMTs. 
The plots show the usual representation of events with the drift time as a function of the coordinates of both views and the colour scale proportional to the signal amplitude for each sample. The channel mapping convention of the TPC is specified in \figref{fig:311-coordinates}. The windows are zoomed around the interaction region and the events are displayed without the application of noise removal algorithms. Even though the TPC is not operated at the nominal electric field settings, one can already observe the good signal to noise ratio for both collection views and the fine spatial resolution which allows to clearly distinguish the vertices and secondary particles. 

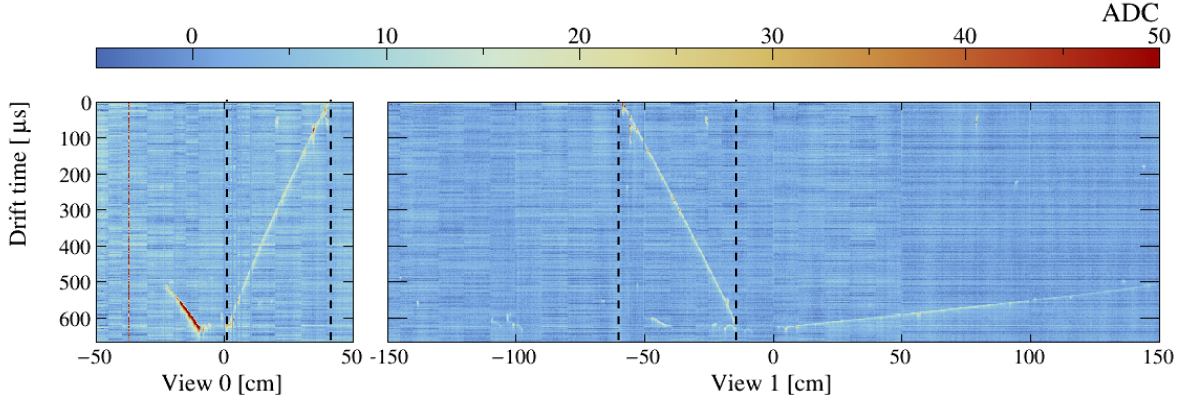
\begin{figure}[ht!]
  \centering
\begin{tikzpicture}
    \node[anchor=south west,inner sep=0] (image) at (0,0) {\includegraphics[width=\textwidth]{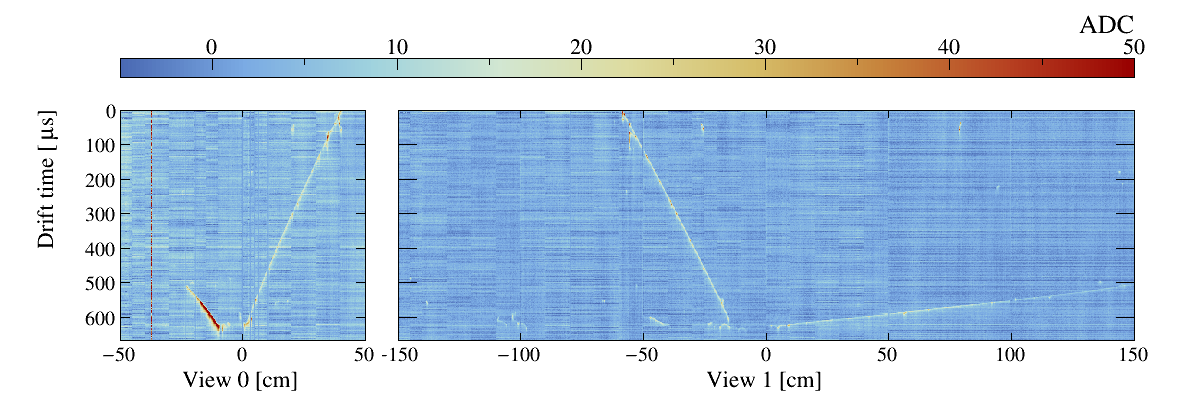}};
    \begin{scope}[x={(image.south east)},y={(image.north west)}]
        \draw[black,dashed, thick] (0.205,0.14) -- (0.205,0.726);
        \draw[black,dashed, thick] (0.288,0.14) -- (0.288,0.726);
        \draw[black,dashed, thick] (0.518,0.14) -- (0.518,0.726);
        \draw[black,dashed, thick] (0.612,0.14) -- (0.612,0.726);
    \end{scope}
\end{tikzpicture}
  \caption{Through going muon crossing the entire drift volume. Dashed lines define the region displayed in \figref{fig:through-muon-waveform}. An off-time track near the cathode is also visible.}\label{fig:ED-through-muon} 
\end{figure}

\begin{figure}[ht!]
\includegraphics[width=\textwidth]{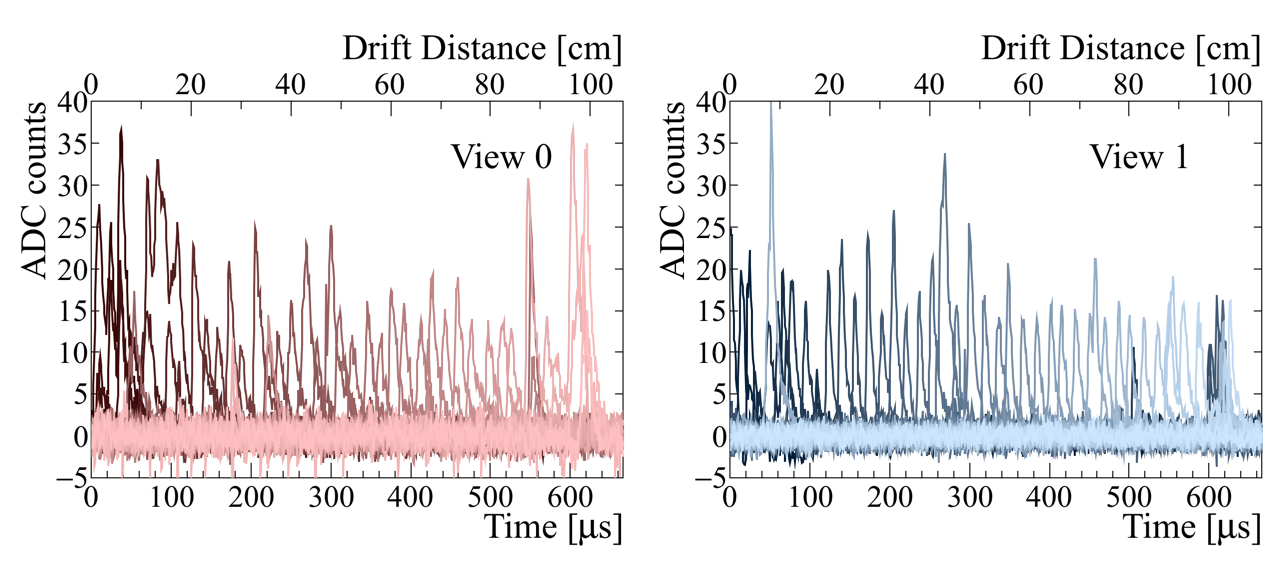}
\caption{Waveforms of the through going muon of \figref{fig:ED-through-muon}. Each channel is represented with a specific colour. The corresponding drift distance is computed assuming an electron velocity of $v=\SI{0.16}{\cm/\micro\second}$ \cite{Li:2015rqa}. For clarity, only one in three  waveforms are displayed.}
    \label{fig:through-muon-waveform}
\end{figure}

\begin{figure}[!h]
\centering 
\includegraphics[width=\textwidth]
  {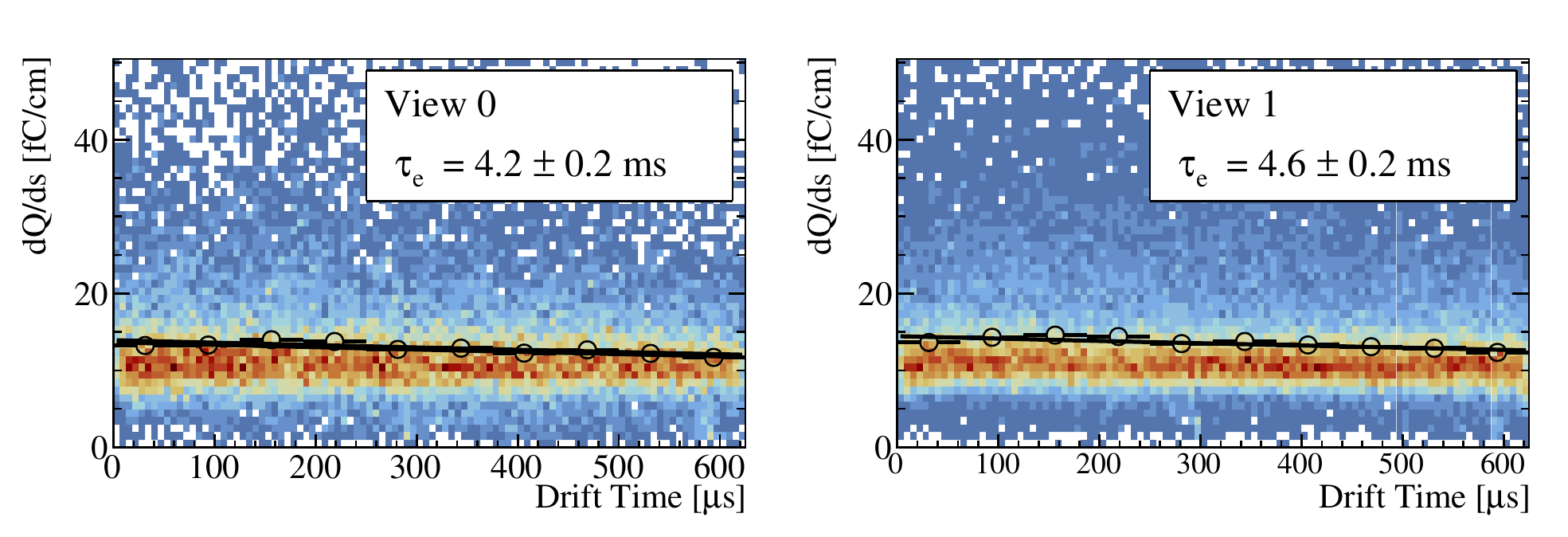}
  \caption{The scatter plot shows the collected charge per unit length in each view as a function of drift time for through-going muons. The collected charge per unit length is averaged in slices of \SI{10}{\cm} drift distance (corresponding to \SI{62.5}{\micro\second} drift time) and represented with the black markers. This distribution is fitted with an exponential function, overlaid in solid black line, the extracted electron lifetime is presented.
  }\label{fig:dqds_drift_fit}
\end{figure}

In order to provide a more quantitative estimation of the detector charge imaging performance at the above-mentioned electric field settings, a sample of cosmic muons that cross the chamber is selected and analysed. Such "through going" muons are minimum ionising particles (MIPs) depositing a known amount of energy of about 2.1 MeV/cm and are therefore used to estimate the free electron lifetime, the amplification in the LEMs and the charge sharing between views. An event showing two muon track candidates is shown in \figref{fig:ED-through-muon}. All the channels from both views are displayed. A muon track crossing the TPC from top to bottom is selected in between the dotted lines and the corresponding digitised waveforms showing the signal amplitudes are displayed in \figref{fig:through-muon-waveform}.
Reconstruction of this specific track in 3D (see later in the text for more details) indicates that it crosses the TPC at an azimuthal angle of $\phi \approx45^\circ$ and therefore deposits similar amount of charge per unit length in each view. From \figref{fig:through-muon-waveform} we observe that the waveforms exhibit near identical amplitudes on both views which are well distinguishable from the noise. One powerful property of the dual-phase TPC, and qualitatively verified here for the first time on readouts at the m$^2$ scale, is indeed the ability to operate with two collection views featuring 50:50 charge sharing. This key property allows to extract physical signals with relative ease. The integral of the signal provides a direct measurement of the local energy deposition of the ionisation track in the liquid argon medium. The clustered hits are fitted with a straight line and the $z$ coordinate of the 3D track is computed by matching the end point drift times of the tracks from both views. More details on the entire procedure used to reconstruct those straight muon tracks can be found in \cite{devis-thesis}. The 3D reconstruction allows to retrieve the length of the track on each strip of view 0 and view 1 ($\Delta s_0$ and $\Delta s_1$), along with the charge collected on the corresponding channels, $\Delta Q_0$ and $\Delta Q_1$. The ratios $\Delta Q_0/\Delta s_0$ and $\Delta Q_1/\Delta s_1$ which are proportional to the energy locally deposited by the track in liquid argon per unit length, are the relevant quantities used to characterise the performance of the \three in terms of purity and effective gain. The distributions of the collected charge per unit length as a function of the electron drift time ($t_{drift}$) for both views are shown in \figref{fig:dqds_drift_fit}.

The black markers indicate the average of the distributions in slices of \SI{62.5}{\micro\second} drift time, which approximately corresponds to a \SI{10}{\cm} drift distance. The mean electron lifetime $\tau_e$ is retrieved by fitting the average energy loss as a function of the drift time with the exponential law $e^{-t_{drift}/\tau_e}$, that accounts for charge losses due to electron attachment to impurities. The fits performed on both views, and shown in black on the figure, consistently indicate an electron lifetime of better than \SI{4}{\milli\second}, which corresponds to an oxygen equivalent impurity of less than \SI{75}{ppt}. This analysis does not account for possible space-charge effects, that were shown to be significant in \cite{Meddage:2017lxo}; however in our configuration they are supposed to be much smaller and their contribution can be neglected.

\begin{figure}[ht!]
  \centering
  \includegraphics[width=\textwidth]{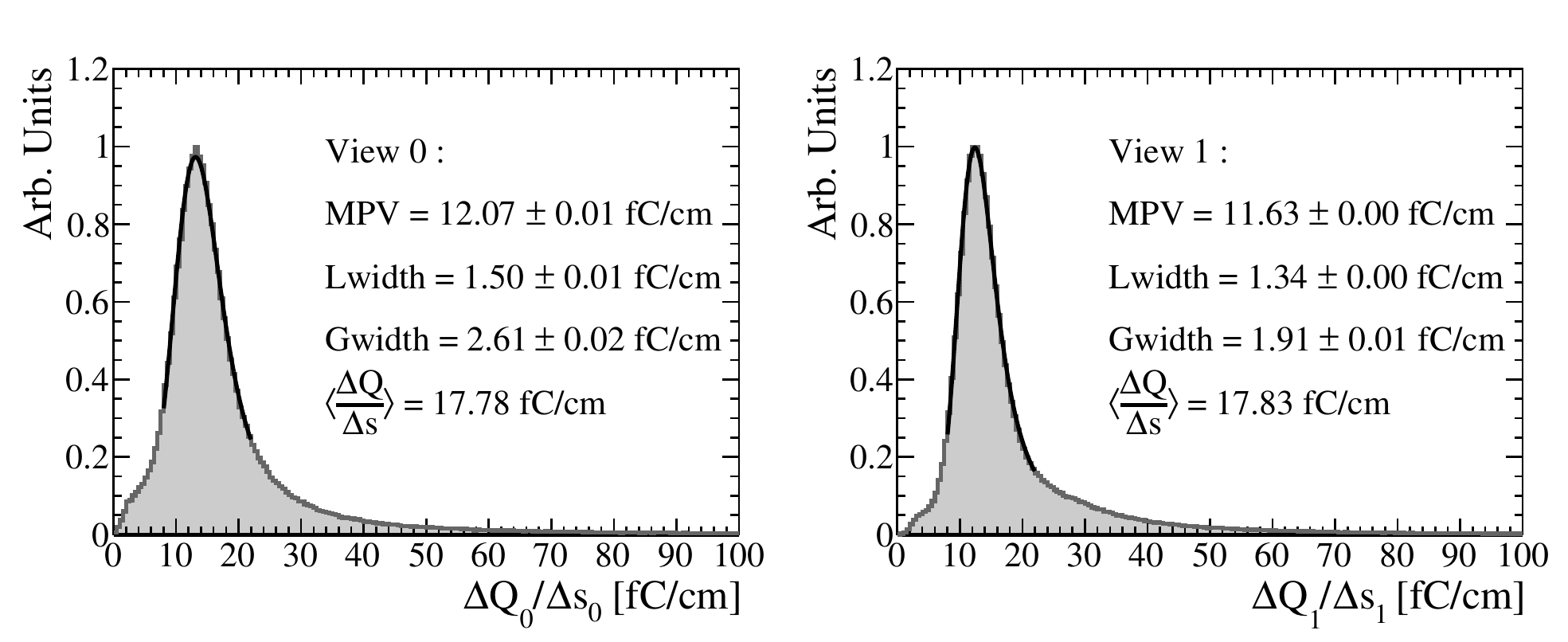}
  \caption{The $\Delta Q_0/\Delta s_0$ and $\Delta Q_1/\Delta s_1$ distributions for both views for selected straight tracks from the run with electric field settings defined in the centre column of \tabref{tab:HV_configurations}. A Landau function convoluted with a Gaussian is fitted to the distributions, the resulting Most Probable Value (MPV) and widths of the Landau (Lwidth) and Gaussian (Gwidth) components are indicated.}
     \label{fig:dqds} 
  \end{figure}

\figref{fig:dqds} shows the $\Delta Q/\Delta s$ distributions for both views corrected for the electron lifetime. Their shapes are described by a Landau function with a Gaussian smearing as expected from the fluctuations of the collected charge per unit length.  Those distributions allow to estimate the effective gain of the chamber which is defined as the sum of the charge collected per unit length in each view divided by the deposited charge of a MIP:
\begin{equation}
    G_{eff}=\frac{\langle\Delta Q_0/\Delta s_0\rangle +\langle\Delta Q_1/\Delta s_1\rangle}{\langle\Delta Q/\Delta s\rangle_\text{MIP}}
\end{equation}
For a MIP track, the average charge deposited per unit length predicted by the Bethe-Bloch formula and accounting for electron-ion recombination is about \SI{10}{\femto\coulomb/\cm}.
The model used for recombination is the modified Birk's law, as discussed in \cite{Amoruso:2004dy}.
By taking the mean of the distributions in \figref{fig:dqds} we obtain a preliminary estimation of $G_{eff}\approx 3.5$, a value obtained before complete charging up of the LEM modules.
We remind the reader that \Geff includes the intrinsic multiplication of electrons inside the LEM holes as well as losses at the liquid-gas interface and on the LEM electrodes (see \secref{sec_intro}). An effective gain larger than one therefore demonstrates Townsend multiplication of electrons inside the \fifty LEMs and operation of the \three TPC with gain. Due to the technical limitations on the CRP high voltage explained in \secref{subsec_op_feedback} the TPC could not be operated stably at the target value of $\Geffinfty\approx20$. Those limitations also prevented us from collecting data continuously during many days in order to quantify the evolution of the effective gain with time and study the impact of the charging up of the LEM dielectric.
 
In addition to verifying the purity of the liquid argon and effective gain of the chamber it is also evident from the distributions in \figref{fig:dqds} that there is no large asymmetry in the charge collected by each view.



\section{Conclusions}
\label{sec_conclusions}
 In this paper we have provided a detailed technical overview and described some aspects of the performance of a large dual-phase liquid argon Time Projection Chamber with a fiducial volume of \three corresponding to more than \SI{4}{\tonne} active detector mass. Its construction and operation represents an important milestone towards the realisation of kilo-tonne LAr TPCs. We have shown that the newly employed membrane cryostat provides a stable cryogenic environment with a flat liquid surface allowing for charge extraction over an area of \SI{3}{\meter\squared}. An excellent liquid argon purity was also achieved with a corresponding electron lifetime of around \SI{4}{\milli\second}. While data collection over brief time periods was possible, the high voltage instabilities in the operation of the extraction grid prevented a proper study of the detector long-term stability and performance. The maximum stable amplification field that could be reached in the LEMs was also lower than envisioned for large dual-phase LAr TPCs. For the next dual-phase LAr TPC prototype, ProtoDUNE-DP, the adopted designs of the extraction grid mounting and its HV connection and LEMs are expected to address these issues. A first look at the data collected at an effective gain of around 3 before complete charging up of the LEMs nevertheless demonstrates the high quality of the dual-phase TPC imaging. We have also demonstrated the charge readout with two collection planes with strips of up to \SI{3}{\meter} length and almost equalised charge collection.  We were able to detect the prompt and electroluminescence scintillation and observe a correlation between electroluminescence in the gas and collected charge. This work represents an important breakthrough in particle detection with liquid argon TPCs and paves the road for deploying this technology on a larger scale.


\section*{Acknowledgements}
This work would not have been possible without the support of the Swiss National Science Foundation, Switzerland; CEA and CNRS/IN2P3, France; KEK, Japan; the Ministerio de Econom\'ia, Industria y Competitividad (SEIDI-MINECO) under grants FPA2016-77347-C2, SEV-2016-0588 and MdM-2015-0509, Spain; the PNCDI III-2015-2020, Programme 5, Module CERN-RO, under contract 9/2017, Romania. This project has received funding from the European Union’s Horizon 2020 Research and Innovation program under Grant Agreement no. 654168. We  acknowledge the important role of CERN in setting up the cryogenic system and the general experimental infrastructure  as well as its continued support throughout the detector operating phase. The authors are also grateful to the LABEX Lyon Institute of Origins (ANR-10-LABX-0066) of the Universit\'e de Lyon for its financial support within the program "Investissements d'Avenir" (ANR-11-IDEX-0007) of the French government operated by the National Research Agency (ANR). We thank the IN2P3 Computing Centre (CC-IN2P3) for providing computing resources for simulations and analyses. We thank the IT division at CERN for providing space and computing resources to store and analyse the data; we are also indebted for having provided the storage servers and the CPUs used to set up the online storage and processing  farm, and the continuous support given during its commissioning and  operations.



\bibliographystyle{elsarticle-num}

\begingroup
    \setlength{\bibsep}{10pt}
    \bibliography{bibfile}
\endgroup

\end{document}